\documentclass[reprint,pra,twocolumn,showpacs,superscriptaddress,aps,longbibliography]{revtex4-2}

\usepackage{graphicx}
\usepackage{epstopdf}

\usepackage[T1]{fontenc}
\usepackage[applemac]{inputenc}
\usepackage{lmodern}
\usepackage[english]{babel}

\usepackage{ae}
\usepackage{siunitx}

\usepackage{amsmath,amssymb,natbib,bm}
\usepackage{psfrag}
\usepackage{subfigure}
\usepackage{amsthm}
\usepackage{algorithm}
\usepackage{algpseudocode}

\usepackage[colorlinks]{hyperref}
\hypersetup{%
        plainpages=true,
        breaklinks=true,
        hypertexnames=false,
        pageanchor=true,
        colorlinks=true,
        linkcolor={blue},
        citecolor={blue},
        urlcolor={blue},
        anchorcolor={black}
      }

\usepackage{mleftright} 

\newcommand{\figref}[1]{\mbox{Fig.~\ref{#1}}}

\newcommand{\secref}[1]{\mbox{Sec.~\ref{#1}}}

\newcommand{\appref}[1]{\mbox{Appendix~\ref{#1}}}
\renewcommand{\eqref}[1]{\mbox{Eq.~(\ref{#1})}}

\newcommand{\figpanel}[2]{Fig.~\hyperref[#1]{\ref*{#1}(#2)}}
\newcommand{\figpanels}[3]{Fig.~\hyperref[#1]{\ref*{#1}(#2)--(#3)}}
\newcommand{\figpanelNoPrefix}[2]{\hyperref[#1]{\ref*{#1}(#2)}}

\newcommand{\bra}[1]{\langle #1|}
\newcommand{\ket}[1]{|#1\rangle}

\newcommand{\be}{\begin{equation}}
\newcommand{\ee}{\end{equation}}
\newcommand{\bea}{\begin{eqnarray}}
\newcommand{\eea}{\end{eqnarray}}

\begin{document}

\author{Akshay Gaikwad}
\thanks{Equal author contributions.}
\affiliation{Department of Microtechnology and Nanoscience, Chalmers University of Technology, 41296 Gothenburg, Sweden}

\author{Manuel Sebastian Torres}
\thanks{Equal author contributions.}
\affiliation{Department of Microtechnology and Nanoscience, Chalmers University of Technology, 41296 Gothenburg, Sweden}
\affiliation{Department of Physics and Astronomy, KU Leuven, Celestijnenlaan 200d, B-3001 Leuven, Belgium}

\author{Shahnawaz Ahmed}
\affiliation{Department of Microtechnology and Nanoscience, Chalmers University of Technology, 41296 Gothenburg, Sweden}

\author{Anton Frisk Kockum}
\email{anton.frisk.kockum@chalmers.se}
\affiliation{Department of Microtechnology and Nanoscience, Chalmers University of Technology, 41296 Gothenburg, Sweden}

\title{Gradient-descent methods for fast quantum state tomography}

\begin{abstract}
Quantum state tomography (QST) is a widely employed technique for characterizing the state of a quantum system. However, it is plagued by two fundamental challenges: computational and experimental complexity grows exponentially with the number of qubits, rendering experimental implementation and data post-processing arduous even for moderately sized systems. Here, we introduce gradient-descent (GD) algorithms for the post-processing step of QST in discrete- and continuous-variable systems. To ensure physically valid state reconstruction at each iteration step of the algorithm, we use various density-matrix parameterizations: Cholesky decomposition, Stiefel manifold, and projective normalization. These parameterizations have the added benefit of enabling a rank-controlled ansatz, which simplifies reconstruction when there is prior information about the system. We benchmark the performance of our GD-QST techniques against state-of-the-art methods, including constrained convex optimization, conditional generative adversarial networks, and iterative maximum likelihood estimation. Our comparison focuses on time complexity, iteration counts, data requirements, state rank, and robustness against noise. We find that rank-controlled ansatzes in our stochastic mini-batch GD-QST algorithms effectively handle noisy and incomplete data sets, yielding significantly higher reconstruction fidelity than other methods. Simulations achieving full-rank seven-qubit QST in under three minutes on a standard laptop, with \qty{18}{\giga\byte} of RAM and no dedicated GPU, highlight that GD-QST is computationally more efficient and outperforms other techniques in most scenarios, offering a promising avenue for characterizing noisy intermediate-scale quantum devices. Our Python code for GD-QST algorithms is publicly available at \href{https://github.com/mstorresh/GD-QST}{github.com/mstorresh/GD-QST}.


\end{abstract}

\date{\today}

\maketitle


\section{Introduction}

Quantum computers have advanced from theoretical concept~\cite{Feynman1982} to practical reality with devices encompassing hundreds of qubits~\cite{madsen-nat-2022, Kim2023, Bluvstein2024, Acharya2025}. Down the line, these quantum machines may deliver substantial advantages over classical ones in, e.g., simulations of physics and chemistry, optimization problems, and machine learning (ML)~\cite{Georgescu2014, Montanaro2016, Wendin2017, Preskill2018, McArdle2020, Bauer2020, Cerezo2021, Cerezo2022, daley-nat-2022, Dalzell2023}. Similarly rapid developments are seen in other areas of quantum technology, where quantum sensing and metrology~\cite{Giovannetti2011, Degen2017} is on track to enable advantages in measurements ranging from medicine to fundamental physics~\cite{Aslam2023, Bass2024, Rovny2024, DeMille2024}, and quantum communication~\cite{Gisin2007} is being scaled up towards a quantum internet for secure communication and distribution of quantum information~\cite{Wehner2018, Azuma2023}.

This remarkable progress in quantum technologies has been facilitated by developments in quantum characterization techniques, i.e., diagnostic tools to analyze, understand, and enhance the performance of quantum devices~\cite{Gebhart2023, robin-arxiv-2024}. Prominent among these tools are tomographic methods such as quantum state and process tomography (QST and QPT), characterizing unknown quantum states and processes, respectively~\cite{james-pra-2001, liu-prb-2005, white-pra-2004, lvo-rmp-2009, chuang-jmo-09}. In fact, QST is a fundamental task, since it is connected to QPT by the Choi--Jamiolkowski isomorphism~\cite{choi-laa-1975, leung-2003, jiang-pra-2013}. Therefore, improving QST strategies is vital to aid the further development of quantum technologies.

Here, we leverage techniques based on gradient descent (GD) to upgrade QST. To see how our GD-QST alleviates challenges for QST, note that QST has two main components: (i) measurements, and (ii) an algorithm converting the measurement results into an estimate of the unknown state, a density matrix $\varrho$. The experimental and computational complexity of QST grows polynomially with the Hilbert-space dimension of the quantum system, and this dimension grows exponentially with the number of qubits, making it practically infeasible to implement full QST even for few-qubit systems~\cite{riofrio-2017, li-pra-2017, cotler-prl-2020}. Furthermore, because of statistical limitations (finite ensemble size/shots) and inevitable systematic errors introducing uncertainty in the measurement data, standard linear-inversion methods often lead to incorrect, and sometimes invalid, density matrices~\cite{ban-pra-1999, miranowicz-pra-2014, wolk-njp-2019}. One advantage of GD-QST is that we use parameterizations that ensure our estimate always is a valid density matrix. Furthermore, these parameterizations enable us to fix the rank $r$ of our ansatz, unlike most current QST protocols. Our rank-controlled ansatz facilitates and speeds up QST of pure and lightly mixed states, even for noisy data sets.

To further see how GD-QST compares to other QST strategies, we briefly review the main approaches. All strategies attempt to address the challenges of computational and experimental complexity. For example, instead of reconstructing the entire density matrix, protocols like selective and direct QST have been proposed~\cite{lundeen-prl-2012, wu-sr-2013, kim-natcom-2017, feng-pra-2021, li-pra-2022, ekert-prl-2022, gaikwad-epjd-2023}. These protocols are designed to only obtain specific density-matrix elements of particular interest, reducing the number of required experiments. Even so, these protocols are still resource-demanding, since they require ancilla qubits and high-dimensional complex operations. A related class of QST schemes utilize prior knowledge about the quantum states to be characterized, reducing complexity of experiments and calculations, but also reducing applicability. Examples include matrix-product-state methods~\cite{cramer-natcom-2010, lanyon-np-2017, han-pra-2022, Kurmapu2023}, permutationally invariant QST~\cite{toth-prl-2010, moroder-njp-2012}, and tensor-network approaches~\cite{kuzmin-pra-2024, torlai-2023}.

Another family of QST protocols, based on convex optimization problems, comprises maximum likelihood estimation~\cite{mle-pra-2007, smolin-prl-2012, shang-pra-2017}, compressed-sensing QST~\cite{david-prl-2010, Steffens_2017, yang-pra-2017, Kyrillidis2018, ahn-prl-2019, teo-pra-2020},  least-squares and linear-regression optimization~\cite{qi-quantum-inf-2017, nehra-prr-2020, gaikwad-qip-2021, ingrid-prapp-2022, Mondal-ls-2023}. These methods work with reduced measurement data sets, but enable reconstruction of the full density matrix. The validity conditions (Hermiticity, unit trace, and positivity of $\varrho$) are included as constraints. Such constrained convex optimization (CCO) problems are solved using convex optimization algorithms, e.g., semi-definite programming~\cite{banchi-npj-2020, meng-rp-2023}. However, these CCO problems are computationally expensive: the number of variables in them increases exponentially [$\mathcal{O}(4^N)$] with the number of qubits $N$. Although tools like YALMIP~\cite{lofberg-2004} and CVX~\cite{diamond2016cvxpy} are useful for solving a wide range of CCO problems, they are limited to systems with few dimensions. 

Finally, algorithms inspired by data-driven approaches and ML methods have been applied to many quantum information processing tasks, including QST and QPT~\cite{carleo-rmp-2019, carleo-science-2017, torlai-arcm-2020, Krenn2023}. 
Examples include variational algorithms~\cite{xin-2020, liu-pra-2020}, adaptive Bayesian tomography~\cite{granade-njp-2016, Mondal-bayesian-2023}, and  deep learning: feed-forward neural networks~\cite{quek-npj-2021, gaikwad-pra-2024}, convolutional neural networks~\cite{lohani-mlst-2020, schmale-2022}, conditional generative adversarial networks (CGANs)~\cite{Ahmed2021, Ahmed2021a, cha-mlst-2021}, restricted Boltzmann machines~\cite{torlai-np-2018, glasser-prx-2018, lange-2023}, and many more~\cite{xin-npj-2019, yu-2019, neugebauer-pra-2020, ghosh-2021, kauth-pra-2024, innan-2024, Palmieri2024}. Unlike strategies reviewed above, some of these algorithms both handle reduced measurement data and offer control over the number of parameters in the ansatz. However, these algorithms generally need large data sets for training (which are hard to collect or generate) and still face the exponential scaling challenges. Also, these algorithms are often more effective only for specific types of quantum states. Some algorithms, e.g., CGAN-QST~\cite{Ahmed2021, Ahmed2021a} do not require prior training, but lack control over the ansatz rank.


Our GD-QST is foremost inspired by the recent GD-QPT protocol~\cite{Ahmed2023}, which enables reconstruction of quantum processes with control over the ansatz rank~\cite{Kervinen2024}. Generally, GD is a widely used optimization procedure, e.g., in ML, to iteratively minimize a loss function using gradient calculations. Recently, GD methods have been applied to QST of moderately sized systems (up to 10-12 qubits)~\cite{bolduc-npj-2017, ferrie-prl-2014, hsu-prl-2024, wang-prr-2024}, at the cost of substantial computational resources (hundreds of GB of RAM, high-end GPUs). In self-guided QST~\cite{ferrie-prl-2014}, an estimate of the quantum state is refined by iteratively updating a trial state through evaluating its distance to the true state~\cite{flammia-prl-2011, silva-prl-2011}, utilizing stochastic GD optimization~\cite{spall-ieee-1992}. However, this approach only works if the state overlap can be computed directly in the experiment. Similarly, Ref.~\cite{wang-prr-2024} employs a factored-GD algorithm with momentum acceleration for QST, using Cholesky decomposition to parameterize the density matrix. However, this method lacks the ability to control the rank of the ansatz. In Ref.~\cite{hsu-prl-2024}, a nonconvex Riemannian gradient descent (RGD) algorithm for QST was proposed, improving factored GD by minimizing the iteration count to achieve a desired approximation error. The RGD algorithm incorporates singular-value decomposition in the optimization, ensuring positivity of $\varrho$. Notably, the RGD algorithm provides control over the rank of the ansatz. 

In this article, we reformulate QST into a \textit{mini-batch} GD-assisted function minimization problem. We propose three different parameterizations of the density matrix: Cholesky decomposition (CD), Stiefel manifold (SM), and projective normalization (PN), each ensuring valid reconstruction at each iterative step. We show that all three parametrizations enables controlling the rank of the ansatz, speeding up computations and enabling determination of any desired rank-$r$ estimation of the state. This includes pure-state tomography as the special case $r=1$. We assess the performance of these algorithms for discrete variables (DVs) up to seven qubits and on continuous-variable (CV) systems. We benchmark against several established techniques, including CCO algorithms for DVs and iterative maximum likelihood estimation (iMLE) and CGANs for CVs.

Our analysis focuses on the key aspects time complexity, iteration counts, data requirements, state rank, and noise robustness. We find that GD-QST consistently outperforms other techniques in most scenarios. Our findings emphasize the importance of selecting an appropriate parameterization. Specifically, GD-QST with CD emerges as the most effective approach for high-rank QST in large systems, SM excels in reconstructing pure states, and PN proves optimal for CV cases where the measurement operators are projectors. We also see that employing a rank-controlled ansatz effectively handles noisy and incomplete datasets, enabling the recovery of original quantum states with significantly higher fidelity than other methods. Simulations achieving full-rank seven-qubit QST in under three minutes on a standard laptop, with \qty{18}{\giga\byte} of RAM and no dedicated GPU, further highlight the computational efficiency of GD-QST. These results underscore the broad applicability of GD-QST, making it highly valuable in a wide range of quantum experiments. We facilitate such applications by making our Python code for GD-QST freely available~\cite{gd-qst-python}.

This article is organized as follows. In \secref{sec:TraditionalMethods}, we give an overview of standard QST methods. Then, in \secref{sec:GD-QST-Methods}, we describe our GD-QST algorithms. In \secref{sec:AlgorithmsBenchmarkAgainst}, we outline the other QST schemes against which we benchmark GD-QST, and in \secref{sec:DataBenchmarking}, we delineate data sets used for the benchmarking. We present the detailed numerical results of the benchmarking in \secref{sec:Results}, with subsections on time complexity with respect to state and ansatz rank (\secref{sec:time}), data requirements (\secref{sec:data}), noise robustness (\secref{sec:noise}), and CV systems (\secref{sec:CV}). We conclude in \secref{sec:conclusion} by summarizing our results and suggesting future research directions. The appendixes provide additional details on results for ansatzes with varying rank (\appref{app:DetailsRankVaryingAnsatz}) and on GD hyperparameters (\appref{app:AdamHyperparameters}).


\section{Methods}
\label{sec:Methods}

In this section, we present a detailed overview of state-of-the-art QST methods and the data sets used to benchmark some of these methods against our proposed GD-QST algorithms. We begin in \secref{sec:TraditionalMethods} with a general description of traditional QST methods. Then, in \secref{sec:GD-QST-Methods}, we detail the formalism of our GD-QST algorithms utilizing different parameterizations of density matrix. In \secref{sec:AlgorithmsBenchmarkAgainst}, we discuss the alternative QST schemes that we compare to our GD-QST methods. Finally, in \secref{sec:DataBenchmarking}, we describe the data sets for the benchmarking (types of quantum states and observables), along with the definition of the fidelity measure employed to evaluate the performance of the QST algorithms.


\subsection{Standard quantum state tomography methods}
\label{sec:TraditionalMethods}

A general $N$-qubit quantum state can be represented as a $2^N \times 2^N$-dimensional density matrix $\varrho$, expressed in terms of a chosen fixed set of basis operators $\{ \ket{i}\bra{j} \}$ formed by the $N$-qubit computational basis set $\{ \ket{i} \}$ as
\begin{equation} \label{eq:rho-def}
\varrho = \sum_{i,j=0}^{2^N-1} \varrho_{ij} \ket{i}\bra{j},\hspace{0.15cm}{\rm s.t.} \hspace{0.15cm} \varrho  = \varrho^{\dagger}, \hspace{0.15cm} {\rm Tr}(\varrho) = 1, \hspace{0.15cm} \& \hspace{0.15cm} \varrho \geq 0 ,
\end{equation}
where $\varrho_{ij}$ is the element of the density matrix in the $i$th row and $j$th column. The $N$-qubit operator basis set has cardinality $4^N$, indicating the number of independent real parameters required to uniquely represent the density matrix (excluding the trace condition; otherwise it is $4^N-1$). In the case of pure states, the number of independent real parameters scales as $\mathcal{O}(2^N)$. However, $\varrho$ can be expressed in multiple different ways (cf.~\secref{sec:GD-QST-Methods}). 

The goal of QST is to reconstruct an unknown $\varrho$ from measurement outcomes $\{ {\mathcal{B}}_i \}$, generally expressed as expectation values of the corresponding measurement operators $\{ {\Pi}_i \}$: ${\mathcal{B}}_i = {\rm Tr}({\Pi}_i \varrho) $. The set $\{ {\Pi}_i \}$ that allows complete and unique estimation of $\varrho$ is said to be informationally complete (IC). By carrying out an IC set of measurements, a system of linear equations can be formed which allows to determine $\varrho$ by simply solving a linear inversion problem:
\begin{equation} \label{eq:linear-inversion}
{\mathcal{A}} \mathcal{X}_\varrho =  {\mathcal{B}} \quad \xrightarrow{\text { linear inversion }} \quad \mathcal{X}_\varrho =  ({\mathcal{A}}^{T} {\mathcal{A}})^{-1}  {\mathcal{A}^{T}}{\mathcal{B}} ,
\end{equation}
where ${\mathcal{A}} $ is an $M \times 4^N$-dimensional matrix with $M$ the cardinality of the IC set. The matrix ${\mathcal{A}} $ is known as a sensing matrix; it is defined by the chosen operator basis set $\{ \ket{i}\bra{j} \}$ and the set of measurement operators $\{ {\Pi}_i \}$ as $\mathbf{\mathcal{A}}_{mn} ={\rm Tr}[{\Pi}_m E_n]$, where $E_{n = i\times2^N+j} = \ket{i}\bra{j}$. The column matrix ${\mathcal{B}}$ consists of measurement outcomes while the column matrix $\mathcal{X}_\varrho$ in \eqref{eq:linear-inversion} is formed by flattening $\varrho$ into a one-dimensional array with entries $\varrho_{ij}$, which are to be determined. This procedure of obtaining the density matrix is termed the standard QST method.  

However, while a density matrix reconstructed by standard QST has unit trace [ensured by including the trace equation into \eqref{eq:linear-inversion}] and is Hermitian (by construction), it is not guaranteed to be positive semi-definite. This problem can be quickly overcome by reformulating \eqref{eq:linear-inversion} into constrained least-squares~\cite{ingrid-prapp-2022} or compressed-sensing optimization problems~\cite{rod-prb-2014}. The widely used convex optimization approach for QST is given by
%
%
\begin{equation}
\underset{{\mathcal{X}_\varrho}}{\operatorname{min}} \left\|\mathbf{\mathcal{A}} \mathcal{X}_\varrho - {\mathcal{B}} \right\|_{l_2}  + \lambda \left\| \mathcal{X}_\varrho \right\|_{l_1} \hspace{0.15cm} \text{s.t.} \hspace{0.15cm} \varrho \geq 0 ,  \label{cvx}
\end{equation}
%
where $\parallel \cdot \parallel_{l_2}$ and $\parallel \cdot \parallel_{l_1} $ denote the L2 and L1 norm, respectively. The L2 norm is the standard Euclidean norm of a vector, while the L1 norm is the sum of the absolute values of all components. The scalar quantity $\lambda$ in \eqref{cvx} controls the sparsity of the solution vector $\mathcal{X}_{\varrho}$, which in this case is a reconstructed density matrix. 

The convex optimization problem in \eqref{cvx} can handle reduced data sets, but lacks control over the number of variables in the optimization, making it experimentally flexible but computationally inadequate. Common tools such as CVXPY~\cite{diamond2016cvxpy}, SDPT$^3$~\cite{sdpt-2004}, and YALMIP~\cite{lofberg-2004}, along with built-in optimizers like self-dual-minimization (SeDuMi), semidefinite programming algorithm (SDPA), splitting conic solver (SCS), cardinal optimizer (COPT), MOSEK, and others~\cite{cvx-solvers}, which primarily use a semidefinite-quadratic-linear programming (SQLP) approach~\cite{toh-sdpt-1999}, are typically employed to solve such problems. However, these tools become insufficient for large-scale systems, often requiring many hours of computation on a moderately configured laptop~\cite{hou-njp-2016}.


\subsection{GD-QST methods}
\label{sec:GD-QST-Methods}

To address the issues outlined in \secref{sec:TraditionalMethods}, we recast the QST task as a GD-assisted function minimization problem and utilize a variety of parameterizations to maintain validity of the density matrix during the GD optimization. The goal of our GD-QST algorithms is to find an optimal estimate of a quantum state, expressed as a density matrix $\varrho$, by minimizing the loss function
\begin{equation} \label{loss}
    \mathcal{L}[\mathcal{P}_{\varrho}(\bm{\theta})] = \sum_i \Bigr[{\mathcal{B}}_i - {\rm Tr}[{\Pi}_i \mathcal{P}_{\varrho}(\bm{\theta})]\Bigr]^2 + \lambda \mleft\|  \mathcal{P}_{\varrho}(\bm{\theta}) \mright\|_{l_1} ,
\end{equation}
where $\mathcal{P}_{\varrho}(\bm{\theta})$ is an abstract representation of $\varrho$ with $\mathcal{P}$ an appropriate parameterization and $\bm{\theta}$ the parameter vector used to describe this \textit{ansatz}. The first part of $ {\mathcal{L}[\mathcal{P}_{\varrho}(\bm{\theta})]}$ is least-squares error loss (equivalent to L2 norm), which minimizes discrepancy between observed data and the estimated state. The second part is L1 norm scaled by a hyperparameter $\lambda$, which primarily controls the sparsity of the estimated density matrix.

\begin{figure}
\centering
\includegraphics[width=\linewidth]{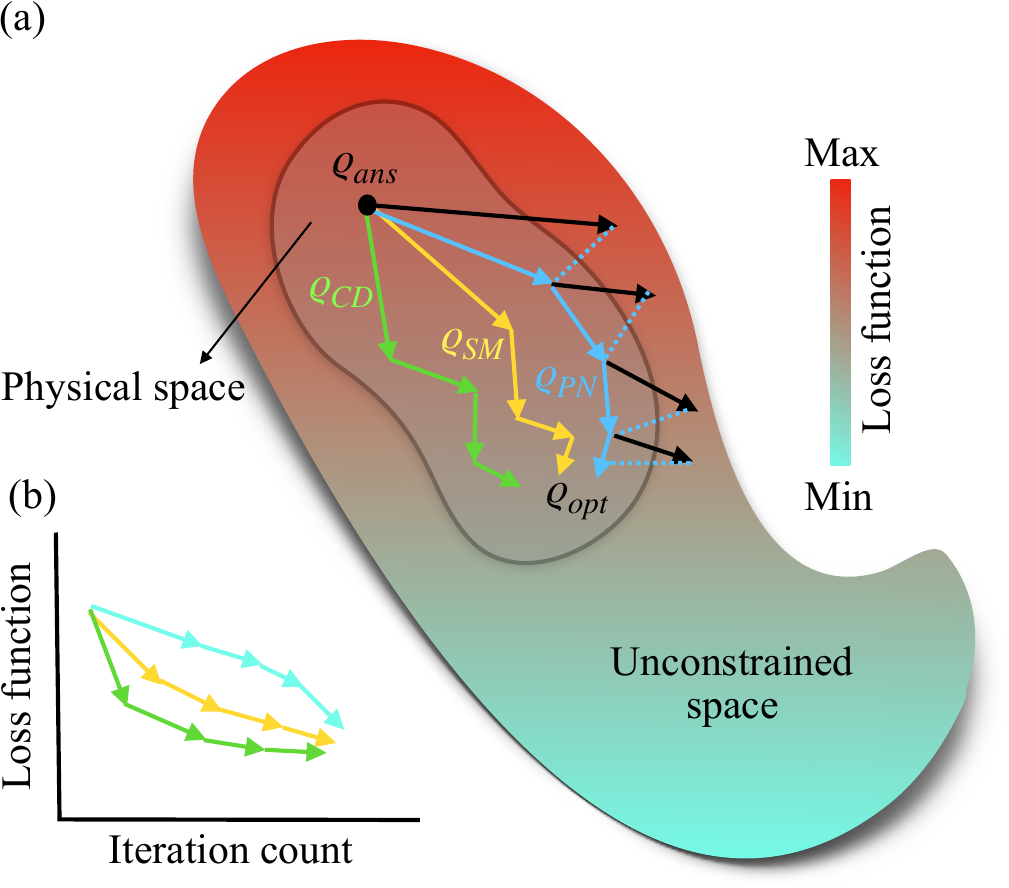}
\caption{Graphical illustration of our GD-QST methods. 
(a) Depiction of GD-QST employing CD (green), SM (yellow), and PN (blue), starting from an ansatz $\varrho_{\text{ans}}$. For CD and SM, GD updates occur within the space of physical density matrices; for PN, the GD updates can go outside the physical space (black arrows) and are then projected back into the physical space (dashed blue lines). All methods approach $\varrho_{\text{opt}}$, the optimal density matrix corresponding to the minimum loss.
(b) Depiction of the behavior of the loss function in a physical space as a function of iteration count.
\label{gd-qst-illustration}}
\end{figure}

We employ vanilla GD (VGD) for the SM parameterization, and the \textit{Adam} optimization algorithm (a hybrid version of the momentum GD algorithm and the root-mean-square propagation algorithm) for the CD and PN parameterizations (see \figref{gd-qst-illustration} for a graphical illustration). The parameter update rules for VGD and Adam are~\cite{ruder2016overview}
\begin{align}
\text{VGD}&: \bm{\theta}_{t} \gets \bm{\theta}_{t-1}-\eta \cdot \nabla \mathcal{L}[\mathcal{P}_{\varrho}(\bm{\theta}_{t-1})] , \label{vgd}\\
\text{Adam}&: \bm{\theta}_{t} \gets \bm{\theta}_{t-1} - \eta \cdot \bm{\hat{m}}_{t} /(\sqrt{\bm{\hat{v}}_{t}} + \epsilon) , \label{adam_gd}
\end{align}
where $\nabla \mathcal{L}[\mathcal{P}_{\varrho}(\bm{\theta}_t)]$ is the gradient of the loss function with respect to $\bm{\theta}$ in step $t$. The vectors $\bm{\hat{m}}_{t}$ and $\bm{\hat{v}}_{t}$ in \eqref{adam_gd} are bias-corrected first and second moments (see \appref{app:AdamHyperparameters} for details). The hyperparameter $\eta$ denotes the step size (learning rate).

For both VGD and Adam, we employ mini-batch stochastic gradient descent, where the dataset is divided into small batches, and gradient updates are performed by randomly selecting a mini-batch at each iteration. This stochastic GD method accelerates convergence by avoiding local optima and saddle points while improving the effectiveness of reaching global minima, with provable guarantees~\cite{dimitri-siam-2000, pana-book-2019, simon-arxiv-2019}.

With Eqs.~(\ref{loss})--(\ref{adam_gd}) as the foundation of the GD-QST algorithms, we next detail the CD, SM, and PN parameterizations in Secs.~\ref{sec:CD-parameterization}, \ref{sec:SM-parameterization}, and \ref{sec:PN-parameterization}, respectively. 





\subsubsection{\textbf{Cholesky decomposition}}
\label{sec:CD-parameterization}

Any arbitrary density matrix $\varrho$ can be parameterized using a Cholesky decomposition (CD) as
\begin{equation}
\label{cd}
\varrho_{\rm CD} = \mathcal{P}_{\varrho}(T_m) = \frac{T_m^{\dagger} T_m}{{\rm Tr}(T_m^{\dagger} T_m)} .
\end{equation}
Here, $T_m$ is an $m \times 2^N$-dimensional arbitrary complex matrix~\cite{riofrio-2017}, which functions as a rank-controlled ansatz. The parameter $1\leq m \leq 2^N$ gives the rank of $\varrho_{CD}$; however, in the literature, $T_m$ is commonly assumed to be a $2^N \times 2^N$-dimensional complex lower triangular matrix~\cite{james-pra-2001, Ahmed2021, Ahmed2021a, Hoshi2025}. From \eqref{cd}, we see that $\varrho_{\rm CD}$ satisfies all three requirements of a density matrix: (i)$\varrho_{\rm CD} = \varrho_{\rm CD}^{\dagger}$, (ii) $\text{Tr}[\varrho_{\rm CD}]=1$, and (iii) $\bra{\psi}\varrho_{\rm CD}\ket{\psi} \geq 0 \hspace{1mm} \forall \psi$. 

In this case, the loss function in \eqref{loss} becomes
\begin{equation} 
\label{cd_loss}
    \begin{split}
    {\mathcal{L}[\mathcal{P}_{\varrho}(T_m)]} = & \sum_i \left[ \mathcal{B}_i - \text{Tr}\left( \Pi_i \frac{T_m^{\dagger} T_m}{\text{Tr}(T_m^{\dagger} T_m)} \right) \right]^2 \\
    & \quad \quad \quad \quad \quad + \lambda \left\|  \frac{T_m^{\dagger} T_m}{{\rm Tr}(T_m^{\dagger} T_m)}  \right\|_{l_1}
    \end{split}
\end{equation}
and the GD updates are computed as
\begin{equation}
\label{cd-update}
    T_m^{t} \xrightarrow{\text { Adam}}  T_m^{t+1} .
\end{equation}
Equation~\ref{cd} ensures the validity of the density matrix throughout the parameter update process in \eqref{cd-update}.
Note that, when applicable and relevant (typically in full-rank QST), we also assess the performance of the CD parameterization with $T_m$ defined as a lower triangular matrix (a full-rank ansatz), which we refer to as CD-tri.


\subsubsection{\textbf{Stiefel manifold}} 
\label{sec:SM-parameterization}

Optimization on the complex Stiefel manifold (SM)~\cite{Tagare2011, Boumal2023} can handle positivity and normalization constraints~\cite{wen-mp-1013, jiang-mp-2015}. Recently it has been employed, e.g., in compressive gate set tomography, where the gate set is parameterized as a rank-constrained tensor~\cite{martin-prx-quant-2023}, in QPT, where the SM is used to compute Kraus operators that constrain the rank of the process~\cite{Ahmed2023}, and in various other problems in quantum physics~\cite{Luchnikov2021}. However, to the best of our knowledge, the SM has not been previously applied to parameterization of the density matrix in QST. Here, drawing inspiration from the use of SM in QPT~\cite{Ahmed2023}, we propose such a parameterization. This approach is flexible: it supports a rank-controlled ansatz and facilitates pure-state tomography (with a rank-1 ansatz) by directly reconstructing the underlying pure state as a vector.

To define a proper parameterization using the SM, consider the density-matrix representation
\begin{equation}
\label{sm1}
\varrho = \sum_{i=1}^m p_i \ket{\psi_i}\bra{\psi_i} ,
\end{equation}
where $\{ \ket{\psi_i} \}$ and $\{ p_i \}$ is a set of $m$ normalized pure states and their corresponding classical probabilities, respectively, such that $ \langle \psi_i \vert \psi_i\rangle  = 1 \:\forall i$ and $\sum_i p_i = 1$. 
Motivated by \eqref{sm1}, we stack the $\{ \ket{\psi_i} \}$ and the corresponding $\{ p_i \}$ into a one-dimensional array
\begin{equation}
\label{sm2}
\mathcal{W}_m = \begin{bmatrix}
\sqrt{p_1}\ket{\psi_1} & \cdots & \sqrt{p_m}\ket{\psi_m}
\end{bmatrix}^T .
\end{equation}
From \eqref{sm2}, it directly follows that
\begin{equation} 
\label{sm3}
    \mathcal{W}^{\dagger}_m \mathcal{W}_m = \begin{bmatrix}
\sqrt{p_1}\bra{\psi_1} & \cdots & \sqrt{p_m}\bra{\psi_m}
\end{bmatrix} \begin{bmatrix}
\sqrt{p_1}\ket{\psi_1} \\
\vdots \\
\sqrt{p_m}\ket{\psi_m} 
\end{bmatrix} = 1 .
\end{equation}

The orthonormality condition in \eqref{sm3} defines the complex SM
\begin{equation}
    St(k,1) = \{ \mathcal{W}_m \in \mathbb{C}^{k \times 1} \mid \mathcal{W}^{\dagger}_m\mathcal{W}_m = I^{1 \times 1} \} ,
\end{equation}
where $k = m\times2^N$ with $N$ the number of qubits. A proper parameterization of the density matrix then becomes
\begin{equation}
    \varrho_{\rm SM} = \mathcal{P}_{\varrho}(\mathcal{W}_m) = \mathcal{W}_m \odot \mathcal{W}^{\dagger}_m ,
\end{equation}
where the vector $\mathcal{W}_m$, an element of $St(k,1)$, acts as a rank-controlled ansatz with rank $m$. Here, 
\begin{equation}
\mathcal{W}_m \odot \mathcal{W}^{\dagger}_m = \sum_i \mathcal{W}_m[i] \mathcal{W}^{\dagger}_m[i]  = \sum_{\alpha}^m p_{\alpha} \ket{\psi_{\alpha}}\bra{\psi_{\alpha}}
\end{equation}
denotes element-wise product followed by summation. 

In this parameterization, the loss function becomes
\begin{equation}
    \begin{split}
        \mathcal{L}[\mathcal{P}_{\varrho}(\mathcal{W}_m)] = & \sum_i \left[ \mathcal{B}_i - {\rm Tr}[\Pi_i (\mathcal{W}_m \odot \mathcal{W}^{\dagger}_m )]  \right]^2 \\ 
        & \quad \quad \quad \quad + \lambda \left\|\mathcal{W}_m \odot \mathcal{W}^{\dagger}_m\right\|_{l_1} .
    \end{split}
\end{equation}
To minimize such a loss function by performing GD updates, $\mathcal{W}_m^t \xrightarrow{\text {VGD}} \mathcal{W}_{m}^{t+1}$, and consistently preserve the orthonormality constraint specified in \eqref{sm3} such that the updated vector remains on the SM [$\mathcal{W}_{m}^{t+1} \in St(k,1) \: \forall t$], is known as \textit{optimization on the SM}. The adherence to the orthonormality constraint can be ensured by a \textit{retraction procedure}~\cite{wen-mp-1013, absil2009optimization}.

The retraction procedure can be described briefly with three new quantities:
\begin{equation}
\label{sm4}
    \Tilde{G} = G/\| G \|_{l_2}, \hspace{2mm} A = \begin{bmatrix} \Tilde{G} & \mathcal{W}_m  \end{bmatrix},  \hspace{2mm} B = \begin{bmatrix} \mathcal{W}_m & -\Tilde{G}   \end{bmatrix} ,
\end{equation}
where $G = \nabla \mathcal{L}[\mathcal{P}_{\varrho}(\mathcal{W}_m)]$ is the standard gradient of the loss function with respect to $\mathcal{W}_m$. Using the Cayley transform and the Sherman-Morrison-Woodbury formula~\cite{jun-arxiv-2020}, we can calculate (following the supplementary material of Ref.~\cite{Ahmed2023}) the conjugate gradient as
\begin{equation}
\label{sm5}
\nabla^{*} \mathcal{L}[\mathcal{P}_{\varrho}(\mathcal{W}_m)]  = A \left(  \mathbb{I} + \frac{\eta}{2} B^{\dagger} A \right)^{-1} B^{\dagger} \mathcal{W}_m
\end{equation}
and compute the updated vector according to the VGD rule in \eqref{vgd} as
\begin{equation}
\mathcal{W}_m^{t+1} = \mathcal{W}_m^{t} - \eta \nabla^{*} \mathcal{L}[\mathcal{P}_{\varrho}(\mathcal{W}_m^t)] .
\end{equation}

Optimization on the SM thus both yields a valid density matrix during each iterative update and gives control over the number of parameters in the optimization process by defining the initial rank of an ansatz $\mathcal{W}_m$.


\subsubsection{\textbf{Projective normalization}}
\label{sec:PN-parameterization}

The parameterization using projective normalization (PN) also starts from the density-matrix representation in \eqref{sm1}. This ansatz is defined by two column vectors: $\mathcal{C}_m = \begin{bmatrix} p_1 & \cdots & p_m \end{bmatrix}^T$ and $\mathcal{Q}_m = \begin{bmatrix} \ket{\psi_1} & \cdots & \ket{\psi_m}  \end{bmatrix}^T$. Using these vectors, the density matrix can be written as
\begin{equation}
    \varrho_{\rm PN} = \mathcal{P}_{\varrho}(\mathcal{C}_m, \mathcal{Q}_m) = \mathcal{C}_m \odot \mathcal{Q}_m \odot  \mathcal{Q}_m^{\dagger} ,
\end{equation}
where 
\begin{equation}
\mathcal{C}_m \odot \mathcal{Q}_m \odot  \mathcal{Q}_m^{\dagger} = \sum_i \mathcal{C}_m[i] \mathcal{Q}_m[i]  \mathcal{Q}_m^{\dagger}[i].
\end{equation}
The loss function then becomes
\begin{equation}
    \begin{split}
        \mathcal{L}[\mathcal{P}_{\varrho}(\mathcal{C}_m, \mathcal{Q}_m) ] = & \sum_i \mleft[ \mathcal{B}_i - {\rm Tr}[\Pi_i (\mathcal{C}_m \odot \mathcal{Q}_m \odot  \mathcal{Q}_m^{\dagger} )]  \mright]^2 \\ 
        & \quad \quad + \lambda \left\| \mathcal{C}_m \odot \mathcal{Q}_m \odot  \mathcal{Q}_m^{\dagger}  \right\|_{l_1} .
    \end{split} 
\end{equation} 

We perform the parameter-update process in two steps: (i) first, GD optimization is performed using the \textit{Adam} optimizer [\eqref{adam_gd}] to obtain updated values $\mathcal{C}_m^{t+1}$ and $\mathcal{Q}_m^{t+1}$, then (ii) we perform PN separately on them to obtain final updated vectors $\tilde{\mathcal{C}}_m^{t+1}$ and $\tilde{\mathcal{Q}}_m^{t+1}$. This method of updating parameters with GD is an example of `projected-gradient' techniques in numerical optimization~\cite{levitin-1966, bruck-1977}. 

Thus GD-QST using PN can be written as
\begin{widetext}
\begin{align}
 \mathcal{C}_m^{t} = \begin{bmatrix} p_1^t & \cdots & p_m^t \end{bmatrix}^T &\xrightarrow{\text {Adam}} \mathcal{C}_m^{t+1} = \begin{bmatrix} p_1^{t+1} & \cdots & p_m^{t+1} \end{bmatrix}^T  \xrightarrow{\text { PN }} \tilde{\mathcal{C}}_m^{t+1} = \begin{bmatrix} \tilde{p}_1^{t+1} & \cdots & \tilde{p}_m^{t+1} \end{bmatrix}^T  \label{pnc} , \\
 \mathcal{Q}_m^{t} = \begin{bmatrix} \ket{\psi_1^t} & \cdots & \ket{\psi_m^t}  \end{bmatrix}^T &\xrightarrow{\text {Adam}} \mathcal{C}_m^{t+1} = \begin{bmatrix} \ket{\psi_1^{t+1}} & \cdots & \ket{\psi_m^{t+1}}  \end{bmatrix}^T \xrightarrow{\text { PN }} \tilde{\mathcal{C}}_m^{t+1} = \begin{bmatrix} \ket{\tilde{\psi}_1^{t+1}} & \cdots & \ket{\tilde{\psi}_m^{t+1}}  \end{bmatrix}^T \label{pnq},
\end{align}
\end{widetext}
where the PN steps in Eqs.~(\ref{pnc}) and (\ref{pnq}) are defined by \textit{softmax} function and norm division, respectively:
\begin{equation}
    \tilde{p}_i^{t+1} = \frac{e^{p_i^{t+1}}}{\sum_{\alpha}p_{\alpha}^{t+1}} \quad {\rm and} \quad \ket{\tilde{\psi}_i^{t+1}} = \frac{\ket{\psi_i^{t+1}}}{\sqrt{\langle  \psi_i^{t+1} \vert \psi_i^{t+1} \rangle}} .
\end{equation}
This equation ensures that the updated values of classical probabilities described by $\mathcal{C}_m$ are positive and sum to one, and that the state vectors in $\mathcal{Q}_m$ are normalized, thereby producing a true density matrix. Moreover, it also enables us to control the number of parameters in the optimization process through the initialization of a rank-controlled ansatz defined by the two vectors $(\mathcal{C}_m, \mathcal{Q}_m)$.

\subsection{Algorithms we benchmark against}
\label{sec:AlgorithmsBenchmarkAgainst}

In this subsection, we briefly outline some other algorithms that can be considered state of the art for QST. These algorithms, CCO-QST using CVX, iMLE-QST, and CGAN-QST, are the ones we benchmark our GD-QST algorithms in \secref{sec:GD-QST-Methods} against. \\


\subsubsection{\textbf{Constrained convex optimization using CVX}}
\label{sec:cco-cvx}

The convex-optimization approach, combining least-squares minimization with L1 regularization, as formulated in \eqref{cvx}, is commonly solved using CVX~\cite{diamond2016cvxpy}. An appropriate value of the parameter $\lambda$ in the L1 regularization allows us to determine a good sparse approximation of the density matrix using heavily reduced data sets, provided the sensing matrix $\mathcal{A}$ in \eqref{cvx} meets the restricted-isometry property conditions~\cite{rod-prb-2014}. However, this approach lacks control over the number of variables in the optimization, which scales as $\mathcal{O}(4^N)$ for an $N$-qubit system, making it infeasible for practical purposes beyond a few qubits.


\subsubsection{\textbf{Iterative maximum likelihood estimation}}
\label{sec:imle}

The original iMLE formalism described in Ref.~\cite{imle} is primarily based on expectation values of positive operator-valued measures (POVMs), typically a set of projection operators, rather than a general set of observables characterized by Hermitian matrices. The objective of the iMLE protocol is to determine the density matrix $\varrho_{\text{MLE}}$ that is most likely to generate the observed dataset $\{ \mathcal{B} \}$ by maximizing the likelihood function
\begin{equation}
    L(\varrho_{\text{MLE}} | \mathcal{B}) = \prod_j \langle P_j \rangle ^{\mathcal{B}_j}, 
\end{equation}
%
where $P_j$ is the projection operator onto the $j$th eigenstate of a measurement apparatus (an eigenstate of a Hermitian observable being measured) and $\langle P_j \rangle = {\text{Tr}(P_j \varrho_{\text{MLE}})} $ is the probability of the system being projected onto this state; ${\mathcal{B}_j}$ is the frequency of occurrences.

Determining $\varrho_{\text{MLE}}$ involves iteratively updating an initially guessed density matrix according to the rule
\begin{equation}
\mathcal{N}[\mathcal{R}(\varrho^t) \varrho^t \mathcal{R}(\varrho^t)] \rightarrow \varrho^{t+1} ,
\end{equation}
where
\begin{equation}
\mathcal{R}(\varrho^t)  = \sum_j \frac{\mathcal{B}_i}{\text{Tr}(P_j \varrho^t)} P_j
\end{equation}
is a positive semi-definite operator and $\mathcal{N}$ is normalization factor.

\subsubsection{\textbf{Conditional generative adversarial networks}}
\label{sec:cgan}

The objective of CGAN-QST is to reconstruct the density matrix using a competitive learning framework involving a generator $G$  and a discriminator $D$~\cite{ian-gan-2014, Ahmed2021, Ahmed2021a}. Both the generator and discriminator are nonlinear functions, typically modeled as multilayer deep neural networks with parameters $(\theta_G, \theta_D)$. In this framework, the discriminator's role is to distinguish between experimental data (real data) and data generated by the generator (fake data). Through an iterative optimization process, the generator and discriminator are trained to refine their respective performances, enabling the generator to produce data that closely resembles the experimental data, allowing density matrix reconstruction.

The optimization is carried out in an iterative manner using standard GD. In each iterative step, $\theta_D$ is updated by maximizing the expectation value
\begin{equation} \label{eq:cgan-discriminator}
\begin{split}
& E_{\mathbf{y} \sim p_{\text {data }}}\left[\ln \left(D\left(\mathbf{x, y} ; \theta_D\right)\right)\right]\\
 + & E_{\mathbf{z} \sim p_z}\left\{\ln \left[1-D\left(G\left(\mathbf{x, z} ; \theta_G\right) ; \theta_D\right)\right]\right\},
\end{split}
\end{equation}
where $\mathbf{x}$ is a conditioning vector~\cite{ian-gan-2014, mirza-arxiv-2014}, $\mathbf{y}$ is sampled from the real data distribution ($ y \sim p_{\text{data}} $), and $D(\mathbf{x, y} ; \theta_D)$ represents the discriminator's output for $\mathbf{y}$ conditioned on $\mathbf{x}$. In the second term,  $\mathbf{z}$  is drawn from the noise distribution  $p_z $ ($\mathbf{z} \sim p_z $), and  $G(\mathbf{x, z} ; \theta_G)$ denotes the generator's output for $\mathbf{z}$ conditioned on $\mathbf{x}$. Next, $\theta_G$ is updated by minimizing
\begin{equation}
\label{eq:cgan-generator}
E_{\mathbf{z} \sim p_z}\left\{\ln \left[1-D\left(G\left(\mathbf{x, z} ; \theta_G\right) ; \theta_D\right)\right]\right\} .
\end{equation}
In this way, the generator learns to mimic the real experiment, effectively functioning as $G: \mathbf{x, z} \rightarrow \mathbf{y} $~\cite{ian-gan-2014, Ahmed2021, Ahmed2021a}. 

In the CGAN-QST approach, as demonstrated in Refs.~\cite{Ahmed2021, Ahmed2021a}, the noise vector $\mathbf{z}$ is omitted, and the conditioning input $\mathbf{x}$ to the generator is defined as the measurement data and measurement operators ($\mathbf{x} \rightarrow \mathcal{B}, \{ \Pi \}$). The generator employs two custom-added layers: the first of these layers computes the density matrix $\varrho_G$ and the second layer generates expectation values as $\mathrm{Tr}(\Pi_i \varrho_G)$. Subsequently, the discriminator takes the experimental data $\mathcal{B}$ (used as the conditioning variable, $\mathbf{x} \rightarrow \mathcal{B}$) along with the generated measurement data as input and evaluates the discrepancy between the experimental and generated data in its output. Through the training process [optimizing Eqs.~(\ref{eq:cgan-discriminator}) and (\ref{eq:cgan-generator})], the generator progressively learns the underlying density matrix, making it harder for the discriminator to distinguish between the actual and generated data. The Python code for implementing CGAN-QST is available in Ref.~\cite{cgan-qst-python}.


\subsection{Data sets for benchmarking}
\label{sec:DataBenchmarking}

Here, we outline the data sets used to evaluate the performance of the QST protocols described in Secs.~\ref{sec:GD-QST-Methods} and \ref{sec:AlgorithmsBenchmarkAgainst}, for both DV and CV systems. The data sets encompass types of quantum states (\secref{sec:Qstates}), observables (\secref{sec:obs}), and a fidelity measure (\secref{sec:fidelity}).


\subsubsection{\textbf{Quantum states}} 
\label{sec:Qstates}

For the DV systems, we use several different types of quantum states. These include (i) a set of pure states, (ii) a set of mixed states with varying rank, and two special states: (iii) the Hadamard state 
\begin{equation}
\ket{\Phi_H} = \mleft[ (|0\rangle+|1\rangle) / \sqrt{2} \mright]^{\otimes N}
\end{equation}
and (iv) the GHZ state
\begin{equation}
\ket{\Phi_{\text{GHZ}}} = \left[\left(|0\rangle^{\otimes N}+|1\rangle^{\otimes N}\right) / \sqrt{2}\right].
\end{equation}

Similarly for CV systems, we apply our algorithms to various single-mode optical cat states, defined as the quantum superposition of two coherent states with opposite sign: 
\begin{equation}
\ket{\text{cat}} \propto \ket{\xi} + \ket{-\xi} ,
\end{equation}
where 
\begin{equation}
\ket{\xi} = e^{-\frac{1}{2}|\xi|^2} \sum_{n=0}^{\infty} \frac{\xi^n}{\sqrt{n!}} \ket{n}
\end{equation}
with $\xi = r e^{i \phi}$ a complex parameter.
In both cases, the quantum states are generated using QuTiP~\cite{Johansson2012, Johansson2013, Lambert2024}.



\subsubsection{\textbf{Measurement operators}} 
\label{sec:obs}

The significance of selecting an appropriate set of measurement operators $\{ \Pi_i \}$ for QST has been thoroughly examined in Ref.~\cite{miranowicz-pra-2014}, where the optimal set $\{ \Pi_i \}$ is defined based on achieving the lowest condition number of the sensing matrix $\mathcal{A}$, provided the set $\{ \Pi_i \}$ is informationally complete. This choice enhances the robustness of the QST protocol against errors.  Therefore, inspired by this choice together with current experimental platforms, we use the $N$-qubit Pauli matrices (easy to measure and also yielding a small condition number) as our set of \textit{near-optimal} measurement operators~\cite{miranowicz-pra-2014}: $\Pi = \{ I, \sigma_x, \sigma_y, \sigma_z \}^{\otimes N}$ for DV (multi-qubit) systems. Similarly, in the case of CV systems, we utilize measurement data obtained from the Husimi $Q$ function by evaluating the expectation value of the operator $\Pi_m = (1/\pi) \ket{\beta_m}\bra{\beta_m}$, where $\ket{\beta_m}$ is a coherent state expressed in the Fock basis. This highlights the versatility of the proposed GD-QST algorithms, which can operate with any choice of measurement-operator set.


\subsubsection{\textbf{State-fidelity measure}} 
\label{sec:fidelity}

Throughout this article, we use the Uhlmann--Jozsa (UJ) fidelity metric to measure closeness between two quantum states $\rho$ and $\sigma$ as~\cite{jozsa-jmp-1994}
\begin{equation}
    \mathcal{F}(\rho, \sigma) = \left(\text{Tr} \sqrt{\sqrt{\rho} \sigma \sqrt{\rho}}\right)^2 .
\end{equation}

With all algorithms, data sets, and the fidelity measure described in detail, we proceed to perform numerical simulations. The numerical results presented in the following section were obtained on a standard laptop with \qty{18}{\giga\byte} of RAM and no dedicated GPU.







\section{Results}
\label{sec:Results}

In this section, we assess the performance of the GD-QST algorithms in different scenarios through numerical simulations on multi-qubit and CV systems. Specifically, we compare the CD, SM, and PN parameterizations, and where applicable, benchmark them against several existing QST methods, including the convex optimization algorithm using CVX, iMLE, and CGAN-QST. For DV systems, GD-QST is compared with CVX, while for CV systems, it is compared with iMLE and CGAN. The original mathematical framework of iMLE~\cite{imle} is limited to POVMs and tends to diverge in the DV case; this is why iMLE is only applied to CV systems here. 

In \secref{sec:time}, we analyze the time complexity of our GD-QST algorithms by examining the number of iterations and time per iteration needed to reconstruct the full density matrix with sufficiently high state fidelity. As a case study, we provide in-depth numerical analysis for five-qubit systems, demonstrating the advantage of selecting an ansatz with an appropriate rank, e.g., by leveraging prior knowledge about the target density matrix such as purity or rank. We also highlight fast, high-dimensional pure-state tomography, a special case of the rank-1 ansatz, which is particularly relevant for quantum computing and information processing experiments. 

In \secref{sec:data}, we show the efficacy of our GD-QST algorithms by implementing them on significantly reduced data sets and demonstrate fast, high-quality reconstruction of the full density matrix. Then, in \secref{sec:noise}, we demonstrate the noise robustness of our GD-QST algorithms by applying them to noisy data sets obtained from depolarizing and Gaussian noise channels.  Finally, in \secref{sec:CV}, we benchmark GD-QST on CV systems by reconstructing the density matrix within a truncated Hilbert space. We plot the Wigner functions derived from the reconstructed density matrices and compare them with the ideal Wigner function. 


\subsection{Time complexity}
\label{sec:time}

\begin{figure}
\centering
\includegraphics[width=0.8\linewidth]{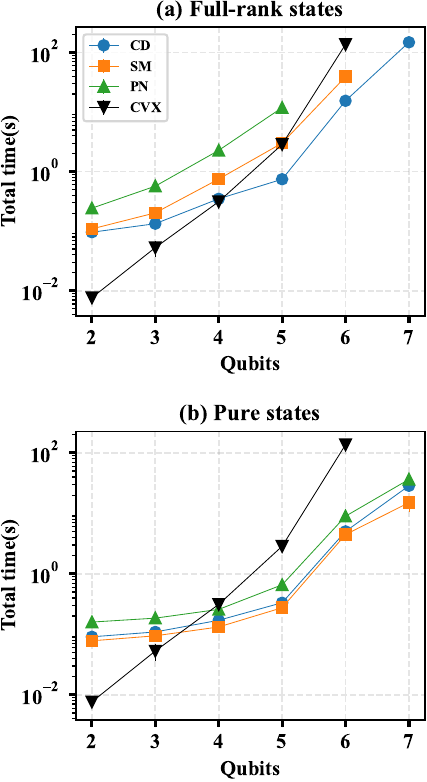}
\caption{Time required for GD-QST algorithms to achieve a state fidelity of $\mathcal{F} > 0.99 $ for (a) full-rank states and (b) pure states, as a function of the number of qubits in the system. The total time in each scenario is averaged over 30 randomly generated full-rank and pure states. The legend indicates different methods: GD-QST with CD (teal circles), SM (orange squares), and PN (green upward triangles), and the CCO tool CVX (black downward triangles). In panel (a), the PN algorithm is excluded for systems with more than five qubits, and SM and CVX are excluded beyond six qubits, while in panel (b) CVX is omitted beyond six qubits, due to their failure to converge within a reasonable time frame.}
\label{qubit_time}
\end{figure}

Here, we evaluate the computational time required for full state reconstruction using the GD-QST methods described in \secref{sec:Methods}, for systems containing up to seven qubits. Figure~\ref{qubit_time} shows the performance of the GD-QST algorithms with three different parameterizations: CD (teal), SM (orange), and PN (green), and compares them to the CVX tool (black). Figures \figpanelNoPrefix{qubit_time}{a} and \figpanelNoPrefix{qubit_time}{b} display results for full-rank and pure states, respectively, with the x axis representing the number of qubits and the y axis indicating the total computational time (in seconds, on a logarithmic scale) required to achieve state reconstruction with fidelity greater than 0.99. A maximum of 800 iterations were used in all cases to ensure high convergence.

For full-rank states [\figpanel{qubit_time}{a}], a full-rank ansatz is used to reflect maximum time complexity. Here, CVX demonstrates superior performance for smaller systems (up to four qubits); however, for higher-dimensional systems (five qubits and beyond), the GD-QST methods with CD and SM parameterizations outperform both PN and CVX. Notably, the PN algorithm is excluded for systems with more than five qubits, and SM and CVX are excluded beyond six qubits, since they take too long to converge then. We find that GD-QST with the CD parameterization emerges as the most effective algorithm for high-rank state tomography in larger systems, with an average reconstruction time of approximately three minutes for seven qubits.

For pure-state tomography [\figpanel{qubit_time}{b}], a rank-1 ansatz is used where possible, i.e., for the GD-QST algorithms. In this case, GD-QST outperforms CVX in systems with more than three qubits. Among the GD-QST parameterizations, SM performs best, reconstructing a seven-qubit pure state in approximately 15 seconds, compared to about 30 seconds for CD and PN. Note that traditional convex optimization methods like CVX struggle with pure-state tomography due to the condition $\text{Tr}(\rho^2) = 1$, which violates the ``disciplined convex programming (DCP) ruleset," rendering them ineffective in reconstructing states with such constraints. Thus, CVX provides no advantage when dealing with different ranks, as its time complexity remains the same for both full-rank and pure states [also illustrated in \figpanel{5q-rank-analysis}{a}].

In summary, GD-QST with CD is the most effective algorithm for high-rank state tomography in larger systems while SM is the most efficient for pure state reconstruction. Here, the total computational time includes both state reconstruction and fidelity computation.


\subsubsection{\textbf{A case study: five qubits}}
\label{sec:CaseStudyFiveQubit}

\begin{figure}
\centering
\includegraphics[width=0.85\linewidth]{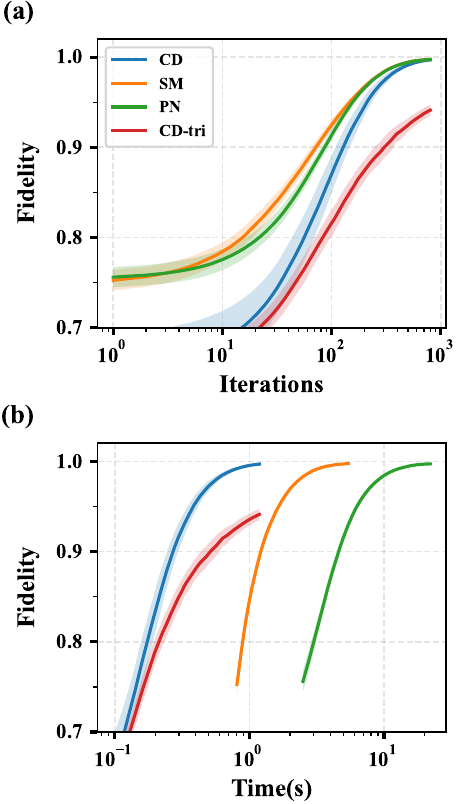}
\caption{GD-QST performance for a five-qubit system. 
(a) Reconstruction fidelity obtained using the parameterizations CD (blue), SM (orange), PN (green), and CD-tri (red), as a function of the number of iterations.
(b) Reconstruction fidelity as a function of cumulative time. The sold lines represent average fidelity values calculated over 30 full-rank random states with full rank-ansatz; shaded areas indicate respective standard deviation. }
\label{5q-iter-time}
\end{figure}

As a case study, we provide an in-depth analysis for a five-qubit system. The results of this study are shown in \figref{5q-iter-time} (iteration and time complexity), \figref{5q-rank-analysis} (complexity w.r.t.~rank-varying states and ansatzes), \figref{data-size} (handling reduced data sets), and \figref{noise} (noise robustness) in the main text, and \figref{5q-rank-vary} in \appref{app:DetailsRankVaryingAnsatz}. 

In particular, \figpanel{5q-iter-time}{a} demonstrates the performance of our GD-QST algorithms for a five-qubit system using CD (teal), SM (orange), PN (green), and CD with a lower-triangular matrix as ansatz (CD-tri, red), in terms of reconstruction fidelity (y axis) as a function of the number of iterations (x axis). Similarly, \figpanel{5q-iter-time}{b} shows fidelity as a function of time (measured in seconds, on the x axis). 

Figure~\ref{5q-iter-time} clearly shows that arbitrary five-qubit states can be tomographed with very high fidelity (infidelity approaches as low as $\approx 10^{-2}$ to $10^{-3}$) within 800 iterations and in a relatively short time: approximately one second for CD, five seconds for SM, and 15 seconds for PN, which also can be seen in \figref{qubit_time}. However, the CD-tri approach fails to achieve high fidelity within 800 iterations; hence it is omitted in most of the analysis in the following sections.


\subsubsection{\textbf{Advantage of rank-controlled ansatz}}
\label{sec:RankControlledAnsatz}

\begin{figure*}
\centering
\includegraphics[width=0.95 \linewidth]{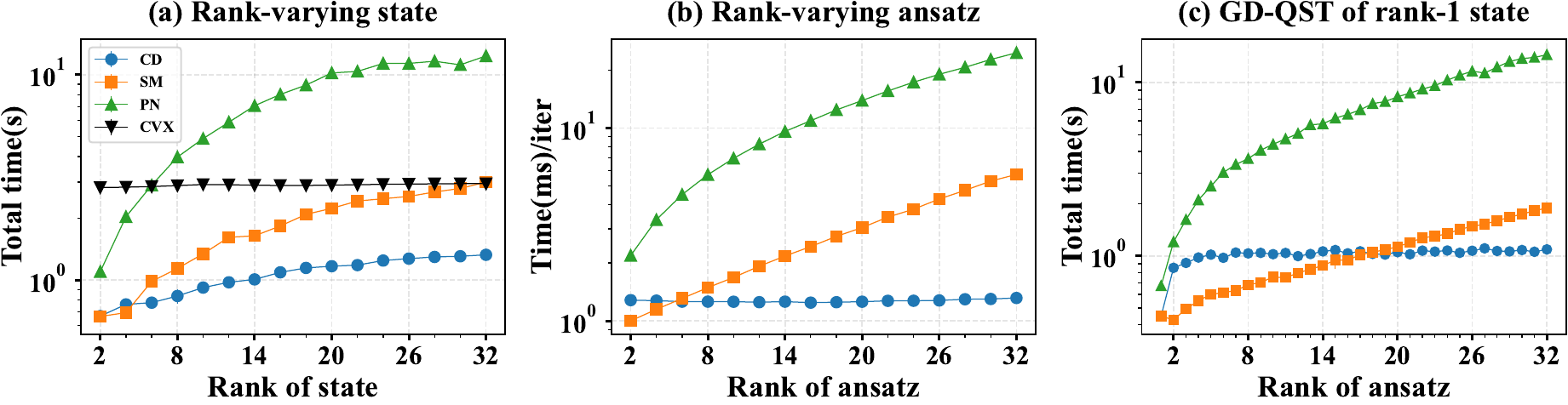}
\caption{Time complexity analysis with respect to rank-varying state and rank-varying ansatz for a five-qubit system. 
(a) Time complexity of GD-QST (with the ansatz having the same rank as the state) and CVX for reconstructing states of specific rank with $\mathcal{F} > 0.99 $.
(b) Time (in milliseconds) per iteration as a function of the rank of the ansatz. 
(c) Time complexity of GD-QST of five-qubit rank-1 states as a function of the ansatz rank.  In all scenarios, the time complexity is averaged over 30 randomly chosen states for each data point. The CVX case is not presented in panels (b) and (c) since it does not support rank-controlled ansatzes. The CD-tri case is also omitted because it does not achieve high fidelity, either within the specified number of iterations or in reasonable time. 
\label{5q-rank-analysis}}
\end{figure*}

In \figref{5q-rank-analysis}, we present a comprehensive investigation of the time complexity of GD-QST algorithms for a five-qubit system, focusing on the ranks of both the target states and the ansatzes. We highlight one of the main features of our GD-QST algorithms --- the flexibility to initialize the algorithms with an ansatz of a specified rank $r$, which significantly reduces the parameter space in the optimization process, resulting in faster computation. This approach yields the optimal $\text{rank} \leq r$ approximation of the density matrix. 

In \figpanel{5q-rank-analysis}{a}, we calculate the total time (y axis) required to reconstruct an arbitrary five-qubit density matrix of a given rank (x axis) when the ansatz has the same rank as the state, a scenario with optimal/minimum time complexity. As expected, it is evident that high-rank quantum states generally require more time for reconstruction than low-rank states. The figure also highlights that low-rank quantum states ($r \leq 7$) are reconstructed more efficiently using GD-QST algorithms than with CVX, whose performance is independent of the rank of the target state. Furthermore, across all rank values, the CD and SM algorithms demonstrate faster reconstructions than both PN and CVX. 

Furthermore, in \figpanel{5q-rank-analysis}{b} we display the time required per iteration (in milliseconds, on the y axis) as a function of the rank of the ansatz (x axis). Our numerical results show that the time required for one iteration using GD-QST with CD is independent of the ansatz rank (also shown in the first column in \figref{5q-rank-vary}), which explains why CD outperforms other algorithms in full-rank reconstruction, as shown in \figref{qubit_time}. 

Additionally, \figpanel{5q-rank-analysis}{b} illustrates that for SM and PN, selecting an appropriate rank for the ansatz can significantly reduce the total reconstruction time, since the time required for one iteration monotonically increases with the rank value (further demonstrated in the second and third columns of \figref{5q-rank-vary}, respectively). For low-rank states ($r < 6$), SM outperforms CD and PN, making it a favorable option for pure-state tomography (also shown in \figpanel{qubit_time}{b}). This is particularly relevant in experimental quantum computing and information processing, where the states of interest are predominantly pure. 

We highlight that the time to achieve faster convergence in the GD optimization process is influenced by the choice of batch size, as shown in the first row of \figref{BatchStepSize} in \appref{app:AdamHyperparameters}. This flexibility in selecting the batch size, commonly referred to as `\textit{mini-batch GD optimization}', offers a key advantage of our GD-QST algorithms, allowing for tailored optimization based on specific computational needs. 

Moreover, in \figpanel{5q-rank-analysis}{c}, we apply our GD-QST algorithms to five-qubit pure (rank-1) states. using ansatzes with different rank values, and measure the average computation time required to achieve a fidelity greater than 0.999 in each case. The numerical results indicate that selecting an appropriate ansatz rank (x axis) significantly reduces reconstruction time (y axis). In contrast, using a higher-rank ansatz offers no additional information and unnecessarily increases both computation time and cost. This feature of being able to select an appropriate ansatz with desired rank is an important advantage over standard QST methods, where the optimization space is a full-rank density matrix. 







\subsection{Reduced data sets}
\label{sec:data}

In the previous subsection, we focused on computational complexity with a complete data set. However, this complexity can be further reduced by using smaller data sets, as long as high-fidelity reconstruction is achieved. Here, we evaluate the performance of our GD-QST algorithms on reduced data sets. This analysis is particularly relevant for experiments, where acquiring tomographically complete (exponentially large) data sets becomes impractical for high-dimensional systems. For reduced data sets, the convex optimization method in \eqref{cvx} with non-zero $\lambda$ is a widely used QST protocol. However, as shown in \figref{qubit_time}, the computational time required for CVX grows exponentially with the size of the system, making it experimentally efficient but computationally inefficient for moderately sized systems. 

\begin{figure}
\centering
\includegraphics[width=0.8\linewidth]{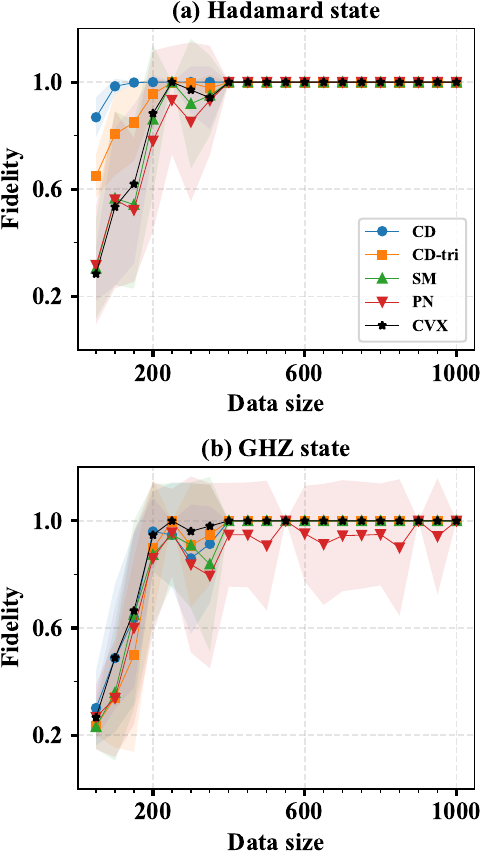}
\caption{Quantum state tomography of five-qubit (a) Hadamard and (b) GHZ states using reduced data sets of varying sizes. In both cases, the average fidelity (solid line) and the corresponding standard deviation (shaded area) are calculated by randomly sampling reduced data sets of given size from the full data set 15 times. For both cases, the rank of the ansatz for the GD-QST methods is set to one. }
\label{data-size}
\end{figure}

In Figs.~\figpanelNoPrefix{data-size}{a} and \figpanelNoPrefix{data-size}{b}, we numerically demonstrate the performance of our GD-QST algorithms using data sets of varying sizes (x axis) for two cases: (i) a five-qubit Hadamard state~\cite{hsu-prl-2024}, and (ii) a five-qubit maximally entangled GHZ state, respectively (cf.~\secref{sec:Qstates}). We show that both states can be efficiently reconstructed using substantially reduced data sets across all parameterizations; the full data set size is $4^5 = 1024$. For the Hadamard state, the CD algorithm performs clearly best, managing with a data set as small as $\approx 150$ points to achieve high-fidelity reconstruction; the other methods require $\approx 400$ data points. For the GHZ state, all methods achieve accurate reconstruction with a reduced data set size of $\approx 400$, except for the PN parameterization, which fails to produce good reconstruction results.

We note that the numerical results in \figref{data-size} are specific to the Hadamard and GHZ states and cannot be immediately generalized to other quantum states, as the data requirements for QST primarily depend on the characteristics of the target state. Low-rank and sufficiently sparse states typically require fewer measurements, whereas high-rank and less sparse states demand more data. Nevertheless, we demonstrate that GD-QST algorithms can perform full QST efficiently, even when dealing with informationally incomplete data sets, making it experimentally as well as computationally practical.

\subsection{Robustness to noise}
\label{sec:noise}

We now turn to analyzing the robustness of our GD-QST algorithms to noise in the data. In practical scenarios, experimental data often becomes corrupted due to various factors such as decoherence processes, noisy channels, imperfect measurements, statistical limitations (finite ensemble size/shots), and hardware imprecision, leading to information loss during the state reconstruction process using traditional tomography methods. In these situations, the reconstruction algorithm must be resilient to errors in the data to recover information accurately and precisely. Here, we apply our GD-QST algorithms to noisy data sets with custom-added Gaussian and depolarizing noise. We focus on the quality of the reconstruction rather than its time complexity.

We also emphasize that pure states are particularly relevant in quantum computing and information-processing tasks, where information is encoded into pure states by initializing all qubits in the ground state and performing unitary gate operations that encode information into the final output state. Ideally, the output state remains pure, but due to decoherence and unavoidable statistical and systematic errors, the target state may deviate from the pure-state space, leading to $\text{Tr}(\varrho^2)<1$, resulting in information loss when using conventional QST methods for reconstruction. In such cases, our GD-QST methods, applied with a rank-1 ansatz, ensure pure-state reconstruction by effectively mitigating and filtering out incoherent errors during the reconstruction process, enabling more accurate recovery of the desired information.

\begin{figure}
\centering
\includegraphics[width=0.8 \linewidth]{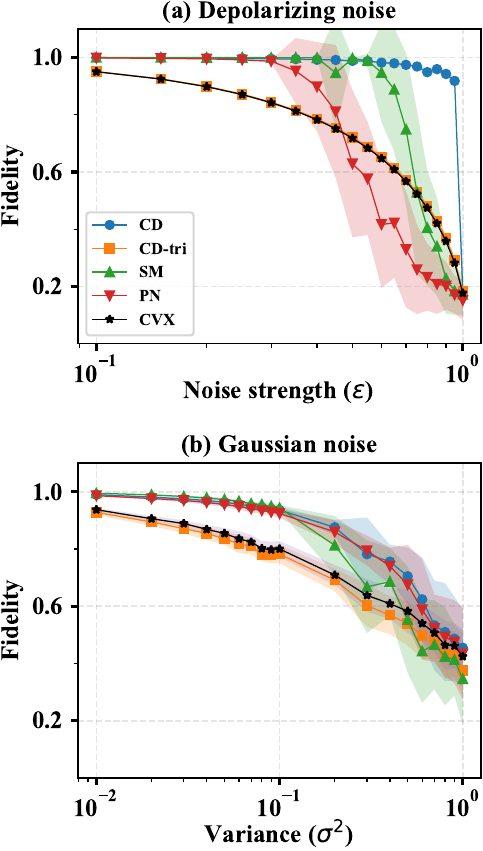}
\caption{Quantum state tomography of five-qubit pure states under (a) depolarizing and (b) Gaussian noise. In both scenarios, the average fidelity and the standard deviation are computed over 30 randomly generated pure states. The x axis in (a) shows the strength of the depolarizing noise, while the x axis in (b) is the variance of the Gaussian noise channel.}
\label{noise}
\end{figure}

In Figs.~\figpanelNoPrefix{noise}{a} and \figpanelNoPrefix{noise}{b}, we demonstrate the robustness of our GD-QST algorithms against depolarizing and Gaussian noise, respectively, for five-qubit pure states. The results highlight the algorithm's ability to recover information with greater accuracy by reconstructing quantum states with high fidelity. 

Figure~\figpanelNoPrefix{noise}{a} illustrates a scenario where the target state, containing the desired information, becomes corrupted due to a depolarizing noisy channel. This situation commonly arises when transmitting a quantum state through a quantum channel, as in quantum state transfer or distributed quantum computing protocols, leading to noisy measurement data: $\tilde{\mathcal{B}}_i = \text{Tr}(\Pi_i \varrho_{\text{depo}})$~\cite{yang-arxiv-2024}. Here, the corrupted quantum state under depolarizing noise is given by $\varrho_{\text{depo}} = (1-\varepsilon) \varrho + \frac{\varepsilon}{2^N} I$, where $\varrho$ is the original quantum state and $\varepsilon$ represents the strength of the depolarizing noise. Figure~\figpanelNoPrefix{noise}{a} shows that the GD-QST algorithms with a rank-1 ansatz are able to recover the original state, enabling more accurate recovery of encoded information than CVX and CD-tri, which lack the flexibility to choose the rank of the ansatz. Within the GD-QST framework, CD outperforms SM and PN in terms of robustness, as it reconstructs the original quantum state with higher fidelity, even at $\varepsilon = 0.9$.

Similarly, in \figpanel{noise}{b}, we consider Gaussian noise, a scenario relevant to faulty measurements caused by statistical and systematic errors. This leads to noisy measurement outcomes, $\tilde{\mathcal{B}}_i$, sampled from a Gaussian distribution with mean $\mathcal{B}_i = \text{Tr}(\Pi_i \varrho)$ and variance $\sigma^2$. The results in \figpanel{noise}{b} demonstrate that, for noise levels with variance (x axis) up to $\mathcal{O}(10^{-1})$, GD-QST with a rank-1 ansatz is able to reconstruct the original pure quantum state with sufficiently high fidelity, whereas CVX and CD-tri fail to achieve high-quality reconstruction. We observe that, in the presence of Gaussian noise, all three parameterizations (CD, SM, and PN) perform equally, with no significant advantage of one over the others.

\subsection{Continuous-variable systems}
\label{sec:CV}

We now turn from DV systems to instead implement our GD-QST algorithms for CV systems and benchmark them against two existing CV QST methods: CGANs~\cite{Ahmed2021} and iMLE~\cite{imle}. We apply our algorithms on single-mode optical cat states, as described in \secref{sec:Qstates}, using measurement data obtained from the Husimi $Q$ function, as described in \secref{sec:obs}. We set $|\xi| = r = 2 $ and choose the phase $\phi$ randomly from the interval $[0, 2\pi]$. Finally, we set the truncated Hilbert-space dimension to 32 and define the probe parameter $\beta_m = x_m + i y_m$, with $x_m, y_m \in [-4, 4]$ in steps of $32$ (a $32 \times 32$ grid), resulting in a total of 1024 measurement operators.

\begin{figure}
\centering
\includegraphics[width=0.85 \linewidth]{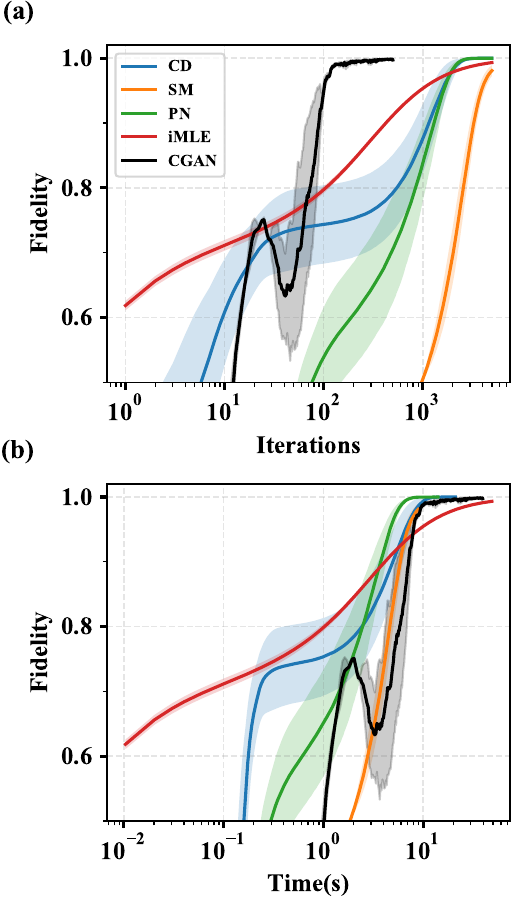}
\caption{Quantum state tomography of single-mode optical cat states. 
(a) Reconstruction fidelity obtained by CD (blue), SM (orange), PN (green), iMLE (red), and CGAN (black) as a function of the number of iterations.
(b) Reconstruction fidelity as a function of cumulative time. The fidelity is averaged over 20 randomly generated single-mode optical cat states $\ket{\text{cat}} \propto \ket{\xi} + \ket{- \xi}$ with $|\xi| =r =2$ and $\phi \in [ 0, 2\pi]$. }
\label{cv-cat-state}
\end{figure}

In \figref{cv-cat-state}, we show that the GD-QST algorithms for CV systems can reconstruct single-mode optical cat states with high fidelity in under 10 seconds. The results in \figpanel{cv-cat-state}{a} indicate that the CGAN requires only $500$ iterations to achieve high reconstruction fidelity, while other methods, including our GD-QST algorithms and iMLE, require approximately $3000-5000$ iterations. However, as depicted in \figpanel{cv-cat-state}{b}, the PN algorithm demonstrates superior performance compared to CD, SM, iMLE, and even CGAN in terms of fidelity and reconstruction time; SM and CD perform similarly to CGAN.

\begin{figure*}
\centering
\includegraphics[width=0.9\linewidth]{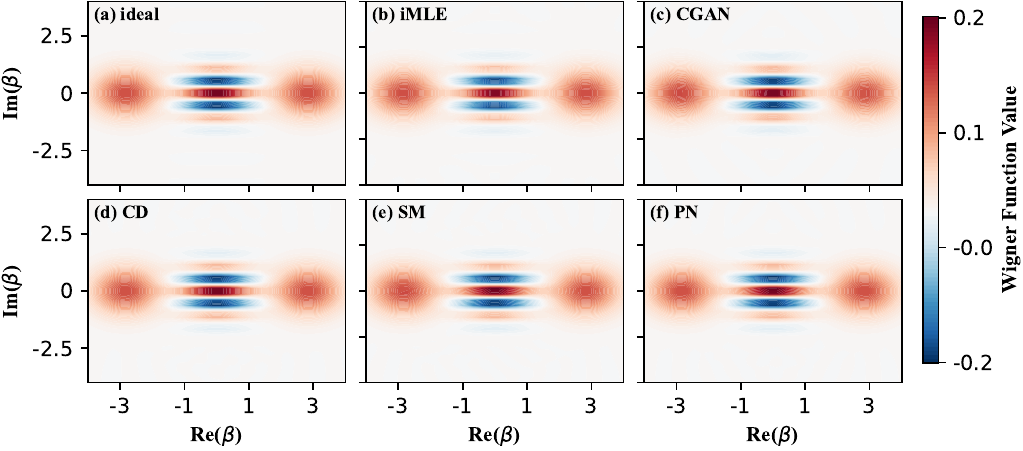}
\caption{Wigner function of the cat state $\ket{\text{cat}} \propto \ket{2} + \ket{-2}$, reconstructed from the density matrix using the (b) iMLE, (c) CGAN, (d) CD, (e) SM, and (f) PN methods. Panel (a) shows the ideal Wigner function computed from the ideal density matrix. In all cases, the state fidelity is $\mathcal{F} > 0.999$, and the color scale is consistent across all plots.}
\label{wigner_tomo}
\end{figure*}

Furthermore, in \figref{wigner_tomo}, we present a comparison between the Wigner functions of a cat state ($\xi = 2$) obtained from reconstructed density matrices employing various QST methods and the ideal Wigner function derived from the ideal density matrix. The reconstructed Wigner functions exhibit remarkable correspondence with the ideal Wigner function, affirming the successful application of the GD-QST algorithms for CV systems. All plots in \figref{wigner_tomo} are on the same color scale.


\section{Conclusion and outlook}
\label{sec:conclusion}

We have introduced several gradient-descent (GD) techniques with tailored density-matrix parameterizations, including Cholesky decomposition (CD), Stiefel manifold (SM), and projective normalization (PN), designed to support rank-controlled ansatzes for efficient quantum state tomography (QST). Through numerical simulations, we demonstrated the benefits of rank-controlled ansatzes in significantly reducing computational time complexity during QST data post-processing. Furthermore, we highlight that the ability to select the rank of an ansatz effectively mitigates decoherence effects in state reconstruction while enabling highly accurate information recovery in pure-state tomography. We demonstrated the effectiveness of our GD-QST algorithms on both discrete- and continuous-variable (DV and CV) systems, achieving full-rank QST of a seven-qubit system in under three minutes on a standard laptop with \qty{18}{\giga\byte} of RAM and no dedicated GPU. 

Moreover, our numerical analysis showed that GD-QST algorithms generally outperform other methods that are standard in the field, emphasizing the importance of selecting the appropriate parameterization for faster convergence. Specifically, for DV systems, SM was the most effective parameterization for pure states, while CD excelled for high-rank states. Interestingly, in CV systems, PN outperformed all other methods, including iMLE and CGAN. 

We further demonstrated that our GD-QST algorithms are robust against noise in data sets and efficiently handle reduced data sets, achieving highly accurate reconstruction of the underlying density matrix with excellent fidelity. Moreover, these algorithms are remarkably versatile in terms of the choice of measurement operator sets. They can seamlessly operate with Hermitian observables, as shown in DV systems, and projection operators, as demonstrated in CV systems. This adaptability contrasts with current methods, such as iMLE or accelerated projected-gradient maximum likelihood estimation~\cite{shang-pra-2017, apg-mle-code}, where existing implementations are restricted to specific types of operators only, typically a set of POVMs or projection operators, rather than a general Hermitian operator set.


The reason for the variation in performance for the different parameterizations is not fully understood, but part of the explanation may lie in the choice of measurement operators.
In the case of PN, we technically have two vectors to compute gradient updates, whereas CD and SM only use one. Intuitively, for an arbitrary set of measurement operators, PN should take more time than SM and CD, which aligns with our observations. However, the key aspect of PN is the projection step, which can project the GD-updated vector anywhere on the space of physical states. The exact projected vector (state) after the GD step depends on the measurement operators we have, particularly their structure (somehow gradient information is erased after projection unless we have some directional preference even during/after the projection step). For instance, projectors have an outer product structure, which provides projection information with a directional preference, but general Hermitian operators may not provide such directional information and partially erase gradient direction. 

Another difference between parameterizations is that the time per iteration for SM varies (it is low for low-rank ansatzes and high for high-rank ansatzes), while for CD, it remains constant. Therefore, one might expect CD to perform better overall, while SM performs better for low-rank states, assuming that the total number of iterations for both CD and SM is the same.

Considering the results presented here, we are optimistic that GD-QST can be of great use in characterizing a large variety of DV and CV systems in the ongoing development of quantum technologies. To facilitate such applications, we have made our codes for GD-QST freely available~\cite{gd-qst-python}.

As an outlook, we note that it remains an open challenge not only to fully understand which parameterization is most suited for a particular situation, but also to find the optimal hyperparameters for a given algorithm and system dimension. As another possible future research direction, the proposed GD-QST algorithms can be extended to GD-QPT for $N$-qubit processes by leveraging its connection with ancilla-assisted QPT, as implied by the state-channel duality theorem. The GD paradigm could potentially be extended also to other types of tomography, e.g., detector tomography.


\begin{acknowledgments}

The numerical calculations were performed using the QuTiP library~\cite{Johansson2012, Johansson2013, Lambert2024}.
We acknowledge support from the Knut and Alice Wallenberg Foundation through the Wallenberg Centre for Quantum Technology (WACQT) and from the Horizon Europe programme HORIZON-CL4-2022-QUANTUM-01-SGA via the project 101113946 OpenSuperQPlus100. 
AFK is also supported by the Swedish Foundation for Strategic Research (grant numbers FFL21-0279 and FUS21-0063).

\end{acknowledgments}


\appendix


\section{Detailed results on rank-varying ansatzes}
\label{app:DetailsRankVaryingAnsatz}

\begin{figure*}
\centering
\includegraphics[width=0.95\linewidth]{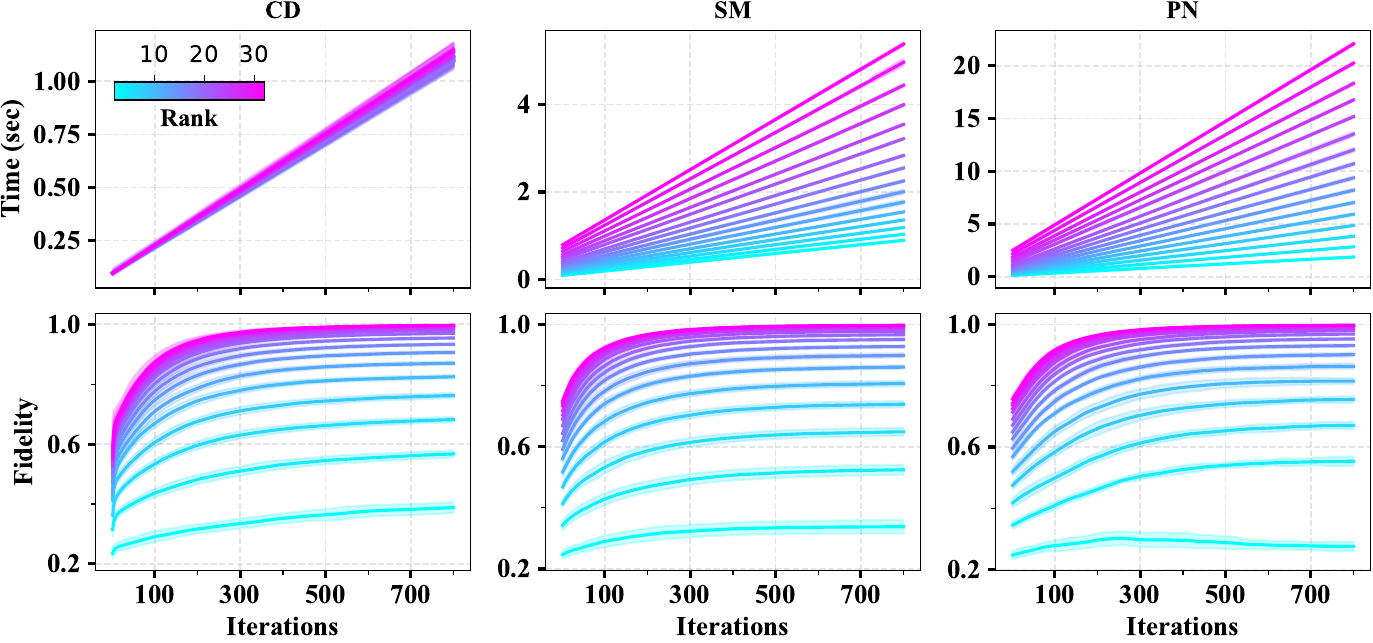}
\caption{Time complexity (time per iteration) and reconstruction quality as a function of the ansatz rank for GD-QST in a 5-qubit system. The first row shows cumulative time as a function of the number of iterations, while the second row displays the average reconstruction fidelity over 30 random states as a function of the number of iterations. The three columns correspond to, from left to right, the CD, SM, and PN parameterizations.}
\label{5q-rank-vary}
\end{figure*}

In \figref{5q-rank-analysis} in \secref{sec:RankControlledAnsatz}, we presented results on time complexity for ansatzes and states of varying rank in a five-qubit DV system. Here, in \figref{5q-rank-vary}, we provide further results for the reconstruction fidelity and the corresponding time complexity as a function of the number of iterations, for different rank values of the ansatz. The first, second, and third columns show results for the CD, SM, and PN parameterizations, respectively, while the first and second rows indicate time complexity and reconstruction fidelity, respectively. The lines follow a color scale where sky blue represents lower-rank ansatzes, while dark pink indicates higher rank. 

In all cases, it is evident that as the rank of the ansatz increases, the reconstruction fidelity also improves, as shown in the second row, which is expected. The first row of the figure shows that the cumulative time grows linearly with the number of iterations for a given ansatz. For SM and PN, our numerical results indicate that the slope (i.e., time per iteration) increases as the rank of the ansatz increases. Interestingly, for CD, the slope remains constant, regardless of the rank of the ansatz.





\section{Adam algorithm for gradient descent and hyperparameters} 
\label{app:AdamHyperparameters}


\algrenewcommand\algorithmicrequire{\textbf{Input:}}
\algrenewcommand\algorithmicensure{\textbf{Output:}}

\newcommand{\algorithmicinitialize}{\textbf{Initialize:}}
\newcommand{\Initialize}{\item[\algorithmicinitialize]}

In this appendix, we provide additional details on the algorithm we used for the GD optimization (see sec:GD-QST-Methods), and on the optimization of the hyperparameters in the GD-QST algorithms. For the case of the CD and PN parameterizations, we used the Adam optimizer~\cite{diederik2014adam}, a hybrid version of the momentum GD algorithm and the root mean square propagation (RMSP) algorithm, with decaying step size, to carry out the GD optimization. The complete parameter update process of the Adam optimization algorithm is given in Algorithm~\ref{adam}.

\begin{algorithm}[H]
\caption{Adam optimization algorithm}
\label{adam}
\begin{algorithmic}[1] 
\Require $\{ \mathcal{B}_i \}$ (data set), $\{ \Pi_i \}$ (observable set), $\lambda$ (L1-regularization), $\eta$ (step size), decay rate $(\alpha)$, batch size $(s)$, \text{max iter}
\Ensure $\varrho$ (Reconstructed density matrix)
\Initialize $\bm{\theta}_0$ (ansatz), $\beta_1 = 0.9$, $\beta_2 = 0.999$, $\bm{m}_0 = 0$, $\bm{v}_0 = 0$, $\epsilon = 10^{-8}$, $\eta_0 = \eta$
\For{$t \gets 1, ...,$ max iter}
\State $\eta_{t}\gets \alpha * \eta_{t-1}$   \Comment{\footnotesize{Decaying step size}}
\State $\bm{G}_{t} \gets \nabla \mathcal{L}[\mathcal{P}_{\varrho}(\bm{\theta}_{t-1})] $   \Comment{\footnotesize{Gradients w.r.t.~$\bm{\theta}_t$  }}
\State $\bm{m}_{t} \gets \beta_1 \cdot \bm{m}_{t-1} + \left(1 - \beta_1\right) \cdot \bm{G}_{t}$  \Comment{\footnotesize{Biased 1st moment}}
\State $\bm{v}_{t} \gets \beta_2 \cdot \bm{v}_{t-1} + \left(1 - \beta_2\right) \cdot \bm{G}_{t}^2$   \Comment{\footnotesize{Biased 2nd moment}}
\State $\bm{\hat{m}}_{t} \gets \bm{m}_{t}/(1 - \beta_1^{t})$ \Comment{\footnotesize{Bias-corrected 1st moment}}
\State $\bm{\hat{v}}_{t} \gets \bm{v}_{t}/(1 - \beta_2^{t})$   \Comment{\footnotesize{Bias-corrected 2nd moment}}
\State $\bm{\theta}_{t} \gets \bm{\theta}_{t-1} - \eta_{t} \cdot \bm{\hat{m}}_{t} /(\sqrt{\bm{\hat{v}}_{t}} + \epsilon)$  \Comment{\footnotesize{Updated ansatz }}
\EndFor
\State \textbf{Return:} $\bm{\theta}_{t},  \mathcal{P}_{\varrho}(\bm{\theta}_{t})$ \Comment{\footnotesize{Resulting ansatz and density matrix}}
\end{algorithmic} 
\end{algorithm}

\begin{figure*}
\centering
\includegraphics[width=0.93\linewidth]{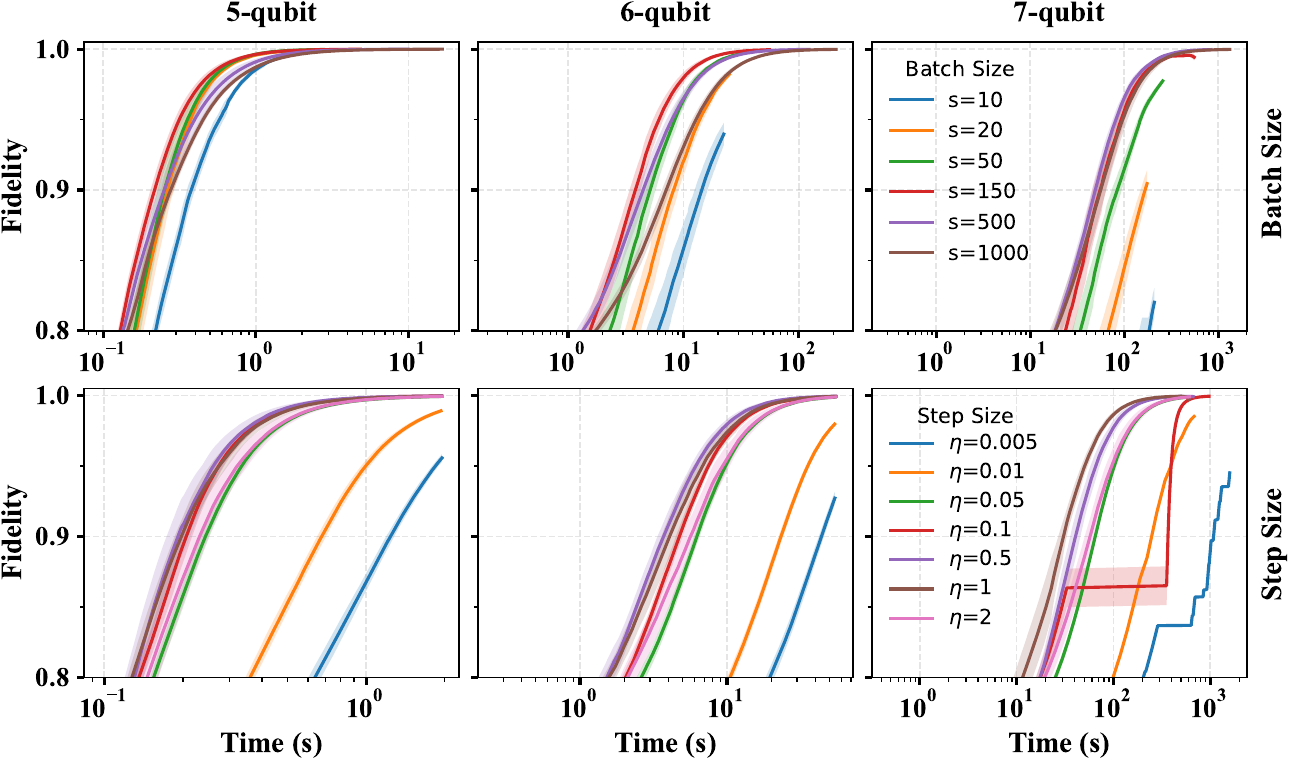}
\caption{Numerical optimization of batch size $s$ (top row) and step size $\eta$ (bottom row) for five- (first column), six- (second column), and seven-qubit (third column) systems using the CD method. The y axis shows the average reconstruction fidelity over 30 random states with a full-rank ansatz, and the x-axis indicates total time (in seconds). The optimal batch size is found to be in the range of 50 to 500, while a suitable step size lies between 0.05 and 2.}
\label{BatchStepSize}
\end{figure*}

Note that $\beta_1^{t}$ and $\beta_2^{t}$ in lines 6 and line 7 of the algorithm denote $\beta_1$ and $\beta_2$ to the power $t$, respectively, not the iteration step count. In the Adam algorithm, all vector operations are performed element-wise. We used the Python-based optimization libraries OPTAX~\cite{optax-2020} and JAX~\cite{jax-github} to perform GD optimization with Adam. 

In our GD-QST algorithms, we focus on optimizing four key hyperparameters: batch size, step size $\eta$, decay rate, and the L1-regularization parameter $\lambda$. Meanwhile, we use default values for the other hyperparameters required for the Adam algorithm, specifically: $\beta_1 = 0.9$, $\beta_2 = 0.999$, and $\epsilon = 10^{-8}$, as recommended in Ref.~\cite{diederik2014adam}. Additionally, we deliberately set the L1-regularization parameter $\lambda$ to zero in all cases, since its value primarily depends on the sparsity of the target density matrix and the target states we consider are entirely random. However, we note that when the sparsity of the target density matrix is known, choosing an appropriate $\lambda \neq 0$ and using a significantly reduced data set can enhance the reconstruction of the density matrix compared to cases with $\lambda = 0$. Similarly, we set the decay rate $(\alpha)$ to $0.999$, keeping it slightly below 1 to maintain stability near the optimum for all cases. 



Furthermore, we use a stochastic mini-batch GD technique as described in \secref{sec:GD-QST-Methods}, which not only enables optimization tailored to specific computational needs but also helps avoid becoming trapped in local minima.  Within the mini-batch gradient framework, larger batch sizes generally lead to convergence in fewer iterations, but each iteration takes longer and demands more memory usage from the computational device. On the other hand, smaller batch sizes generally require more iterations to reach convergence, but each iteration is faster. Therefore, optimizing the batch size is tricky, as the total time to achieve convergence depends on the interplay between batch size, the number of iterations, and the time per iteration. 


With other parameter values established, we optimize the batch size $s$ and the step size $\eta$. Note that the maximum batch size is the size of the entire data set, which in our case is $4^N$. In \figref{BatchStepSize}, we present results for ranges for $s$ in the top row and $\eta$ in the bottom row for a five- to seven-qubit system using the CD parameterization. It appears that a batch size between approximately 150 and 500 performs better than very small $(< 150)$ or very large $(> 500)$ values, while the optimal step size is found to be in the range of around 0.5 to 2. These values are determined statistically by applying the GD-QST algorithm to randomly selected states, aiming to achieve faster reconstruction (minimizing computation time) and improved reconstruction fidelity. However, Ref.~\cite{diederik2014adam} recommends using a batch size between $50$ and $250$, which also overlaps with the range we obtained for mini-batch GD. 

Similarly, for the SM case, the optimal range of the step size is $0.1$ to $0.5$, while for the PN case, the optimal range is on the order of $10^{-3}$. Nevertheless, one can further optimize these values by varying $\eta$ in conjunction with $\alpha$. In any case, our GD-QST algorithm, which employs the mini-batch gradient method, can be efficiently executed on a low-end computational device without requiring extensive memory resources (a standard laptop with as low as \qty{18}{\giga\byte} of RAM and no dedicated GPU is sufficient to perform full-rank seven-qubit GD-QST in under three minutes). This contrasts with the batch gradient method (which processes the entire dataset simultaneously), where devices with similar configurations are unable to perform the GD computations within a reasonable time frame.

\bibliography{References}

\begin{thebibliography}{141}%
\makeatletter
\providecommand \@ifxundefined [1]{%
 \@ifx{#1\undefined}
}%
\providecommand \@ifnum [1]{%
 \ifnum #1\expandafter \@firstoftwo
 \else \expandafter \@secondoftwo
 \fi
}%
\providecommand \@ifx [1]{%
 \ifx #1\expandafter \@firstoftwo
 \else \expandafter \@secondoftwo
 \fi
}%
\providecommand \natexlab [1]{#1}%
\providecommand \enquote  [1]{``#1''}%
\providecommand \bibnamefont  [1]{#1}%
\providecommand \bibfnamefont [1]{#1}%
\providecommand \citenamefont [1]{#1}%
\providecommand \href@noop [0]{\@secondoftwo}%
\providecommand \href [0]{\begingroup \@sanitize@url \@href}%
\providecommand \@href[1]{\@@startlink{#1}\@@href}%
\providecommand \@@href[1]{\endgroup#1\@@endlink}%
\providecommand \@sanitize@url [0]{\catcode `\\12\catcode `\$12\catcode
  `\&12\catcode `\#12\catcode `\^12\catcode `\_12\catcode `\%12\relax}%
\providecommand \@@startlink[1]{}%
\providecommand \@@endlink[0]{}%
\providecommand \url  [0]{\begingroup\@sanitize@url \@url }%
\providecommand \@url [1]{\endgroup\@href {#1}{\urlprefix }}%
\providecommand \urlprefix  [0]{URL }%
\providecommand \Eprint [0]{\href }%
\providecommand \doibase [0]{https://doi.org/}%
\providecommand \selectlanguage [0]{\@gobble}%
\providecommand \bibinfo  [0]{\@secondoftwo}%
\providecommand \bibfield  [0]{\@secondoftwo}%
\providecommand \translation [1]{[#1]}%
\providecommand \BibitemOpen [0]{}%
\providecommand \bibitemStop [0]{}%
\providecommand \bibitemNoStop [0]{.\EOS\space}%
\providecommand \EOS [0]{\spacefactor3000\relax}%
\providecommand \BibitemShut  [1]{\csname bibitem#1\endcsname}%
\let\auto@bib@innerbib\@empty
\bibitem [{\citenamefont {Feynman}(1982)}]{Feynman1982}%
  \BibitemOpen
  \bibfield  {author} {\bibinfo {author} {\bibfnamefont {R.~P.}\ \bibnamefont
  {Feynman}},\ }\bibfield  {title} {\bibinfo {title} {Simulating physics with
  computers},\ }\href {https://doi.org/10.1007/BF02650179} {\bibfield
  {journal} {\bibinfo  {journal} {International Journal of Theoretical
  Physics}\ }\textbf {\bibinfo {volume} {21}},\ \bibinfo {pages} {467}
  (\bibinfo {year} {1982})}\BibitemShut {NoStop}%
\bibitem [{\citenamefont {Madsen}\ \emph {et~al.}(2022)\citenamefont {Madsen},
  \citenamefont {Laudenbach}, \citenamefont {Askarani}, \citenamefont
  {Rortais}, \citenamefont {Vincent}, \citenamefont {Bulmer}, \citenamefont
  {Miatto}, \citenamefont {Neuhaus}, \citenamefont {Helt}, \citenamefont
  {Collins}, \citenamefont {Lita}, \citenamefont {Gerrits}, \citenamefont
  {Nam}, \citenamefont {Vaidya}, \citenamefont {Menotti}, \citenamefont
  {Dhand}, \citenamefont {Vernon}, \citenamefont {Quesada},\ and\ \citenamefont
  {Lavoie}}]{madsen-nat-2022}%
  \BibitemOpen
  \bibfield  {author} {\bibinfo {author} {\bibfnamefont {L.~S.}\ \bibnamefont
  {Madsen}}, \bibinfo {author} {\bibfnamefont {F.}~\bibnamefont {Laudenbach}},
  \bibinfo {author} {\bibfnamefont {M.~F.}\ \bibnamefont {Askarani}}, \bibinfo
  {author} {\bibfnamefont {F.}~\bibnamefont {Rortais}}, \bibinfo {author}
  {\bibfnamefont {T.}~\bibnamefont {Vincent}}, \bibinfo {author} {\bibfnamefont
  {J.~F.~F.}\ \bibnamefont {Bulmer}}, \bibinfo {author} {\bibfnamefont {F.~M.}\
  \bibnamefont {Miatto}}, \bibinfo {author} {\bibfnamefont {L.}~\bibnamefont
  {Neuhaus}}, \bibinfo {author} {\bibfnamefont {L.~G.}\ \bibnamefont {Helt}},
  \bibinfo {author} {\bibfnamefont {M.~J.}\ \bibnamefont {Collins}}, \bibinfo
  {author} {\bibfnamefont {A.~E.}\ \bibnamefont {Lita}}, \bibinfo {author}
  {\bibfnamefont {T.}~\bibnamefont {Gerrits}}, \bibinfo {author} {\bibfnamefont
  {S.~W.}\ \bibnamefont {Nam}}, \bibinfo {author} {\bibfnamefont {V.~D.}\
  \bibnamefont {Vaidya}}, \bibinfo {author} {\bibfnamefont {M.}~\bibnamefont
  {Menotti}}, \bibinfo {author} {\bibfnamefont {I.}~\bibnamefont {Dhand}},
  \bibinfo {author} {\bibfnamefont {Z.}~\bibnamefont {Vernon}}, \bibinfo
  {author} {\bibfnamefont {N.}~\bibnamefont {Quesada}},\ and\ \bibinfo {author}
  {\bibfnamefont {J.}~\bibnamefont {Lavoie}},\ }\bibfield  {title} {\bibinfo
  {title} {Quantum computational advantage with a programmable photonic
  processor},\ }\href {https://doi.org/10.1038/s41586-022-04725-x} {\bibfield
  {journal} {\bibinfo  {journal} {Nature}\ }\textbf {\bibinfo {volume} {606}},\
  \bibinfo {pages} {75} (\bibinfo {year} {2022})}\BibitemShut {NoStop}%
\bibitem [{\citenamefont {Kim}\ \emph {et~al.}(2023)\citenamefont {Kim},
  \citenamefont {Eddins}, \citenamefont {Anand}, \citenamefont {Wei},
  \citenamefont {van~den Berg}, \citenamefont {Rosenblatt}, \citenamefont
  {Nayfeh}, \citenamefont {Wu}, \citenamefont {Zaletel}, \citenamefont
  {Temme},\ and\ \citenamefont {Kandala}}]{Kim2023}%
  \BibitemOpen
  \bibfield  {author} {\bibinfo {author} {\bibfnamefont {Y.}~\bibnamefont
  {Kim}}, \bibinfo {author} {\bibfnamefont {A.}~\bibnamefont {Eddins}},
  \bibinfo {author} {\bibfnamefont {S.}~\bibnamefont {Anand}}, \bibinfo
  {author} {\bibfnamefont {K.~X.}\ \bibnamefont {Wei}}, \bibinfo {author}
  {\bibfnamefont {E.}~\bibnamefont {van~den Berg}}, \bibinfo {author}
  {\bibfnamefont {S.}~\bibnamefont {Rosenblatt}}, \bibinfo {author}
  {\bibfnamefont {H.}~\bibnamefont {Nayfeh}}, \bibinfo {author} {\bibfnamefont
  {Y.}~\bibnamefont {Wu}}, \bibinfo {author} {\bibfnamefont {M.}~\bibnamefont
  {Zaletel}}, \bibinfo {author} {\bibfnamefont {K.}~\bibnamefont {Temme}},\
  and\ \bibinfo {author} {\bibfnamefont {A.}~\bibnamefont {Kandala}},\
  }\bibfield  {title} {\bibinfo {title} {{Evidence for the utility of quantum
  computing before fault tolerance}},\ }\href
  {https://doi.org/10.1038/s41586-023-06096-3} {\bibfield  {journal} {\bibinfo
  {journal} {Nature}\ }\textbf {\bibinfo {volume} {618}},\ \bibinfo {pages}
  {500} (\bibinfo {year} {2023})}\BibitemShut {NoStop}%
\bibitem [{\citenamefont {Bluvstein}\ \emph {et~al.}(2024)\citenamefont
  {Bluvstein}, \citenamefont {Evered}, \citenamefont {Geim}, \citenamefont
  {Li}, \citenamefont {Zhou}, \citenamefont {Manovitz}, \citenamefont {Ebadi},
  \citenamefont {Cain}, \citenamefont {Kalinowski}, \citenamefont {Hangleiter},
  \citenamefont {{Bonilla Ataides}}, \citenamefont {Maskara}, \citenamefont
  {Cong}, \citenamefont {Gao}, \citenamefont {{Sales Rodriguez}}, \citenamefont
  {Karolyshyn}, \citenamefont {Semeghini}, \citenamefont {Gullans},
  \citenamefont {Greiner}, \citenamefont {Vuleti{\'{c}}},\ and\ \citenamefont
  {Lukin}}]{Bluvstein2024}%
  \BibitemOpen
  \bibfield  {author} {\bibinfo {author} {\bibfnamefont {D.}~\bibnamefont
  {Bluvstein}}, \bibinfo {author} {\bibfnamefont {S.~J.}\ \bibnamefont
  {Evered}}, \bibinfo {author} {\bibfnamefont {A.~A.}\ \bibnamefont {Geim}},
  \bibinfo {author} {\bibfnamefont {S.~H.}\ \bibnamefont {Li}}, \bibinfo
  {author} {\bibfnamefont {H.}~\bibnamefont {Zhou}}, \bibinfo {author}
  {\bibfnamefont {T.}~\bibnamefont {Manovitz}}, \bibinfo {author}
  {\bibfnamefont {S.}~\bibnamefont {Ebadi}}, \bibinfo {author} {\bibfnamefont
  {M.}~\bibnamefont {Cain}}, \bibinfo {author} {\bibfnamefont {M.}~\bibnamefont
  {Kalinowski}}, \bibinfo {author} {\bibfnamefont {D.}~\bibnamefont
  {Hangleiter}}, \bibinfo {author} {\bibfnamefont {J.~P.}\ \bibnamefont
  {{Bonilla Ataides}}}, \bibinfo {author} {\bibfnamefont {N.}~\bibnamefont
  {Maskara}}, \bibinfo {author} {\bibfnamefont {I.}~\bibnamefont {Cong}},
  \bibinfo {author} {\bibfnamefont {X.}~\bibnamefont {Gao}}, \bibinfo {author}
  {\bibfnamefont {P.}~\bibnamefont {{Sales Rodriguez}}}, \bibinfo {author}
  {\bibfnamefont {T.}~\bibnamefont {Karolyshyn}}, \bibinfo {author}
  {\bibfnamefont {G.}~\bibnamefont {Semeghini}}, \bibinfo {author}
  {\bibfnamefont {M.~J.}\ \bibnamefont {Gullans}}, \bibinfo {author}
  {\bibfnamefont {M.}~\bibnamefont {Greiner}}, \bibinfo {author} {\bibfnamefont
  {V.}~\bibnamefont {Vuleti{\'{c}}}},\ and\ \bibinfo {author} {\bibfnamefont
  {M.~D.}\ \bibnamefont {Lukin}},\ }\bibfield  {title} {\bibinfo {title}
  {{Logical quantum processor based on reconfigurable atom arrays}},\ }\href
  {https://doi.org/10.1038/s41586-023-06927-3} {\bibfield  {journal} {\bibinfo
  {journal} {Nature}\ }\textbf {\bibinfo {volume} {626}},\ \bibinfo {pages}
  {58} (\bibinfo {year} {2024})}\BibitemShut {NoStop}%
\bibitem [{\citenamefont {Acharya}\ \emph {et~al.}(2025)\citenamefont {Acharya}
  \emph {et~al.}}]{Acharya2025}%
  \BibitemOpen
  \bibfield  {author} {\bibinfo {author} {\bibfnamefont {R.}~\bibnamefont
  {Acharya}} \emph {et~al.},\ }\bibfield  {title} {\bibinfo {title} {{Quantum
  error correction below the surface code threshold}},\ }\href
  {https://doi.org/10.1038/s41586-024-08449-y} {\bibfield  {journal} {\bibinfo
  {journal} {Nature}\ }\textbf {\bibinfo {volume} {638}},\ \bibinfo {pages}
  {920} (\bibinfo {year} {2025})}\BibitemShut {NoStop}%
\bibitem [{\citenamefont {Georgescu}\ \emph {et~al.}(2014)\citenamefont
  {Georgescu}, \citenamefont {Ashhab},\ and\ \citenamefont
  {Nori}}]{Georgescu2014}%
  \BibitemOpen
  \bibfield  {author} {\bibinfo {author} {\bibfnamefont {I.~M.}\ \bibnamefont
  {Georgescu}}, \bibinfo {author} {\bibfnamefont {S.}~\bibnamefont {Ashhab}},\
  and\ \bibinfo {author} {\bibfnamefont {F.}~\bibnamefont {Nori}},\ }\bibfield
  {title} {\bibinfo {title} {{Quantum simulation}},\ }\href
  {https://doi.org/10.1103/RevModPhys.86.153} {\bibfield  {journal} {\bibinfo
  {journal} {Reviews of Modern Physics}\ }\textbf {\bibinfo {volume} {86}},\
  \bibinfo {pages} {153} (\bibinfo {year} {2014})}\BibitemShut {NoStop}%
\bibitem [{\citenamefont {Montanaro}(2016)}]{Montanaro2016}%
  \BibitemOpen
  \bibfield  {author} {\bibinfo {author} {\bibfnamefont {A.}~\bibnamefont
  {Montanaro}},\ }\bibfield  {title} {\bibinfo {title} {{Quantum algorithms: an
  overview}},\ }\href {https://doi.org/10.1038/npjqi.2015.23} {\bibfield
  {journal} {\bibinfo  {journal} {npj Quantum Information}\ }\textbf {\bibinfo
  {volume} {2}},\ \bibinfo {pages} {15023} (\bibinfo {year}
  {2016})}\BibitemShut {NoStop}%
\bibitem [{\citenamefont {Wendin}(2017)}]{Wendin2017}%
  \BibitemOpen
  \bibfield  {author} {\bibinfo {author} {\bibfnamefont {G.}~\bibnamefont
  {Wendin}},\ }\bibfield  {title} {\bibinfo {title} {{Quantum information
  processing with superconducting circuits: a review}},\ }\href
  {https://doi.org/10.1088/1361-6633/aa7e1a} {\bibfield  {journal} {\bibinfo
  {journal} {Reports on Progress in Physics}\ }\textbf {\bibinfo {volume}
  {80}},\ \bibinfo {pages} {106001} (\bibinfo {year} {2017})}\BibitemShut
  {NoStop}%
\bibitem [{\citenamefont {Preskill}(2018)}]{Preskill2018}%
  \BibitemOpen
  \bibfield  {author} {\bibinfo {author} {\bibfnamefont {J.}~\bibnamefont
  {Preskill}},\ }\bibfield  {title} {\bibinfo {title} {{Quantum Computing in
  the NISQ era and beyond}},\ }\href {https://doi.org/10.22331/q-2018-08-06-79}
  {\bibfield  {journal} {\bibinfo  {journal} {Quantum}\ }\textbf {\bibinfo
  {volume} {2}},\ \bibinfo {pages} {79} (\bibinfo {year} {2018})}\BibitemShut
  {NoStop}%
\bibitem [{\citenamefont {McArdle}\ \emph {et~al.}(2020)\citenamefont
  {McArdle}, \citenamefont {Endo}, \citenamefont {Aspuru-Guzik}, \citenamefont
  {Benjamin},\ and\ \citenamefont {Yuan}}]{McArdle2020}%
  \BibitemOpen
  \bibfield  {author} {\bibinfo {author} {\bibfnamefont {S.}~\bibnamefont
  {McArdle}}, \bibinfo {author} {\bibfnamefont {S.}~\bibnamefont {Endo}},
  \bibinfo {author} {\bibfnamefont {A.}~\bibnamefont {Aspuru-Guzik}}, \bibinfo
  {author} {\bibfnamefont {S.~C.}\ \bibnamefont {Benjamin}},\ and\ \bibinfo
  {author} {\bibfnamefont {X.}~\bibnamefont {Yuan}},\ }\bibfield  {title}
  {\bibinfo {title} {{Quantum computational chemistry}},\ }\href
  {https://doi.org/10.1103/RevModPhys.92.015003} {\bibfield  {journal}
  {\bibinfo  {journal} {Reviews of Modern Physics}\ }\textbf {\bibinfo {volume}
  {92}},\ \bibinfo {pages} {015003} (\bibinfo {year} {2020})}\BibitemShut
  {NoStop}%
\bibitem [{\citenamefont {Bauer}\ \emph {et~al.}(2020)\citenamefont {Bauer},
  \citenamefont {Bravyi}, \citenamefont {Motta},\ and\ \citenamefont {{Kin-Lic
  Chan}}}]{Bauer2020}%
  \BibitemOpen
  \bibfield  {author} {\bibinfo {author} {\bibfnamefont {B.}~\bibnamefont
  {Bauer}}, \bibinfo {author} {\bibfnamefont {S.}~\bibnamefont {Bravyi}},
  \bibinfo {author} {\bibfnamefont {M.}~\bibnamefont {Motta}},\ and\ \bibinfo
  {author} {\bibfnamefont {G.}~\bibnamefont {{Kin-Lic Chan}}},\ }\bibfield
  {title} {\bibinfo {title} {{Quantum Algorithms for Quantum Chemistry and
  Quantum Materials Science}},\ }\href
  {https://doi.org/10.1021/acs.chemrev.9b00829} {\bibfield  {journal} {\bibinfo
   {journal} {Chemical Reviews}\ }\textbf {\bibinfo {volume} {120}},\ \bibinfo
  {pages} {12685} (\bibinfo {year} {2020})}\BibitemShut {NoStop}%
\bibitem [{\citenamefont {Cerezo}\ \emph {et~al.}(2021)\citenamefont {Cerezo},
  \citenamefont {Arrasmith}, \citenamefont {Babbush}, \citenamefont {Benjamin},
  \citenamefont {Endo}, \citenamefont {Fujii}, \citenamefont {McClean},
  \citenamefont {Mitarai}, \citenamefont {Yuan}, \citenamefont {Cincio},\ and\
  \citenamefont {Coles}}]{Cerezo2021}%
  \BibitemOpen
  \bibfield  {author} {\bibinfo {author} {\bibfnamefont {M.}~\bibnamefont
  {Cerezo}}, \bibinfo {author} {\bibfnamefont {A.}~\bibnamefont {Arrasmith}},
  \bibinfo {author} {\bibfnamefont {R.}~\bibnamefont {Babbush}}, \bibinfo
  {author} {\bibfnamefont {S.~C.}\ \bibnamefont {Benjamin}}, \bibinfo {author}
  {\bibfnamefont {S.}~\bibnamefont {Endo}}, \bibinfo {author} {\bibfnamefont
  {K.}~\bibnamefont {Fujii}}, \bibinfo {author} {\bibfnamefont {J.~R.}\
  \bibnamefont {McClean}}, \bibinfo {author} {\bibfnamefont {K.}~\bibnamefont
  {Mitarai}}, \bibinfo {author} {\bibfnamefont {X.}~\bibnamefont {Yuan}},
  \bibinfo {author} {\bibfnamefont {L.}~\bibnamefont {Cincio}},\ and\ \bibinfo
  {author} {\bibfnamefont {P.~J.}\ \bibnamefont {Coles}},\ }\bibfield  {title}
  {\bibinfo {title} {{Variational quantum algorithms}},\ }\href
  {https://doi.org/10.1038/s42254-021-00348-9} {\bibfield  {journal} {\bibinfo
  {journal} {Nature Reviews Physics}\ }\textbf {\bibinfo {volume} {3}},\
  \bibinfo {pages} {625} (\bibinfo {year} {2021})}\BibitemShut {NoStop}%
\bibitem [{\citenamefont {Cerezo}\ \emph {et~al.}(2022)\citenamefont {Cerezo},
  \citenamefont {Verdon}, \citenamefont {Huang}, \citenamefont {Cincio},\ and\
  \citenamefont {Coles}}]{Cerezo2022}%
  \BibitemOpen
  \bibfield  {author} {\bibinfo {author} {\bibfnamefont {M.}~\bibnamefont
  {Cerezo}}, \bibinfo {author} {\bibfnamefont {G.}~\bibnamefont {Verdon}},
  \bibinfo {author} {\bibfnamefont {H.-Y.}\ \bibnamefont {Huang}}, \bibinfo
  {author} {\bibfnamefont {L.}~\bibnamefont {Cincio}},\ and\ \bibinfo {author}
  {\bibfnamefont {P.~J.}\ \bibnamefont {Coles}},\ }\bibfield  {title} {\bibinfo
  {title} {{Challenges and opportunities in quantum machine learning}},\ }\href
  {https://doi.org/10.1038/s43588-022-00311-3} {\bibfield  {journal} {\bibinfo
  {journal} {Nature Computational Science}\ }\textbf {\bibinfo {volume} {2}},\
  \bibinfo {pages} {567} (\bibinfo {year} {2022})}\BibitemShut {NoStop}%
\bibitem [{\citenamefont {Daley}\ \emph {et~al.}(2022)\citenamefont {Daley},
  \citenamefont {Bloch}, \citenamefont {Kokail}, \citenamefont {Flannigan},
  \citenamefont {Pearson}, \citenamefont {Troyer},\ and\ \citenamefont
  {Zoller}}]{daley-nat-2022}%
  \BibitemOpen
  \bibfield  {author} {\bibinfo {author} {\bibfnamefont {A.~J.}\ \bibnamefont
  {Daley}}, \bibinfo {author} {\bibfnamefont {I.}~\bibnamefont {Bloch}},
  \bibinfo {author} {\bibfnamefont {C.}~\bibnamefont {Kokail}}, \bibinfo
  {author} {\bibfnamefont {S.}~\bibnamefont {Flannigan}}, \bibinfo {author}
  {\bibfnamefont {N.}~\bibnamefont {Pearson}}, \bibinfo {author} {\bibfnamefont
  {M.}~\bibnamefont {Troyer}},\ and\ \bibinfo {author} {\bibfnamefont
  {P.}~\bibnamefont {Zoller}},\ }\bibfield  {title} {\bibinfo {title}
  {Practical quantum advantage in quantum simulation},\ }\href
  {https://doi.org/10.1038/s41586-022-04940-6} {\bibfield  {journal} {\bibinfo
  {journal} {Nature}\ }\textbf {\bibinfo {volume} {607}},\ \bibinfo {pages}
  {667} (\bibinfo {year} {2022})}\BibitemShut {NoStop}%
\bibitem [{\citenamefont {Dalzell}\ \emph {et~al.}(2023)\citenamefont
  {Dalzell}, \citenamefont {McArdle}, \citenamefont {Berta}, \citenamefont
  {Bienias}, \citenamefont {Chen}, \citenamefont {Gily{\'{e}}n}, \citenamefont
  {Hann}, \citenamefont {Kastoryano}, \citenamefont {Khabiboulline},
  \citenamefont {Kubica}, \citenamefont {Salton}, \citenamefont {Wang},\ and\
  \citenamefont {Brand{\~{a}}o}}]{Dalzell2023}%
  \BibitemOpen
  \bibfield  {author} {\bibinfo {author} {\bibfnamefont {A.~M.}\ \bibnamefont
  {Dalzell}}, \bibinfo {author} {\bibfnamefont {S.}~\bibnamefont {McArdle}},
  \bibinfo {author} {\bibfnamefont {M.}~\bibnamefont {Berta}}, \bibinfo
  {author} {\bibfnamefont {P.}~\bibnamefont {Bienias}}, \bibinfo {author}
  {\bibfnamefont {C.-F.}\ \bibnamefont {Chen}}, \bibinfo {author}
  {\bibfnamefont {A.}~\bibnamefont {Gily{\'{e}}n}}, \bibinfo {author}
  {\bibfnamefont {C.~T.}\ \bibnamefont {Hann}}, \bibinfo {author}
  {\bibfnamefont {M.~J.}\ \bibnamefont {Kastoryano}}, \bibinfo {author}
  {\bibfnamefont {E.~T.}\ \bibnamefont {Khabiboulline}}, \bibinfo {author}
  {\bibfnamefont {A.}~\bibnamefont {Kubica}}, \bibinfo {author} {\bibfnamefont
  {G.}~\bibnamefont {Salton}}, \bibinfo {author} {\bibfnamefont
  {S.}~\bibnamefont {Wang}},\ and\ \bibinfo {author} {\bibfnamefont {F.~G.
  S.~L.}\ \bibnamefont {Brand{\~{a}}o}},\ }\href@noop {} {\bibinfo {title}
  {{Quantum algorithms: A survey of applications and end-to-end complexities}}}
  (\bibinfo {year} {2023}),\ \Eprint {https://arxiv.org/abs/2310.03011}
  {arXiv:2310.03011} \BibitemShut {NoStop}%
\bibitem [{\citenamefont {Giovannetti}\ \emph {et~al.}(2011)\citenamefont
  {Giovannetti}, \citenamefont {Lloyd},\ and\ \citenamefont
  {Maccone}}]{Giovannetti2011}%
  \BibitemOpen
  \bibfield  {author} {\bibinfo {author} {\bibfnamefont {V.}~\bibnamefont
  {Giovannetti}}, \bibinfo {author} {\bibfnamefont {S.}~\bibnamefont {Lloyd}},\
  and\ \bibinfo {author} {\bibfnamefont {L.}~\bibnamefont {Maccone}},\
  }\bibfield  {title} {\bibinfo {title} {{Advances in quantum metrology}},\
  }\href {https://doi.org/10.1038/nphoton.2011.35} {\bibfield  {journal}
  {\bibinfo  {journal} {Nature Photonics}\ }\textbf {\bibinfo {volume} {5}},\
  \bibinfo {pages} {222} (\bibinfo {year} {2011})}\BibitemShut {NoStop}%
\bibitem [{\citenamefont {Degen}\ \emph {et~al.}(2017)\citenamefont {Degen},
  \citenamefont {Reinhard},\ and\ \citenamefont {Cappellaro}}]{Degen2017}%
  \BibitemOpen
  \bibfield  {author} {\bibinfo {author} {\bibfnamefont {C.~L.}\ \bibnamefont
  {Degen}}, \bibinfo {author} {\bibfnamefont {F.}~\bibnamefont {Reinhard}},\
  and\ \bibinfo {author} {\bibfnamefont {P.}~\bibnamefont {Cappellaro}},\
  }\bibfield  {title} {\bibinfo {title} {{Quantum sensing}},\ }\href
  {https://doi.org/10.1103/RevModPhys.89.035002} {\bibfield  {journal}
  {\bibinfo  {journal} {Reviews of Modern Physics}\ }\textbf {\bibinfo {volume}
  {89}},\ \bibinfo {pages} {035002} (\bibinfo {year} {2017})}\BibitemShut
  {NoStop}%
\bibitem [{\citenamefont {Aslam}\ \emph {et~al.}(2023)\citenamefont {Aslam},
  \citenamefont {Zhou}, \citenamefont {Urbach}, \citenamefont {Turner},
  \citenamefont {Walsworth}, \citenamefont {Lukin},\ and\ \citenamefont
  {Park}}]{Aslam2023}%
  \BibitemOpen
  \bibfield  {author} {\bibinfo {author} {\bibfnamefont {N.}~\bibnamefont
  {Aslam}}, \bibinfo {author} {\bibfnamefont {H.}~\bibnamefont {Zhou}},
  \bibinfo {author} {\bibfnamefont {E.~K.}\ \bibnamefont {Urbach}}, \bibinfo
  {author} {\bibfnamefont {M.~J.}\ \bibnamefont {Turner}}, \bibinfo {author}
  {\bibfnamefont {R.~L.}\ \bibnamefont {Walsworth}}, \bibinfo {author}
  {\bibfnamefont {M.~D.}\ \bibnamefont {Lukin}},\ and\ \bibinfo {author}
  {\bibfnamefont {H.}~\bibnamefont {Park}},\ }\bibfield  {title} {\bibinfo
  {title} {{Quantum sensors for biomedical applications}},\ }\href
  {https://doi.org/10.1038/s42254-023-00558-3} {\bibfield  {journal} {\bibinfo
  {journal} {Nature Reviews Physics}\ }\textbf {\bibinfo {volume} {5}},\
  \bibinfo {pages} {157} (\bibinfo {year} {2023})}\BibitemShut {NoStop}%
\bibitem [{\citenamefont {Bass}\ and\ \citenamefont {Doser}(2024)}]{Bass2024}%
  \BibitemOpen
  \bibfield  {author} {\bibinfo {author} {\bibfnamefont {S.~D.}\ \bibnamefont
  {Bass}}\ and\ \bibinfo {author} {\bibfnamefont {M.}~\bibnamefont {Doser}},\
  }\bibfield  {title} {\bibinfo {title} {{Quantum sensing for particle
  physics}},\ }\href {https://doi.org/10.1038/s42254-024-00714-3} {\bibfield
  {journal} {\bibinfo  {journal} {Nature Reviews Physics}\ }\textbf {\bibinfo
  {volume} {6}},\ \bibinfo {pages} {329} (\bibinfo {year} {2024})}\BibitemShut
  {NoStop}%
\bibitem [{\citenamefont {Rovny}\ \emph {et~al.}(2024)\citenamefont {Rovny},
  \citenamefont {Gopalakrishnan}, \citenamefont {Jayich}, \citenamefont
  {Maletinsky}, \citenamefont {Demler},\ and\ \citenamefont
  {de~Leon}}]{Rovny2024}%
  \BibitemOpen
  \bibfield  {author} {\bibinfo {author} {\bibfnamefont {J.}~\bibnamefont
  {Rovny}}, \bibinfo {author} {\bibfnamefont {S.}~\bibnamefont
  {Gopalakrishnan}}, \bibinfo {author} {\bibfnamefont {A.~C.~B.}\ \bibnamefont
  {Jayich}}, \bibinfo {author} {\bibfnamefont {P.}~\bibnamefont {Maletinsky}},
  \bibinfo {author} {\bibfnamefont {E.}~\bibnamefont {Demler}},\ and\ \bibinfo
  {author} {\bibfnamefont {N.~P.}\ \bibnamefont {de~Leon}},\ }\bibfield
  {title} {\bibinfo {title} {{Nanoscale diamond quantum sensors for many-body
  physics}},\ }\href {https://doi.org/10.1038/s42254-024-00775-4} {\bibfield
  {journal} {\bibinfo  {journal} {Nature Reviews Physics}\ }\textbf {\bibinfo
  {volume} {6}},\ \bibinfo {pages} {753} (\bibinfo {year} {2024})}\BibitemShut
  {NoStop}%
\bibitem [{\citenamefont {DeMille}\ \emph {et~al.}(2024)\citenamefont
  {DeMille}, \citenamefont {Hutzler}, \citenamefont {Rey},\ and\ \citenamefont
  {Zelevinsky}}]{DeMille2024}%
  \BibitemOpen
  \bibfield  {author} {\bibinfo {author} {\bibfnamefont {D.}~\bibnamefont
  {DeMille}}, \bibinfo {author} {\bibfnamefont {N.~R.}\ \bibnamefont
  {Hutzler}}, \bibinfo {author} {\bibfnamefont {A.~M.}\ \bibnamefont {Rey}},\
  and\ \bibinfo {author} {\bibfnamefont {T.}~\bibnamefont {Zelevinsky}},\
  }\bibfield  {title} {\bibinfo {title} {{Quantum sensing and metrology for
  fundamental physics with molecules}},\ }\href
  {https://doi.org/10.1038/s41567-024-02499-9} {\bibfield  {journal} {\bibinfo
  {journal} {Nature Physics}\ }\textbf {\bibinfo {volume} {20}},\ \bibinfo
  {pages} {741} (\bibinfo {year} {2024})}\BibitemShut {NoStop}%
\bibitem [{\citenamefont {Gisin}\ and\ \citenamefont {Thew}(2007)}]{Gisin2007}%
  \BibitemOpen
  \bibfield  {author} {\bibinfo {author} {\bibfnamefont {N.}~\bibnamefont
  {Gisin}}\ and\ \bibinfo {author} {\bibfnamefont {R.}~\bibnamefont {Thew}},\
  }\bibfield  {title} {\bibinfo {title} {{Quantum communication}},\ }\href
  {https://doi.org/10.1038/nphoton.2007.22} {\bibfield  {journal} {\bibinfo
  {journal} {Nature Photonics}\ }\textbf {\bibinfo {volume} {1}},\ \bibinfo
  {pages} {165} (\bibinfo {year} {2007})}\BibitemShut {NoStop}%
\bibitem [{\citenamefont {Wehner}\ \emph {et~al.}(2018)\citenamefont {Wehner},
  \citenamefont {Elkouss},\ and\ \citenamefont {Hanson}}]{Wehner2018}%
  \BibitemOpen
  \bibfield  {author} {\bibinfo {author} {\bibfnamefont {S.}~\bibnamefont
  {Wehner}}, \bibinfo {author} {\bibfnamefont {D.}~\bibnamefont {Elkouss}},\
  and\ \bibinfo {author} {\bibfnamefont {R.}~\bibnamefont {Hanson}},\
  }\bibfield  {title} {\bibinfo {title} {{Quantum internet: A vision for the
  road ahead}},\ }\href {https://doi.org/10.1126/science.aam9288} {\bibfield
  {journal} {\bibinfo  {journal} {Science}\ }\textbf {\bibinfo {volume}
  {362}},\ \bibinfo {pages} {eaam9288} (\bibinfo {year} {2018})}\BibitemShut
  {NoStop}%
\bibitem [{\citenamefont {Azuma}\ \emph {et~al.}(2023)\citenamefont {Azuma},
  \citenamefont {Economou}, \citenamefont {Elkouss}, \citenamefont {Hilaire},
  \citenamefont {Jiang}, \citenamefont {Lo},\ and\ \citenamefont
  {Tzitrin}}]{Azuma2023}%
  \BibitemOpen
  \bibfield  {author} {\bibinfo {author} {\bibfnamefont {K.}~\bibnamefont
  {Azuma}}, \bibinfo {author} {\bibfnamefont {S.~E.}\ \bibnamefont {Economou}},
  \bibinfo {author} {\bibfnamefont {D.}~\bibnamefont {Elkouss}}, \bibinfo
  {author} {\bibfnamefont {P.}~\bibnamefont {Hilaire}}, \bibinfo {author}
  {\bibfnamefont {L.}~\bibnamefont {Jiang}}, \bibinfo {author} {\bibfnamefont
  {H.-K.}\ \bibnamefont {Lo}},\ and\ \bibinfo {author} {\bibfnamefont
  {I.}~\bibnamefont {Tzitrin}},\ }\bibfield  {title} {\bibinfo {title}
  {{Quantum repeaters: From quantum networks to the quantum internet}},\ }\href
  {https://doi.org/10.1103/RevModPhys.95.045006} {\bibfield  {journal}
  {\bibinfo  {journal} {Reviews of Modern Physics}\ }\textbf {\bibinfo {volume}
  {95}},\ \bibinfo {pages} {045006} (\bibinfo {year} {2023})}\BibitemShut
  {NoStop}%
\bibitem [{\citenamefont {Gebhart}\ \emph {et~al.}(2023)\citenamefont
  {Gebhart}, \citenamefont {Santagati}, \citenamefont {Gentile}, \citenamefont
  {Gauger}, \citenamefont {Craig}, \citenamefont {Ares}, \citenamefont
  {Banchi}, \citenamefont {Marquardt}, \citenamefont {Pezz{\`{e}}},\ and\
  \citenamefont {Bonato}}]{Gebhart2023}%
  \BibitemOpen
  \bibfield  {author} {\bibinfo {author} {\bibfnamefont {V.}~\bibnamefont
  {Gebhart}}, \bibinfo {author} {\bibfnamefont {R.}~\bibnamefont {Santagati}},
  \bibinfo {author} {\bibfnamefont {A.~A.}\ \bibnamefont {Gentile}}, \bibinfo
  {author} {\bibfnamefont {E.~M.}\ \bibnamefont {Gauger}}, \bibinfo {author}
  {\bibfnamefont {D.}~\bibnamefont {Craig}}, \bibinfo {author} {\bibfnamefont
  {N.}~\bibnamefont {Ares}}, \bibinfo {author} {\bibfnamefont {L.}~\bibnamefont
  {Banchi}}, \bibinfo {author} {\bibfnamefont {F.}~\bibnamefont {Marquardt}},
  \bibinfo {author} {\bibfnamefont {L.}~\bibnamefont {Pezz{\`{e}}}},\ and\
  \bibinfo {author} {\bibfnamefont {C.}~\bibnamefont {Bonato}},\ }\bibfield
  {title} {\bibinfo {title} {{Learning quantum systems}},\ }\href
  {https://doi.org/10.1038/s42254-022-00552-1} {\bibfield  {journal} {\bibinfo
  {journal} {Nature Reviews Physics}\ }\textbf {\bibinfo {volume} {5}},\
  \bibinfo {pages} {141} (\bibinfo {year} {2023})}\BibitemShut {NoStop}%
\bibitem [{\citenamefont {Hashim}\ \emph {et~al.}(2024)\citenamefont {Hashim},
  \citenamefont {Nguyen}, \citenamefont {Goss}, \citenamefont {Marinelli},
  \citenamefont {Naik}, \citenamefont {Chistolini}, \citenamefont {Hines},
  \citenamefont {Marceaux}, \citenamefont {Kim}, \citenamefont {Gokhale},
  \citenamefont {Tomesh}, \citenamefont {Chen}, \citenamefont {Jiang},
  \citenamefont {Ferracin}, \citenamefont {Rudinger}, \citenamefont {Proctor},
  \citenamefont {Young}, \citenamefont {Blume-Kohout},\ and\ \citenamefont
  {Siddiqi}}]{robin-arxiv-2024}%
  \BibitemOpen
  \bibfield  {author} {\bibinfo {author} {\bibfnamefont {A.}~\bibnamefont
  {Hashim}}, \bibinfo {author} {\bibfnamefont {L.~B.}\ \bibnamefont {Nguyen}},
  \bibinfo {author} {\bibfnamefont {N.}~\bibnamefont {Goss}}, \bibinfo {author}
  {\bibfnamefont {B.}~\bibnamefont {Marinelli}}, \bibinfo {author}
  {\bibfnamefont {R.~K.}\ \bibnamefont {Naik}}, \bibinfo {author}
  {\bibfnamefont {T.}~\bibnamefont {Chistolini}}, \bibinfo {author}
  {\bibfnamefont {J.}~\bibnamefont {Hines}}, \bibinfo {author} {\bibfnamefont
  {J.~P.}\ \bibnamefont {Marceaux}}, \bibinfo {author} {\bibfnamefont
  {Y.}~\bibnamefont {Kim}}, \bibinfo {author} {\bibfnamefont {P.}~\bibnamefont
  {Gokhale}}, \bibinfo {author} {\bibfnamefont {T.}~\bibnamefont {Tomesh}},
  \bibinfo {author} {\bibfnamefont {S.}~\bibnamefont {Chen}}, \bibinfo {author}
  {\bibfnamefont {L.}~\bibnamefont {Jiang}}, \bibinfo {author} {\bibfnamefont
  {S.}~\bibnamefont {Ferracin}}, \bibinfo {author} {\bibfnamefont
  {K.}~\bibnamefont {Rudinger}}, \bibinfo {author} {\bibfnamefont
  {T.}~\bibnamefont {Proctor}}, \bibinfo {author} {\bibfnamefont {K.~C.}\
  \bibnamefont {Young}}, \bibinfo {author} {\bibfnamefont {R.}~\bibnamefont
  {Blume-Kohout}},\ and\ \bibinfo {author} {\bibfnamefont {I.}~\bibnamefont
  {Siddiqi}},\ }\href@noop {} {\bibinfo {title} {A practical introduction to
  benchmarking and characterization of quantum computers}} (\bibinfo {year}
  {2024}),\ \Eprint {https://arxiv.org/abs/2408.12064} {arXiv:2408.12064}
  \BibitemShut {NoStop}%
\bibitem [{\citenamefont {James}\ \emph {et~al.}(2001)\citenamefont {James},
  \citenamefont {Kwiat}, \citenamefont {Munro},\ and\ \citenamefont
  {White}}]{james-pra-2001}%
  \BibitemOpen
  \bibfield  {author} {\bibinfo {author} {\bibfnamefont {D.~F.~V.}\
  \bibnamefont {James}}, \bibinfo {author} {\bibfnamefont {P.~G.}\ \bibnamefont
  {Kwiat}}, \bibinfo {author} {\bibfnamefont {W.~J.}\ \bibnamefont {Munro}},\
  and\ \bibinfo {author} {\bibfnamefont {A.~G.}\ \bibnamefont {White}},\
  }\bibfield  {title} {\bibinfo {title} {Measurement of qubits},\ }\href
  {https://doi.org/10.1103/PhysRevA.64.052312} {\bibfield  {journal} {\bibinfo
  {journal} {Physical Review A}\ }\textbf {\bibinfo {volume} {64}},\ \bibinfo
  {pages} {052312} (\bibinfo {year} {2001})}\BibitemShut {NoStop}%
\bibitem [{\citenamefont {Liu}\ \emph {et~al.}(2005)\citenamefont {Liu},
  \citenamefont {Wei},\ and\ \citenamefont {Nori}}]{liu-prb-2005}%
  \BibitemOpen
  \bibfield  {author} {\bibinfo {author} {\bibfnamefont {Y.-x.}\ \bibnamefont
  {Liu}}, \bibinfo {author} {\bibfnamefont {L.~F.}\ \bibnamefont {Wei}},\ and\
  \bibinfo {author} {\bibfnamefont {F.}~\bibnamefont {Nori}},\ }\bibfield
  {title} {\bibinfo {title} {Tomographic measurements on superconducting qubit
  states},\ }\href {https://doi.org/10.1103/PhysRevB.72.014547} {\bibfield
  {journal} {\bibinfo  {journal} {Physical Review B}\ }\textbf {\bibinfo
  {volume} {72}},\ \bibinfo {pages} {014547} (\bibinfo {year}
  {2005})}\BibitemShut {NoStop}%
\bibitem [{\citenamefont {O'Brien}\ \emph {et~al.}(2004)\citenamefont
  {O'Brien}, \citenamefont {Pryde}, \citenamefont {Gilchrist}, \citenamefont
  {James}, \citenamefont {Langford}, \citenamefont {Ralph},\ and\ \citenamefont
  {White}}]{white-pra-2004}%
  \BibitemOpen
  \bibfield  {author} {\bibinfo {author} {\bibfnamefont {J.~L.}\ \bibnamefont
  {O'Brien}}, \bibinfo {author} {\bibfnamefont {G.~J.}\ \bibnamefont {Pryde}},
  \bibinfo {author} {\bibfnamefont {A.}~\bibnamefont {Gilchrist}}, \bibinfo
  {author} {\bibfnamefont {D.~F.~V.}\ \bibnamefont {James}}, \bibinfo {author}
  {\bibfnamefont {N.~K.}\ \bibnamefont {Langford}}, \bibinfo {author}
  {\bibfnamefont {T.~C.}\ \bibnamefont {Ralph}},\ and\ \bibinfo {author}
  {\bibfnamefont {A.~G.}\ \bibnamefont {White}},\ }\bibfield  {title} {\bibinfo
  {title} {Quantum process tomography of a controlled-not gate},\ }\href
  {https://doi.org/10.1103/PhysRevLett.93.080502} {\bibfield  {journal}
  {\bibinfo  {journal} {Physical Review Letters}\ }\textbf {\bibinfo {volume}
  {93}},\ \bibinfo {pages} {080502} (\bibinfo {year} {2004})}\BibitemShut
  {NoStop}%
\bibitem [{\citenamefont {Lvovsky}\ and\ \citenamefont
  {Raymer}(2009)}]{lvo-rmp-2009}%
  \BibitemOpen
  \bibfield  {author} {\bibinfo {author} {\bibfnamefont {A.~I.}\ \bibnamefont
  {Lvovsky}}\ and\ \bibinfo {author} {\bibfnamefont {M.~G.}\ \bibnamefont
  {Raymer}},\ }\bibfield  {title} {\bibinfo {title} {Continuous-variable
  optical quantum-state tomography},\ }\href
  {https://doi.org/10.1103/RevModPhys.81.299} {\bibfield  {journal} {\bibinfo
  {journal} {Reviews of Modern Physics}\ }\textbf {\bibinfo {volume} {81}},\
  \bibinfo {pages} {299} (\bibinfo {year} {2009})}\BibitemShut {NoStop}%
\bibitem [{\citenamefont {Chuang}\ and\ \citenamefont
  {Nielsen}(1997)}]{chuang-jmo-09}%
  \BibitemOpen
  \bibfield  {author} {\bibinfo {author} {\bibfnamefont {I.~L.}\ \bibnamefont
  {Chuang}}\ and\ \bibinfo {author} {\bibfnamefont {M.~A.}\ \bibnamefont
  {Nielsen}},\ }\bibfield  {title} {\bibinfo {title} {Prescription for
  experimental determination of the dynamics of a quantum black box},\ }\href
  {https://doi.org/10.1080/09500349708231894} {\bibfield  {journal} {\bibinfo
  {journal} {Journal of Modern Optics}\ }\textbf {\bibinfo {volume} {44}},\
  \bibinfo {pages} {2455} (\bibinfo {year} {1997})}\BibitemShut {NoStop}%
\bibitem [{\citenamefont {Choi}(1975)}]{choi-laa-1975}%
  \BibitemOpen
  \bibfield  {author} {\bibinfo {author} {\bibfnamefont {M.-D.}\ \bibnamefont
  {Choi}},\ }\bibfield  {title} {\bibinfo {title} {Completely positive linear
  maps on complex matrices},\ }\href
  {https://doi.org/10.1016/0024-3795(75)90075-0} {\bibfield  {journal}
  {\bibinfo  {journal} {Linear Algebra and its Applications}\ }\textbf
  {\bibinfo {volume} {10}},\ \bibinfo {pages} {285} (\bibinfo {year}
  {1975})}\BibitemShut {NoStop}%
\bibitem [{\citenamefont {Leung}(2003)}]{leung-2003}%
  \BibitemOpen
  \bibfield  {author} {\bibinfo {author} {\bibfnamefont {D.~W.}\ \bibnamefont
  {Leung}},\ }\bibfield  {title} {\bibinfo {title} {{Choi's proof as a recipe
  for quantum process tomography}},\ }\href {https://doi.org/10.1063/1.1518554}
  {\bibfield  {journal} {\bibinfo  {journal} {Journal of Mathematical Physics}\
  }\textbf {\bibinfo {volume} {44}},\ \bibinfo {pages} {528} (\bibinfo {year}
  {2003})}\BibitemShut {NoStop}%
\bibitem [{\citenamefont {Jiang}\ \emph {et~al.}(2013)\citenamefont {Jiang},
  \citenamefont {Luo},\ and\ \citenamefont {Fu}}]{jiang-pra-2013}%
  \BibitemOpen
  \bibfield  {author} {\bibinfo {author} {\bibfnamefont {M.}~\bibnamefont
  {Jiang}}, \bibinfo {author} {\bibfnamefont {S.}~\bibnamefont {Luo}},\ and\
  \bibinfo {author} {\bibfnamefont {S.}~\bibnamefont {Fu}},\ }\bibfield
  {title} {\bibinfo {title} {Channel-state duality},\ }\href
  {https://doi.org/10.1103/PhysRevA.87.022310} {\bibfield  {journal} {\bibinfo
  {journal} {Physical Review A}\ }\textbf {\bibinfo {volume} {87}},\ \bibinfo
  {pages} {022310} (\bibinfo {year} {2013})}\BibitemShut {NoStop}%
\bibitem [{\citenamefont {Riofr{\'i}o}\ \emph {et~al.}(2017)\citenamefont
  {Riofr{\'i}o}, \citenamefont {Gross}, \citenamefont {Flammia}, \citenamefont
  {Monz}, \citenamefont {Nigg}, \citenamefont {Blatt},\ and\ \citenamefont
  {Eisert}}]{riofrio-2017}%
  \BibitemOpen
  \bibfield  {author} {\bibinfo {author} {\bibfnamefont {C.~A.}\ \bibnamefont
  {Riofr{\'i}o}}, \bibinfo {author} {\bibfnamefont {D.}~\bibnamefont {Gross}},
  \bibinfo {author} {\bibfnamefont {S.~T.}\ \bibnamefont {Flammia}}, \bibinfo
  {author} {\bibfnamefont {T.}~\bibnamefont {Monz}}, \bibinfo {author}
  {\bibfnamefont {D.}~\bibnamefont {Nigg}}, \bibinfo {author} {\bibfnamefont
  {R.}~\bibnamefont {Blatt}},\ and\ \bibinfo {author} {\bibfnamefont
  {J.}~\bibnamefont {Eisert}},\ }\bibfield  {title} {\bibinfo {title}
  {Experimental quantum compressed sensing for a seven-qubit system},\ }\href
  {https://doi.org/10.1038/ncomms15305} {\bibfield  {journal} {\bibinfo
  {journal} {Nature Communications}\ }\textbf {\bibinfo {volume} {8}},\
  \bibinfo {pages} {15305} (\bibinfo {year} {2017})}\BibitemShut {NoStop}%
\bibitem [{\citenamefont {Li}\ \emph {et~al.}(2017)\citenamefont {Li},
  \citenamefont {Huang}, \citenamefont {Luo}, \citenamefont {Li}, \citenamefont
  {Lu},\ and\ \citenamefont {Zeng}}]{li-pra-2017}%
  \BibitemOpen
  \bibfield  {author} {\bibinfo {author} {\bibfnamefont {J.}~\bibnamefont
  {Li}}, \bibinfo {author} {\bibfnamefont {S.}~\bibnamefont {Huang}}, \bibinfo
  {author} {\bibfnamefont {Z.}~\bibnamefont {Luo}}, \bibinfo {author}
  {\bibfnamefont {K.}~\bibnamefont {Li}}, \bibinfo {author} {\bibfnamefont
  {D.}~\bibnamefont {Lu}},\ and\ \bibinfo {author} {\bibfnamefont
  {B.}~\bibnamefont {Zeng}},\ }\bibfield  {title} {\bibinfo {title} {Optimal
  design of measurement settings for quantum-state-tomography experiments},\
  }\href {https://doi.org/10.1103/PhysRevA.96.032307} {\bibfield  {journal}
  {\bibinfo  {journal} {Physical Review A}\ }\textbf {\bibinfo {volume} {96}},\
  \bibinfo {pages} {032307} (\bibinfo {year} {2017})}\BibitemShut {NoStop}%
\bibitem [{\citenamefont {Cotler}\ and\ \citenamefont
  {Wilczek}(2020)}]{cotler-prl-2020}%
  \BibitemOpen
  \bibfield  {author} {\bibinfo {author} {\bibfnamefont {J.}~\bibnamefont
  {Cotler}}\ and\ \bibinfo {author} {\bibfnamefont {F.}~\bibnamefont
  {Wilczek}},\ }\bibfield  {title} {\bibinfo {title} {{Quantum Overlapping
  Tomography}},\ }\href {https://doi.org/10.1103/PhysRevLett.124.100401}
  {\bibfield  {journal} {\bibinfo  {journal} {Physical Review Letters}\
  }\textbf {\bibinfo {volume} {124}},\ \bibinfo {pages} {100401} (\bibinfo
  {year} {2020})}\BibitemShut {NoStop}%
\bibitem [{\citenamefont {Banaszek}\ \emph {et~al.}(1999)\citenamefont
  {Banaszek}, \citenamefont {D'Ariano}, \citenamefont {Paris},\ and\
  \citenamefont {Sacchi}}]{ban-pra-1999}%
  \BibitemOpen
  \bibfield  {author} {\bibinfo {author} {\bibfnamefont {K.}~\bibnamefont
  {Banaszek}}, \bibinfo {author} {\bibfnamefont {G.~M.}\ \bibnamefont
  {D'Ariano}}, \bibinfo {author} {\bibfnamefont {M.~G.~A.}\ \bibnamefont
  {Paris}},\ and\ \bibinfo {author} {\bibfnamefont {M.~F.}\ \bibnamefont
  {Sacchi}},\ }\bibfield  {title} {\bibinfo {title} {Maximum-likelihood
  estimation of the density matrix},\ }\href
  {https://doi.org/10.1103/PhysRevA.61.010304} {\bibfield  {journal} {\bibinfo
  {journal} {Physical Review A}\ }\textbf {\bibinfo {volume} {61}},\ \bibinfo
  {pages} {010304} (\bibinfo {year} {1999})}\BibitemShut {NoStop}%
\bibitem [{\citenamefont {Miranowicz}\ \emph {et~al.}(2014)\citenamefont
  {Miranowicz}, \citenamefont {Bartkiewicz}, \citenamefont
  {Pe\ifmmode~\check{r}\else \v{r}\fi{}ina}, \citenamefont {Koashi},
  \citenamefont {Imoto},\ and\ \citenamefont {Nori}}]{miranowicz-pra-2014}%
  \BibitemOpen
  \bibfield  {author} {\bibinfo {author} {\bibfnamefont {A.}~\bibnamefont
  {Miranowicz}}, \bibinfo {author} {\bibfnamefont {K.}~\bibnamefont
  {Bartkiewicz}}, \bibinfo {author} {\bibfnamefont {J.}~\bibnamefont
  {Pe\ifmmode~\check{r}\else \v{r}\fi{}ina}}, \bibinfo {author} {\bibfnamefont
  {M.}~\bibnamefont {Koashi}}, \bibinfo {author} {\bibfnamefont
  {N.}~\bibnamefont {Imoto}},\ and\ \bibinfo {author} {\bibfnamefont
  {F.}~\bibnamefont {Nori}},\ }\bibfield  {title} {\bibinfo {title} {{Optimal
  two-qubit tomography based on local and global measurements: Maximal
  robustness against errors as described by condition numbers}},\ }\href
  {https://doi.org/10.1103/PhysRevA.90.062123} {\bibfield  {journal} {\bibinfo
  {journal} {Physical Review A}\ }\textbf {\bibinfo {volume} {90}},\ \bibinfo
  {pages} {062123} (\bibinfo {year} {2014})}\BibitemShut {NoStop}%
\bibitem [{\citenamefont {W{\"o}lk}\ \emph {et~al.}(2019)\citenamefont
  {W{\"o}lk}, \citenamefont {Sriarunothai}, \citenamefont {Giri},\ and\
  \citenamefont {Wunderlich}}]{wolk-njp-2019}%
  \BibitemOpen
  \bibfield  {author} {\bibinfo {author} {\bibfnamefont {S.}~\bibnamefont
  {W{\"o}lk}}, \bibinfo {author} {\bibfnamefont {T.}~\bibnamefont
  {Sriarunothai}}, \bibinfo {author} {\bibfnamefont {G.~S.}\ \bibnamefont
  {Giri}},\ and\ \bibinfo {author} {\bibfnamefont {C.}~\bibnamefont
  {Wunderlich}},\ }\bibfield  {title} {\bibinfo {title} {Distinguishing between
  statistical and systematic errors in quantum process tomography},\ }\href
  {https://doi.org/10.1088/1367-2630/aaf5f2} {\bibfield  {journal} {\bibinfo
  {journal} {New Journal of Physics}\ }\textbf {\bibinfo {volume} {21}},\
  \bibinfo {pages} {013015} (\bibinfo {year} {2019})}\BibitemShut {NoStop}%
\bibitem [{\citenamefont {Lundeen}\ and\ \citenamefont
  {Bamber}(2012)}]{lundeen-prl-2012}%
  \BibitemOpen
  \bibfield  {author} {\bibinfo {author} {\bibfnamefont {J.~S.}\ \bibnamefont
  {Lundeen}}\ and\ \bibinfo {author} {\bibfnamefont {C.}~\bibnamefont
  {Bamber}},\ }\bibfield  {title} {\bibinfo {title} {{Procedure for Direct
  Measurement of General Quantum States Using Weak Measurement}},\ }\href
  {https://doi.org/10.1103/PhysRevLett.108.070402} {\bibfield  {journal}
  {\bibinfo  {journal} {Physical Review Letters}\ }\textbf {\bibinfo {volume}
  {108}},\ \bibinfo {pages} {070402} (\bibinfo {year} {2012})}\BibitemShut
  {NoStop}%
\bibitem [{\citenamefont {Wu}(2013)}]{wu-sr-2013}%
  \BibitemOpen
  \bibfield  {author} {\bibinfo {author} {\bibfnamefont {S.}~\bibnamefont
  {Wu}},\ }\bibfield  {title} {\bibinfo {title} {State tomography via weak
  measurements},\ }\href {https://doi.org/10.1038/srep01193} {\bibfield
  {journal} {\bibinfo  {journal} {Scientific Reports}\ }\textbf {\bibinfo
  {volume} {3}},\ \bibinfo {pages} {1193} (\bibinfo {year} {2013})}\BibitemShut
  {NoStop}%
\bibitem [{\citenamefont {Kim}\ \emph {et~al.}(2018)\citenamefont {Kim},
  \citenamefont {Kim}, \citenamefont {Lee}, \citenamefont {Han}, \citenamefont
  {Moon}, \citenamefont {Kim},\ and\ \citenamefont {Cho}}]{kim-natcom-2017}%
  \BibitemOpen
  \bibfield  {author} {\bibinfo {author} {\bibfnamefont {Y.}~\bibnamefont
  {Kim}}, \bibinfo {author} {\bibfnamefont {Y.-S.}\ \bibnamefont {Kim}},
  \bibinfo {author} {\bibfnamefont {S.-Y.}\ \bibnamefont {Lee}}, \bibinfo
  {author} {\bibfnamefont {S.-W.}\ \bibnamefont {Han}}, \bibinfo {author}
  {\bibfnamefont {S.}~\bibnamefont {Moon}}, \bibinfo {author} {\bibfnamefont
  {Y.-H.}\ \bibnamefont {Kim}},\ and\ \bibinfo {author} {\bibfnamefont {Y.-W.}\
  \bibnamefont {Cho}},\ }\bibfield  {title} {\bibinfo {title} {Direct quantum
  process tomography via measuring sequential weak values of incompatible
  observables},\ }\href {https://doi.org/10.1038/s41467-017-02511-2} {\bibfield
   {journal} {\bibinfo  {journal} {Nature Communications}\ }\textbf {\bibinfo
  {volume} {9}},\ \bibinfo {pages} {192} (\bibinfo {year} {2018})}\BibitemShut
  {NoStop}%
\bibitem [{\citenamefont {Feng}\ \emph {et~al.}(2021)\citenamefont {Feng},
  \citenamefont {Ren},\ and\ \citenamefont {Zhou}}]{feng-pra-2021}%
  \BibitemOpen
  \bibfield  {author} {\bibinfo {author} {\bibfnamefont {T.}~\bibnamefont
  {Feng}}, \bibinfo {author} {\bibfnamefont {C.}~\bibnamefont {Ren}},\ and\
  \bibinfo {author} {\bibfnamefont {X.}~\bibnamefont {Zhou}},\ }\bibfield
  {title} {\bibinfo {title} {Direct measurement of density-matrix elements
  using a phase-shifting technique},\ }\href
  {https://doi.org/10.1103/PhysRevA.104.042403} {\bibfield  {journal} {\bibinfo
   {journal} {Physical Review A}\ }\textbf {\bibinfo {volume} {104}},\ \bibinfo
  {pages} {042403} (\bibinfo {year} {2021})}\BibitemShut {NoStop}%
\bibitem [{\citenamefont {Li}\ \emph {et~al.}(2022)\citenamefont {Li},
  \citenamefont {Wang}, \citenamefont {Feng}, \citenamefont {Li}, \citenamefont
  {Ren},\ and\ \citenamefont {Zhou}}]{li-pra-2022}%
  \BibitemOpen
  \bibfield  {author} {\bibinfo {author} {\bibfnamefont {C.}~\bibnamefont
  {Li}}, \bibinfo {author} {\bibfnamefont {Y.}~\bibnamefont {Wang}}, \bibinfo
  {author} {\bibfnamefont {T.}~\bibnamefont {Feng}}, \bibinfo {author}
  {\bibfnamefont {Z.}~\bibnamefont {Li}}, \bibinfo {author} {\bibfnamefont
  {C.}~\bibnamefont {Ren}},\ and\ \bibinfo {author} {\bibfnamefont
  {X.}~\bibnamefont {Zhou}},\ }\bibfield  {title} {\bibinfo {title} {Direct
  measurement of density-matrix elements with a phase-shifting technique on a
  quantum photonic chip},\ }\href {https://doi.org/10.1103/PhysRevA.105.062414}
  {\bibfield  {journal} {\bibinfo  {journal} {Physical Review A}\ }\textbf
  {\bibinfo {volume} {105}},\ \bibinfo {pages} {062414} (\bibinfo {year}
  {2022})}\BibitemShut {NoStop}%
\bibitem [{\citenamefont {Ekert}\ \emph {et~al.}(2002)\citenamefont {Ekert},
  \citenamefont {Alves}, \citenamefont {Oi}, \citenamefont {Horodecki},
  \citenamefont {Horodecki},\ and\ \citenamefont {Kwek}}]{ekert-prl-2022}%
  \BibitemOpen
  \bibfield  {author} {\bibinfo {author} {\bibfnamefont {A.~K.}\ \bibnamefont
  {Ekert}}, \bibinfo {author} {\bibfnamefont {C.~M.}\ \bibnamefont {Alves}},
  \bibinfo {author} {\bibfnamefont {D.~K.~L.}\ \bibnamefont {Oi}}, \bibinfo
  {author} {\bibfnamefont {M.}~\bibnamefont {Horodecki}}, \bibinfo {author}
  {\bibfnamefont {P.}~\bibnamefont {Horodecki}},\ and\ \bibinfo {author}
  {\bibfnamefont {L.~C.}\ \bibnamefont {Kwek}},\ }\bibfield  {title} {\bibinfo
  {title} {{Direct Estimations of Linear and Nonlinear Functionals of a Quantum
  State}},\ }\href {https://doi.org/10.1103/PhysRevLett.88.217901} {\bibfield
  {journal} {\bibinfo  {journal} {Physical Review Letters}\ }\textbf {\bibinfo
  {volume} {88}},\ \bibinfo {pages} {217901} (\bibinfo {year}
  {2002})}\BibitemShut {NoStop}%
\bibitem [{\citenamefont {{A. Gaikwad}}\ \emph {et~al.}(2023)\citenamefont {{A.
  Gaikwad}}, \citenamefont {{G. Singh}}, \citenamefont {{K. Dorai}},\ and\
  \citenamefont {{Arvind}}}]{gaikwad-epjd-2023}%
  \BibitemOpen
  \bibfield  {author} {\bibinfo {author} {\bibnamefont {{A. Gaikwad}}},
  \bibinfo {author} {\bibnamefont {{G. Singh}}}, \bibinfo {author}
  {\bibnamefont {{K. Dorai}}},\ and\ \bibinfo {author} {\bibnamefont
  {{Arvind}}},\ }\bibfield  {title} {\bibinfo {title} {{Direct tomography of
  quantum states and processes via weak measurements of Pauli spin operators on
  an NMR quantum processor}},\ }\href
  {https://doi.org/10.1140/epjd/s10053-023-00791-6} {\bibfield  {journal}
  {\bibinfo  {journal} {The European Physical Journal D}\ }\textbf {\bibinfo
  {volume} {77}},\ \bibinfo {pages} {209} (\bibinfo {year} {2023})}\BibitemShut
  {NoStop}%
\bibitem [{\citenamefont {Cramer}\ \emph {et~al.}(2010)\citenamefont {Cramer},
  \citenamefont {Plenio}, \citenamefont {Flammia}, \citenamefont {Somma},
  \citenamefont {Gross}, \citenamefont {Bartlett}, \citenamefont
  {Landon-Cardinal}, \citenamefont {Poulin},\ and\ \citenamefont
  {Liu}}]{cramer-natcom-2010}%
  \BibitemOpen
  \bibfield  {author} {\bibinfo {author} {\bibfnamefont {M.}~\bibnamefont
  {Cramer}}, \bibinfo {author} {\bibfnamefont {M.~B.}\ \bibnamefont {Plenio}},
  \bibinfo {author} {\bibfnamefont {S.~T.}\ \bibnamefont {Flammia}}, \bibinfo
  {author} {\bibfnamefont {R.}~\bibnamefont {Somma}}, \bibinfo {author}
  {\bibfnamefont {D.}~\bibnamefont {Gross}}, \bibinfo {author} {\bibfnamefont
  {S.~D.}\ \bibnamefont {Bartlett}}, \bibinfo {author} {\bibfnamefont
  {O.}~\bibnamefont {Landon-Cardinal}}, \bibinfo {author} {\bibfnamefont
  {D.}~\bibnamefont {Poulin}},\ and\ \bibinfo {author} {\bibfnamefont {Y.-K.}\
  \bibnamefont {Liu}},\ }\bibfield  {title} {\bibinfo {title} {Efficient
  quantum state tomography},\ }\href {https://doi.org/10.1038/ncomms1147}
  {\bibfield  {journal} {\bibinfo  {journal} {Nature Communications}\ }\textbf
  {\bibinfo {volume} {1}},\ \bibinfo {pages} {149} (\bibinfo {year}
  {2010})}\BibitemShut {NoStop}%
\bibitem [{\citenamefont {Lanyon}\ \emph {et~al.}(2017)\citenamefont {Lanyon},
  \citenamefont {Maier}, \citenamefont {Holz{\"a}pfel}, \citenamefont
  {Baumgratz}, \citenamefont {Hempel}, \citenamefont {Jurcevic}, \citenamefont
  {Dhand}, \citenamefont {Buyskikh}, \citenamefont {Daley}, \citenamefont
  {Cramer}, \citenamefont {Plenio}, \citenamefont {Blatt},\ and\ \citenamefont
  {Roos}}]{lanyon-np-2017}%
  \BibitemOpen
  \bibfield  {author} {\bibinfo {author} {\bibfnamefont {B.~P.}\ \bibnamefont
  {Lanyon}}, \bibinfo {author} {\bibfnamefont {C.}~\bibnamefont {Maier}},
  \bibinfo {author} {\bibfnamefont {M.}~\bibnamefont {Holz{\"a}pfel}}, \bibinfo
  {author} {\bibfnamefont {T.}~\bibnamefont {Baumgratz}}, \bibinfo {author}
  {\bibfnamefont {C.}~\bibnamefont {Hempel}}, \bibinfo {author} {\bibfnamefont
  {P.}~\bibnamefont {Jurcevic}}, \bibinfo {author} {\bibfnamefont
  {I.}~\bibnamefont {Dhand}}, \bibinfo {author} {\bibfnamefont {A.~S.}\
  \bibnamefont {Buyskikh}}, \bibinfo {author} {\bibfnamefont {A.~J.}\
  \bibnamefont {Daley}}, \bibinfo {author} {\bibfnamefont {M.}~\bibnamefont
  {Cramer}}, \bibinfo {author} {\bibfnamefont {M.~B.}\ \bibnamefont {Plenio}},
  \bibinfo {author} {\bibfnamefont {R.}~\bibnamefont {Blatt}},\ and\ \bibinfo
  {author} {\bibfnamefont {C.~F.}\ \bibnamefont {Roos}},\ }\bibfield  {title}
  {\bibinfo {title} {Efficient tomography of a quantum many-body system},\
  }\href {https://doi.org/10.1038/nphys4244} {\bibfield  {journal} {\bibinfo
  {journal} {Nature Physics}\ }\textbf {\bibinfo {volume} {13}},\ \bibinfo
  {pages} {1158} (\bibinfo {year} {2017})}\BibitemShut {NoStop}%
\bibitem [{\citenamefont {Han}\ \emph {et~al.}(2022)\citenamefont {Han},
  \citenamefont {Guo},\ and\ \citenamefont {Wang}}]{han-pra-2022}%
  \BibitemOpen
  \bibfield  {author} {\bibinfo {author} {\bibfnamefont {D.}~\bibnamefont
  {Han}}, \bibinfo {author} {\bibfnamefont {C.}~\bibnamefont {Guo}},\ and\
  \bibinfo {author} {\bibfnamefont {X.}~\bibnamefont {Wang}},\ }\bibfield
  {title} {\bibinfo {title} {Density matrix reconstruction using non-negative
  matrix product states},\ }\href {https://doi.org/10.1103/PhysRevA.106.042435}
  {\bibfield  {journal} {\bibinfo  {journal} {Physical Review A}\ }\textbf
  {\bibinfo {volume} {106}},\ \bibinfo {pages} {042435} (\bibinfo {year}
  {2022})}\BibitemShut {NoStop}%
\bibitem [{\citenamefont {Kurmapu}\ \emph {et~al.}(2023)\citenamefont
  {Kurmapu}, \citenamefont {Tiunova}, \citenamefont {Tiunov}, \citenamefont
  {Ringbauer}, \citenamefont {Maier}, \citenamefont {Blatt}, \citenamefont
  {Monz}, \citenamefont {Fedorov},\ and\ \citenamefont
  {Lvovsky}}]{Kurmapu2023}%
  \BibitemOpen
  \bibfield  {author} {\bibinfo {author} {\bibfnamefont {M.~K.}\ \bibnamefont
  {Kurmapu}}, \bibinfo {author} {\bibfnamefont {V.}~\bibnamefont {Tiunova}},
  \bibinfo {author} {\bibfnamefont {E.}~\bibnamefont {Tiunov}}, \bibinfo
  {author} {\bibfnamefont {M.}~\bibnamefont {Ringbauer}}, \bibinfo {author}
  {\bibfnamefont {C.}~\bibnamefont {Maier}}, \bibinfo {author} {\bibfnamefont
  {R.}~\bibnamefont {Blatt}}, \bibinfo {author} {\bibfnamefont
  {T.}~\bibnamefont {Monz}}, \bibinfo {author} {\bibfnamefont {A.~K.}\
  \bibnamefont {Fedorov}},\ and\ \bibinfo {author} {\bibfnamefont
  {A.}~\bibnamefont {Lvovsky}},\ }\bibfield  {title} {\bibinfo {title}
  {{Reconstructing Complex States of a 20-Qubit Quantum Simulator}},\ }\href
  {https://doi.org/10.1103/PRXQuantum.4.040345} {\bibfield  {journal} {\bibinfo
   {journal} {PRX Quantum}\ }\textbf {\bibinfo {volume} {4}},\ \bibinfo {pages}
  {040345} (\bibinfo {year} {2023})}\BibitemShut {NoStop}%
\bibitem [{\citenamefont {T\'{o}th}\ \emph {et~al.}(2010)\citenamefont
  {T\'{o}th}, \citenamefont {Wieczorek}, \citenamefont {Gross}, \citenamefont
  {Krischek}, \citenamefont {Schwemmer},\ and\ \citenamefont
  {Weinfurter}}]{toth-prl-2010}%
  \BibitemOpen
  \bibfield  {author} {\bibinfo {author} {\bibfnamefont {G.}~\bibnamefont
  {T\'{o}th}}, \bibinfo {author} {\bibfnamefont {W.}~\bibnamefont {Wieczorek}},
  \bibinfo {author} {\bibfnamefont {D.}~\bibnamefont {Gross}}, \bibinfo
  {author} {\bibfnamefont {R.}~\bibnamefont {Krischek}}, \bibinfo {author}
  {\bibfnamefont {C.}~\bibnamefont {Schwemmer}},\ and\ \bibinfo {author}
  {\bibfnamefont {H.}~\bibnamefont {Weinfurter}},\ }\bibfield  {title}
  {\bibinfo {title} {{Permutationally Invariant Quantum Tomography}},\ }\href
  {https://doi.org/10.1103/PhysRevLett.105.250403} {\bibfield  {journal}
  {\bibinfo  {journal} {Physical Review Letters}\ }\textbf {\bibinfo {volume}
  {105}},\ \bibinfo {pages} {250403} (\bibinfo {year} {2010})}\BibitemShut
  {NoStop}%
\bibitem [{\citenamefont {Moroder}\ \emph {et~al.}(2012)\citenamefont
  {Moroder}, \citenamefont {Hyllus}, \citenamefont {T\'{o}th}, \citenamefont
  {Schwemmer}, \citenamefont {Niggebaum}, \citenamefont {Gaile}, \citenamefont
  {G\"{u}hne},\ and\ \citenamefont {Weinfurter}}]{moroder-njp-2012}%
  \BibitemOpen
  \bibfield  {author} {\bibinfo {author} {\bibfnamefont {T.}~\bibnamefont
  {Moroder}}, \bibinfo {author} {\bibfnamefont {P.}~\bibnamefont {Hyllus}},
  \bibinfo {author} {\bibfnamefont {G.}~\bibnamefont {T\'{o}th}}, \bibinfo
  {author} {\bibfnamefont {C.}~\bibnamefont {Schwemmer}}, \bibinfo {author}
  {\bibfnamefont {A.}~\bibnamefont {Niggebaum}}, \bibinfo {author}
  {\bibfnamefont {S.}~\bibnamefont {Gaile}}, \bibinfo {author} {\bibfnamefont
  {O.}~\bibnamefont {G\"{u}hne}},\ and\ \bibinfo {author} {\bibfnamefont
  {H.}~\bibnamefont {Weinfurter}},\ }\bibfield  {title} {\bibinfo {title}
  {Permutationally invariant state reconstruction},\ }\href
  {https://doi.org/10.1088/1367-2630/14/10/105001} {\bibfield  {journal}
  {\bibinfo  {journal} {New Journal of Physics}\ }\textbf {\bibinfo {volume}
  {14}},\ \bibinfo {pages} {105001} (\bibinfo {year} {2012})}\BibitemShut
  {NoStop}%
\bibitem [{\citenamefont {Kuzmin}\ \emph {et~al.}(2024)\citenamefont {Kuzmin},
  \citenamefont {Mikhailova}, \citenamefont {Dyakonov},\ and\ \citenamefont
  {Straupe}}]{kuzmin-pra-2024}%
  \BibitemOpen
  \bibfield  {author} {\bibinfo {author} {\bibfnamefont {S.}~\bibnamefont
  {Kuzmin}}, \bibinfo {author} {\bibfnamefont {V.}~\bibnamefont {Mikhailova}},
  \bibinfo {author} {\bibfnamefont {I.}~\bibnamefont {Dyakonov}},\ and\
  \bibinfo {author} {\bibfnamefont {S.}~\bibnamefont {Straupe}},\ }\bibfield
  {title} {\bibinfo {title} {Learning the tensor network model of a quantum
  state using a few single-qubit measurements},\ }\href
  {https://doi.org/10.1103/PhysRevA.109.052616} {\bibfield  {journal} {\bibinfo
   {journal} {Physical Review A}\ }\textbf {\bibinfo {volume} {109}},\ \bibinfo
  {pages} {052616} (\bibinfo {year} {2024})}\BibitemShut {NoStop}%
\bibitem [{\citenamefont {Torlai}\ \emph {et~al.}(2023)\citenamefont {Torlai},
  \citenamefont {Wood}, \citenamefont {Acharya}, \citenamefont {Carleo},
  \citenamefont {Carrasquilla},\ and\ \citenamefont {Aolita}}]{torlai-2023}%
  \BibitemOpen
  \bibfield  {author} {\bibinfo {author} {\bibfnamefont {G.}~\bibnamefont
  {Torlai}}, \bibinfo {author} {\bibfnamefont {C.~J.}\ \bibnamefont {Wood}},
  \bibinfo {author} {\bibfnamefont {A.}~\bibnamefont {Acharya}}, \bibinfo
  {author} {\bibfnamefont {G.}~\bibnamefont {Carleo}}, \bibinfo {author}
  {\bibfnamefont {J.}~\bibnamefont {Carrasquilla}},\ and\ \bibinfo {author}
  {\bibfnamefont {L.}~\bibnamefont {Aolita}},\ }\bibfield  {title} {\bibinfo
  {title} {Quantum process tomography with unsupervised learning and tensor
  networks},\ }\href {https://doi.org/10.1038/s41467-023-38332-9} {\bibfield
  {journal} {\bibinfo  {journal} {Nature Communications}\ }\textbf {\bibinfo
  {volume} {14}},\ \bibinfo {pages} {2858} (\bibinfo {year}
  {2023})}\BibitemShut {NoStop}%
\bibitem [{\citenamefont {\ifmmode \check{R}\else
  \v{R}\fi{}eh\'a\ifmmode~\check{c}\else \v{c}\fi{}ek}\ \emph
  {et~al.}(2007)\citenamefont {\ifmmode \check{R}\else
  \v{R}\fi{}eh\'a\ifmmode~\check{c}\else \v{c}\fi{}ek}, \citenamefont {Hradil},
  \citenamefont {Knill},\ and\ \citenamefont {Lvovsky}}]{mle-pra-2007}%
  \BibitemOpen
  \bibfield  {author} {\bibinfo {author} {\bibfnamefont {J.}~\bibnamefont
  {\ifmmode \check{R}\else \v{R}\fi{}eh\'a\ifmmode~\check{c}\else
  \v{c}\fi{}ek}}, \bibinfo {author} {\bibfnamefont {Z.}~\bibnamefont {Hradil}},
  \bibinfo {author} {\bibfnamefont {E.}~\bibnamefont {Knill}},\ and\ \bibinfo
  {author} {\bibfnamefont {A.~I.}\ \bibnamefont {Lvovsky}},\ }\bibfield
  {title} {\bibinfo {title} {Diluted maximum-likelihood algorithm for quantum
  tomography},\ }\href {https://doi.org/10.1103/PhysRevA.75.042108} {\bibfield
  {journal} {\bibinfo  {journal} {Physical Review A}\ }\textbf {\bibinfo
  {volume} {75}},\ \bibinfo {pages} {042108} (\bibinfo {year}
  {2007})}\BibitemShut {NoStop}%
\bibitem [{\citenamefont {Smolin}\ \emph {et~al.}(2012)\citenamefont {Smolin},
  \citenamefont {Gambetta},\ and\ \citenamefont {Smith}}]{smolin-prl-2012}%
  \BibitemOpen
  \bibfield  {author} {\bibinfo {author} {\bibfnamefont {J.~A.}\ \bibnamefont
  {Smolin}}, \bibinfo {author} {\bibfnamefont {J.~M.}\ \bibnamefont
  {Gambetta}},\ and\ \bibinfo {author} {\bibfnamefont {G.}~\bibnamefont
  {Smith}},\ }\bibfield  {title} {\bibinfo {title} {{Efficient Method for
  Computing the Maximum-Likelihood Quantum State from Measurements with
  Additive Gaussian Noise}},\ }\href
  {https://doi.org/10.1103/PhysRevLett.108.070502} {\bibfield  {journal}
  {\bibinfo  {journal} {Physical Review Letters}\ }\textbf {\bibinfo {volume}
  {108}},\ \bibinfo {pages} {070502} (\bibinfo {year} {2012})}\BibitemShut
  {NoStop}%
\bibitem [{\citenamefont {Shang}\ \emph {et~al.}(2017)\citenamefont {Shang},
  \citenamefont {Zhang},\ and\ \citenamefont {Ng}}]{shang-pra-2017}%
  \BibitemOpen
  \bibfield  {author} {\bibinfo {author} {\bibfnamefont {J.}~\bibnamefont
  {Shang}}, \bibinfo {author} {\bibfnamefont {Z.}~\bibnamefont {Zhang}},\ and\
  \bibinfo {author} {\bibfnamefont {H.~K.}\ \bibnamefont {Ng}},\ }\bibfield
  {title} {\bibinfo {title} {Superfast maximum-likelihood reconstruction for
  quantum tomography},\ }\href {https://doi.org/10.1103/PhysRevA.95.062336}
  {\bibfield  {journal} {\bibinfo  {journal} {Physical Review A}\ }\textbf
  {\bibinfo {volume} {95}},\ \bibinfo {pages} {062336} (\bibinfo {year}
  {2017})}\BibitemShut {NoStop}%
\bibitem [{\citenamefont {Gross}\ \emph {et~al.}(2010)\citenamefont {Gross},
  \citenamefont {Liu}, \citenamefont {Flammia}, \citenamefont {Becker},\ and\
  \citenamefont {Eisert}}]{david-prl-2010}%
  \BibitemOpen
  \bibfield  {author} {\bibinfo {author} {\bibfnamefont {D.}~\bibnamefont
  {Gross}}, \bibinfo {author} {\bibfnamefont {Y.-K.}\ \bibnamefont {Liu}},
  \bibinfo {author} {\bibfnamefont {S.~T.}\ \bibnamefont {Flammia}}, \bibinfo
  {author} {\bibfnamefont {S.}~\bibnamefont {Becker}},\ and\ \bibinfo {author}
  {\bibfnamefont {J.}~\bibnamefont {Eisert}},\ }\bibfield  {title} {\bibinfo
  {title} {{Quantum State Tomography via Compressed Sensing}},\ }\href
  {https://doi.org/10.1103/PhysRevLett.105.150401} {\bibfield  {journal}
  {\bibinfo  {journal} {Physical Review Letters}\ }\textbf {\bibinfo {volume}
  {105}},\ \bibinfo {pages} {150401} (\bibinfo {year} {2010})}\BibitemShut
  {NoStop}%
\bibitem [{\citenamefont {Steffens}\ \emph {et~al.}(2017)\citenamefont
  {Steffens}, \citenamefont {Riofrao}, \citenamefont {McCutcheon},
  \citenamefont {Roth}, \citenamefont {Bell}, \citenamefont {McMillan},
  \citenamefont {Tame}, \citenamefont {Rarity},\ and\ \citenamefont
  {Eisert}}]{Steffens_2017}%
  \BibitemOpen
  \bibfield  {author} {\bibinfo {author} {\bibfnamefont {A.}~\bibnamefont
  {Steffens}}, \bibinfo {author} {\bibfnamefont {C.~A.}\ \bibnamefont
  {Riofrao}}, \bibinfo {author} {\bibfnamefont {W.}~\bibnamefont {McCutcheon}},
  \bibinfo {author} {\bibfnamefont {I.}~\bibnamefont {Roth}}, \bibinfo {author}
  {\bibfnamefont {B.~A.}\ \bibnamefont {Bell}}, \bibinfo {author}
  {\bibfnamefont {A.}~\bibnamefont {McMillan}}, \bibinfo {author}
  {\bibfnamefont {M.~S.}\ \bibnamefont {Tame}}, \bibinfo {author}
  {\bibfnamefont {J.~G.}\ \bibnamefont {Rarity}},\ and\ \bibinfo {author}
  {\bibfnamefont {J.}~\bibnamefont {Eisert}},\ }\bibfield  {title} {\bibinfo
  {title} {Experimentally exploring compressed sensing quantum tomography},\
  }\href {https://doi.org/10.1088/2058-9565/aa6ae2} {\bibfield  {journal}
  {\bibinfo  {journal} {Quantum Science and Technology}\ }\textbf {\bibinfo
  {volume} {2}},\ \bibinfo {pages} {025005} (\bibinfo {year}
  {2017})}\BibitemShut {NoStop}%
\bibitem [{\citenamefont {Yang}\ \emph {et~al.}(2017)\citenamefont {Yang},
  \citenamefont {Cong}, \citenamefont {Liu}, \citenamefont {Li},\ and\
  \citenamefont {Li}}]{yang-pra-2017}%
  \BibitemOpen
  \bibfield  {author} {\bibinfo {author} {\bibfnamefont {J.}~\bibnamefont
  {Yang}}, \bibinfo {author} {\bibfnamefont {S.}~\bibnamefont {Cong}}, \bibinfo
  {author} {\bibfnamefont {X.}~\bibnamefont {Liu}}, \bibinfo {author}
  {\bibfnamefont {Z.}~\bibnamefont {Li}},\ and\ \bibinfo {author}
  {\bibfnamefont {K.}~\bibnamefont {Li}},\ }\bibfield  {title} {\bibinfo
  {title} {{Effective quantum state reconstruction using compressed sensing in
  NMR quantum computing}},\ }\href {https://doi.org/10.1103/PhysRevA.96.052101}
  {\bibfield  {journal} {\bibinfo  {journal} {Physical Review A}\ }\textbf
  {\bibinfo {volume} {96}},\ \bibinfo {pages} {052101} (\bibinfo {year}
  {2017})}\BibitemShut {NoStop}%
\bibitem [{\citenamefont {Kyrillidis}\ \emph {et~al.}(2018)\citenamefont
  {Kyrillidis}, \citenamefont {Kalev}, \citenamefont {Park}, \citenamefont
  {Bhojanapalli}, \citenamefont {Caramanis},\ and\ \citenamefont
  {Sanghavi}}]{Kyrillidis2018}%
  \BibitemOpen
  \bibfield  {author} {\bibinfo {author} {\bibfnamefont {A.}~\bibnamefont
  {Kyrillidis}}, \bibinfo {author} {\bibfnamefont {A.}~\bibnamefont {Kalev}},
  \bibinfo {author} {\bibfnamefont {D.}~\bibnamefont {Park}}, \bibinfo {author}
  {\bibfnamefont {S.}~\bibnamefont {Bhojanapalli}}, \bibinfo {author}
  {\bibfnamefont {C.}~\bibnamefont {Caramanis}},\ and\ \bibinfo {author}
  {\bibfnamefont {S.}~\bibnamefont {Sanghavi}},\ }\bibfield  {title} {\bibinfo
  {title} {Provable compressed sensing quantum state tomography via non-convex
  methods},\ }\href {https://doi.org/10.1038/s41534-018-0080-4} {\bibfield
  {journal} {\bibinfo  {journal} {npj Quantum Information}\ }\textbf {\bibinfo
  {volume} {4}},\ \bibinfo {pages} {36} (\bibinfo {year} {2018})}\BibitemShut
  {NoStop}%
\bibitem [{\citenamefont {Ahn}\ \emph {et~al.}(2019)\citenamefont {Ahn},
  \citenamefont {Teo}, \citenamefont {Jeong}, \citenamefont {Bouchard},
  \citenamefont {Hufnagel}, \citenamefont {Karimi}, \citenamefont {Koutn\'y},
  \citenamefont {\ifmmode \check{R}\else \v{R}\fi{}eh\'a\ifmmode~\check{c}\else
  \v{c}\fi{}ek}, \citenamefont {Hradil}, \citenamefont {Leuchs},\ and\
  \citenamefont {S\'anchez-Soto}}]{ahn-prl-2019}%
  \BibitemOpen
  \bibfield  {author} {\bibinfo {author} {\bibfnamefont {D.}~\bibnamefont
  {Ahn}}, \bibinfo {author} {\bibfnamefont {Y.~S.}\ \bibnamefont {Teo}},
  \bibinfo {author} {\bibfnamefont {H.}~\bibnamefont {Jeong}}, \bibinfo
  {author} {\bibfnamefont {F.}~\bibnamefont {Bouchard}}, \bibinfo {author}
  {\bibfnamefont {F.}~\bibnamefont {Hufnagel}}, \bibinfo {author}
  {\bibfnamefont {E.}~\bibnamefont {Karimi}}, \bibinfo {author} {\bibfnamefont
  {D.}~\bibnamefont {Koutn\'y}}, \bibinfo {author} {\bibfnamefont
  {J.}~\bibnamefont {\ifmmode \check{R}\else
  \v{R}\fi{}eh\'a\ifmmode~\check{c}\else \v{c}\fi{}ek}}, \bibinfo {author}
  {\bibfnamefont {Z.}~\bibnamefont {Hradil}}, \bibinfo {author} {\bibfnamefont
  {G.}~\bibnamefont {Leuchs}},\ and\ \bibinfo {author} {\bibfnamefont {L.~L.}\
  \bibnamefont {S\'anchez-Soto}},\ }\bibfield  {title} {\bibinfo {title}
  {{Adaptive Compressive Tomography with No a priori Information}},\ }\href
  {https://doi.org/10.1103/PhysRevLett.122.100404} {\bibfield  {journal}
  {\bibinfo  {journal} {Physical Review Letters}\ }\textbf {\bibinfo {volume}
  {122}},\ \bibinfo {pages} {100404} (\bibinfo {year} {2019})}\BibitemShut
  {NoStop}%
\bibitem [{\citenamefont {Teo}\ \emph {et~al.}(2020)\citenamefont {Teo},
  \citenamefont {Struchalin}, \citenamefont {Kovlakov}, \citenamefont {Ahn},
  \citenamefont {Jeong}, \citenamefont {Straupe}, \citenamefont {Kulik},
  \citenamefont {Leuchs},\ and\ \citenamefont {S\'anchez-Soto}}]{teo-pra-2020}%
  \BibitemOpen
  \bibfield  {author} {\bibinfo {author} {\bibfnamefont {Y.~S.}\ \bibnamefont
  {Teo}}, \bibinfo {author} {\bibfnamefont {G.~I.}\ \bibnamefont {Struchalin}},
  \bibinfo {author} {\bibfnamefont {E.~V.}\ \bibnamefont {Kovlakov}}, \bibinfo
  {author} {\bibfnamefont {D.}~\bibnamefont {Ahn}}, \bibinfo {author}
  {\bibfnamefont {H.}~\bibnamefont {Jeong}}, \bibinfo {author} {\bibfnamefont
  {S.~S.}\ \bibnamefont {Straupe}}, \bibinfo {author} {\bibfnamefont {S.~P.}\
  \bibnamefont {Kulik}}, \bibinfo {author} {\bibfnamefont {G.}~\bibnamefont
  {Leuchs}},\ and\ \bibinfo {author} {\bibfnamefont {L.~L.}\ \bibnamefont
  {S\'anchez-Soto}},\ }\bibfield  {title} {\bibinfo {title} {Objective
  compressive quantum process tomography},\ }\href
  {https://doi.org/10.1103/PhysRevA.101.022334} {\bibfield  {journal} {\bibinfo
   {journal} {Physical Review A}\ }\textbf {\bibinfo {volume} {101}},\ \bibinfo
  {pages} {022334} (\bibinfo {year} {2020})}\BibitemShut {NoStop}%
\bibitem [{\citenamefont {Qi}\ \emph {et~al.}(2017)\citenamefont {Qi},
  \citenamefont {Hou}, \citenamefont {Wang}, \citenamefont {Dong},
  \citenamefont {Zhong}, \citenamefont {Li}, \citenamefont {Xiang},
  \citenamefont {Wiseman}, \citenamefont {Li},\ and\ \citenamefont
  {Guo}}]{qi-quantum-inf-2017}%
  \BibitemOpen
  \bibfield  {author} {\bibinfo {author} {\bibfnamefont {B.}~\bibnamefont
  {Qi}}, \bibinfo {author} {\bibfnamefont {Z.}~\bibnamefont {Hou}}, \bibinfo
  {author} {\bibfnamefont {Y.}~\bibnamefont {Wang}}, \bibinfo {author}
  {\bibfnamefont {D.}~\bibnamefont {Dong}}, \bibinfo {author} {\bibfnamefont
  {H.-S.}\ \bibnamefont {Zhong}}, \bibinfo {author} {\bibfnamefont
  {L.}~\bibnamefont {Li}}, \bibinfo {author} {\bibfnamefont {G.-Y.}\
  \bibnamefont {Xiang}}, \bibinfo {author} {\bibfnamefont {H.~M.}\ \bibnamefont
  {Wiseman}}, \bibinfo {author} {\bibfnamefont {C.-F.}\ \bibnamefont {Li}},\
  and\ \bibinfo {author} {\bibfnamefont {G.-C.}\ \bibnamefont {Guo}},\
  }\bibfield  {title} {\bibinfo {title} {{Adaptive quantum state tomography via
  linear regression estimation: Theory and two-qubit experiment}},\ }\href
  {https://doi.org/10.1038/s41534-017-0016-4} {\bibfield  {journal} {\bibinfo
  {journal} {Quantum Information Processing}\ }\textbf {\bibinfo {volume}
  {3}},\ \bibinfo {pages} {19} (\bibinfo {year} {2017})}\BibitemShut {NoStop}%
\bibitem [{\citenamefont {Nehra}\ \emph {et~al.}(2020)\citenamefont {Nehra},
  \citenamefont {Eaton}, \citenamefont {Gonz\'alez-Arciniegas}, \citenamefont
  {Kim}, \citenamefont {Gerrits}, \citenamefont {Lita}, \citenamefont {Nam},\
  and\ \citenamefont {Pfister}}]{nehra-prr-2020}%
  \BibitemOpen
  \bibfield  {author} {\bibinfo {author} {\bibfnamefont {R.}~\bibnamefont
  {Nehra}}, \bibinfo {author} {\bibfnamefont {M.}~\bibnamefont {Eaton}},
  \bibinfo {author} {\bibfnamefont {C.}~\bibnamefont {Gonz\'alez-Arciniegas}},
  \bibinfo {author} {\bibfnamefont {M.~S.}\ \bibnamefont {Kim}}, \bibinfo
  {author} {\bibfnamefont {T.}~\bibnamefont {Gerrits}}, \bibinfo {author}
  {\bibfnamefont {A.}~\bibnamefont {Lita}}, \bibinfo {author} {\bibfnamefont
  {S.~W.}\ \bibnamefont {Nam}},\ and\ \bibinfo {author} {\bibfnamefont
  {O.}~\bibnamefont {Pfister}},\ }\bibfield  {title} {\bibinfo {title}
  {Generalized overlap quantum state tomography},\ }\href
  {https://doi.org/10.1103/PhysRevResearch.2.042002} {\bibfield  {journal}
  {\bibinfo  {journal} {Physical Review Research}\ }\textbf {\bibinfo {volume}
  {2}},\ \bibinfo {pages} {042002} (\bibinfo {year} {2020})}\BibitemShut
  {NoStop}%
\bibitem [{\citenamefont {Gaikwad}\ \emph {et~al.}(2021)\citenamefont
  {Gaikwad}, \citenamefont {{Arvind}},\ and\ \citenamefont
  {Dorai}}]{gaikwad-qip-2021}%
  \BibitemOpen
  \bibfield  {author} {\bibinfo {author} {\bibfnamefont {A.}~\bibnamefont
  {Gaikwad}}, \bibinfo {author} {\bibnamefont {{Arvind}}},\ and\ \bibinfo
  {author} {\bibfnamefont {K.}~\bibnamefont {Dorai}},\ }\bibfield  {title}
  {\bibinfo {title} {True experimental reconstruction of quantum states and
  processes via convex optimization},\ }\href
  {https://doi.org/10.1007/s11128-020-02930-z} {\bibfield  {journal} {\bibinfo
  {journal} {Quantum Information Processing}\ }\textbf {\bibinfo {volume}
  {20}},\ \bibinfo {pages} {19} (\bibinfo {year} {2021})}\BibitemShut {NoStop}%
\bibitem [{\citenamefont {Strandberg}(2022)}]{ingrid-prapp-2022}%
  \BibitemOpen
  \bibfield  {author} {\bibinfo {author} {\bibfnamefont {I.}~\bibnamefont
  {Strandberg}},\ }\bibfield  {title} {\bibinfo {title} {{Simple, Reliable, and
  Noise-Resilient Continuous-Variable Quantum State Tomography with Convex
  Optimization}},\ }\href {https://doi.org/10.1103/PhysRevApplied.18.044041}
  {\bibfield  {journal} {\bibinfo  {journal} {Physical Review Applied}\
  }\textbf {\bibinfo {volume} {18}},\ \bibinfo {pages} {044041} (\bibinfo
  {year} {2022})}\BibitemShut {NoStop}%
\bibitem [{\citenamefont {Mondal}\ and\ \citenamefont
  {Dutta}(2023{\natexlab{a}})}]{Mondal-ls-2023}%
  \BibitemOpen
  \bibfield  {author} {\bibinfo {author} {\bibfnamefont {S.}~\bibnamefont
  {Mondal}}\ and\ \bibinfo {author} {\bibfnamefont {A.~K.}\ \bibnamefont
  {Dutta}},\ }\bibfield  {title} {\bibinfo {title} {A modified least
  squares-based tomography with density matrix perturbation and linear entropy
  consideration along with performance analysis},\ }\href
  {https://doi.org/10.1088/1367-2630/acf187} {\bibfield  {journal} {\bibinfo
  {journal} {New Journal of Physics}\ }\textbf {\bibinfo {volume} {25}},\
  \bibinfo {pages} {083051} (\bibinfo {year} {2023}{\natexlab{a}})}\BibitemShut
  {NoStop}%
\bibitem [{\citenamefont {Banchi}\ \emph {et~al.}(2020)\citenamefont {Banchi},
  \citenamefont {Pereira}, \citenamefont {Lloyd},\ and\ \citenamefont
  {Pirandola}}]{banchi-npj-2020}%
  \BibitemOpen
  \bibfield  {author} {\bibinfo {author} {\bibfnamefont {L.}~\bibnamefont
  {Banchi}}, \bibinfo {author} {\bibfnamefont {J.}~\bibnamefont {Pereira}},
  \bibinfo {author} {\bibfnamefont {S.}~\bibnamefont {Lloyd}},\ and\ \bibinfo
  {author} {\bibfnamefont {S.}~\bibnamefont {Pirandola}},\ }\bibfield  {title}
  {\bibinfo {title} {Convex optimization of programmable quantum computers},\
  }\href {https://doi.org/10.1038/s41534-020-0268-2} {\bibfield  {journal}
  {\bibinfo  {journal} {npj Quantum Information}\ }\textbf {\bibinfo {volume}
  {6}},\ \bibinfo {pages} {42} (\bibinfo {year} {2020})}\BibitemShut {NoStop}%
\bibitem [{\citenamefont {Meng}\ \emph {et~al.}(2023)\citenamefont {Meng},
  \citenamefont {Han}, \citenamefont {Cong},\ and\ \citenamefont
  {Guo}}]{meng-rp-2023}%
  \BibitemOpen
  \bibfield  {author} {\bibinfo {author} {\bibfnamefont {X.}~\bibnamefont
  {Meng}}, \bibinfo {author} {\bibfnamefont {Z.}~\bibnamefont {Han}}, \bibinfo
  {author} {\bibfnamefont {J.}~\bibnamefont {Cong}},\ and\ \bibinfo {author}
  {\bibfnamefont {X.}~\bibnamefont {Guo}},\ }\bibfield  {title} {\bibinfo
  {title} {Intelligent optimization based density matrix reconstruction method
  with semi-positive constraint},\ }\href
  {https://doi.org/https://doi.org/10.1016/j.rinp.2023.106661} {\bibfield
  {journal} {\bibinfo  {journal} {Results in Physics}\ }\textbf {\bibinfo
  {volume} {51}},\ \bibinfo {pages} {106661} (\bibinfo {year}
  {2023})}\BibitemShut {NoStop}%
\bibitem [{\citenamefont {Lofberg}(2004)}]{lofberg-2004}%
  \BibitemOpen
  \bibfield  {author} {\bibinfo {author} {\bibfnamefont {J.}~\bibnamefont
  {Lofberg}},\ }\bibfield  {title} {\bibinfo {title} {Yalmip : a toolbox for
  modeling and optimization in matlab},\ }in\ \href
  {https://doi.org/10.1109/CACSD.2004.1393890} {\emph {\bibinfo {booktitle}
  {2004 IEEE International Conference on Robotics and Automation (IEEE Cat.
  No.04CH37508)}}}\ (\bibinfo {year} {2004})\ pp.\ \bibinfo {pages}
  {284--289}\BibitemShut {NoStop}%
\bibitem [{\citenamefont {Diamond}\ and\ \citenamefont
  {Boyd}(2016)}]{diamond2016cvxpy}%
  \BibitemOpen
  \bibfield  {author} {\bibinfo {author} {\bibfnamefont {S.}~\bibnamefont
  {Diamond}}\ and\ \bibinfo {author} {\bibfnamefont {S.}~\bibnamefont {Boyd}},\
  }\bibfield  {title} {\bibinfo {title} {{CVXPY}: {A} {P}ython-embedded
  modeling language for convex optimization},\ }\href
  {https://doi.org/https://dl.acm.org/doi/10.5555/2946645.3007036} {\bibfield
  {journal} {\bibinfo  {journal} {Journal of Machine Learning Research}\
  }\textbf {\bibinfo {volume} {17}},\ \bibinfo {pages} {2909} (\bibinfo {year}
  {2016})}\BibitemShut {NoStop}%
\bibitem [{\citenamefont {Carleo}\ \emph {et~al.}(2019)\citenamefont {Carleo},
  \citenamefont {Cirac}, \citenamefont {Cranmer}, \citenamefont {Daudet},
  \citenamefont {Schuld}, \citenamefont {Tishby}, \citenamefont
  {Vogt-Maranto},\ and\ \citenamefont {Zdeborov\'a}}]{carleo-rmp-2019}%
  \BibitemOpen
  \bibfield  {author} {\bibinfo {author} {\bibfnamefont {G.}~\bibnamefont
  {Carleo}}, \bibinfo {author} {\bibfnamefont {I.}~\bibnamefont {Cirac}},
  \bibinfo {author} {\bibfnamefont {K.}~\bibnamefont {Cranmer}}, \bibinfo
  {author} {\bibfnamefont {L.}~\bibnamefont {Daudet}}, \bibinfo {author}
  {\bibfnamefont {M.}~\bibnamefont {Schuld}}, \bibinfo {author} {\bibfnamefont
  {N.}~\bibnamefont {Tishby}}, \bibinfo {author} {\bibfnamefont
  {L.}~\bibnamefont {Vogt-Maranto}},\ and\ \bibinfo {author} {\bibfnamefont
  {L.}~\bibnamefont {Zdeborov\'a}},\ }\bibfield  {title} {\bibinfo {title}
  {Machine learning and the physical sciences},\ }\href
  {https://doi.org/10.1103/RevModPhys.91.045002} {\bibfield  {journal}
  {\bibinfo  {journal} {Reviews of Modern Physics}\ }\textbf {\bibinfo {volume}
  {91}},\ \bibinfo {pages} {045002} (\bibinfo {year} {2019})}\BibitemShut
  {NoStop}%
\bibitem [{\citenamefont {Carleo}\ and\ \citenamefont
  {Troyer}(2017)}]{carleo-science-2017}%
  \BibitemOpen
  \bibfield  {author} {\bibinfo {author} {\bibfnamefont {G.}~\bibnamefont
  {Carleo}}\ and\ \bibinfo {author} {\bibfnamefont {M.}~\bibnamefont
  {Troyer}},\ }\bibfield  {title} {\bibinfo {title} {Solving the quantum
  many-body problem with artificial neural networks},\ }\href
  {https://doi.org/10.1126/science.aag2302} {\bibfield  {journal} {\bibinfo
  {journal} {Science}\ }\textbf {\bibinfo {volume} {355}},\ \bibinfo {pages}
  {602} (\bibinfo {year} {2017})}\BibitemShut {NoStop}%
\bibitem [{\citenamefont {Torlai}\ and\ \citenamefont
  {Melko}(2020)}]{torlai-arcm-2020}%
  \BibitemOpen
  \bibfield  {author} {\bibinfo {author} {\bibfnamefont {G.}~\bibnamefont
  {Torlai}}\ and\ \bibinfo {author} {\bibfnamefont {R.~G.}\ \bibnamefont
  {Melko}},\ }\bibfield  {title} {\bibinfo {title} {{Machine-Learning Quantum
  States in the NISQ Era}},\ }\href
  {https://doi.org/10.1146/annurev-conmatphys-031119-050651} {\bibfield
  {journal} {\bibinfo  {journal} {Annual Review of Condensed Matter Physics}\
  }\textbf {\bibinfo {volume} {11}},\ \bibinfo {pages} {325} (\bibinfo {year}
  {2020})}\BibitemShut {NoStop}%
\bibitem [{\citenamefont {Krenn}\ \emph {et~al.}(2023)\citenamefont {Krenn},
  \citenamefont {Landgraf}, \citenamefont {Foesel},\ and\ \citenamefont
  {Marquardt}}]{Krenn2023}%
  \BibitemOpen
  \bibfield  {author} {\bibinfo {author} {\bibfnamefont {M.}~\bibnamefont
  {Krenn}}, \bibinfo {author} {\bibfnamefont {J.}~\bibnamefont {Landgraf}},
  \bibinfo {author} {\bibfnamefont {T.}~\bibnamefont {Foesel}},\ and\ \bibinfo
  {author} {\bibfnamefont {F.}~\bibnamefont {Marquardt}},\ }\bibfield  {title}
  {\bibinfo {title} {{Artificial intelligence and machine learning for quantum
  technologies}},\ }\href {https://doi.org/10.1103/PhysRevA.107.010101}
  {\bibfield  {journal} {\bibinfo  {journal} {Physical Review A}\ }\textbf
  {\bibinfo {volume} {107}},\ \bibinfo {pages} {010101} (\bibinfo {year}
  {2023})}\BibitemShut {NoStop}%
\bibitem [{\citenamefont {Xin}\ \emph {et~al.}(2020)\citenamefont {Xin},
  \citenamefont {Nie}, \citenamefont {Kong}, \citenamefont {Wen}, \citenamefont
  {Lu},\ and\ \citenamefont {Li}}]{xin-2020}%
  \BibitemOpen
  \bibfield  {author} {\bibinfo {author} {\bibfnamefont {T.}~\bibnamefont
  {Xin}}, \bibinfo {author} {\bibfnamefont {X.}~\bibnamefont {Nie}}, \bibinfo
  {author} {\bibfnamefont {X.}~\bibnamefont {Kong}}, \bibinfo {author}
  {\bibfnamefont {J.}~\bibnamefont {Wen}}, \bibinfo {author} {\bibfnamefont
  {D.}~\bibnamefont {Lu}},\ and\ \bibinfo {author} {\bibfnamefont
  {J.}~\bibnamefont {Li}},\ }\bibfield  {title} {\bibinfo {title} {{Quantum
  Pure State Tomography via Variational Hybrid Quantum-Classical Method}},\
  }\href {https://doi.org/10.1103/PhysRevApplied.13.024013} {\bibfield
  {journal} {\bibinfo  {journal} {Physical Review Applied}\ }\textbf {\bibinfo
  {volume} {13}},\ \bibinfo {pages} {024013} (\bibinfo {year}
  {2020})}\BibitemShut {NoStop}%
\bibitem [{\citenamefont {Liu}\ \emph {et~al.}(2020)\citenamefont {Liu},
  \citenamefont {Wang}, \citenamefont {Xue}, \citenamefont {Huang},
  \citenamefont {Fu}, \citenamefont {Qiang}, \citenamefont {Xu}, \citenamefont
  {Huang}, \citenamefont {Deng}, \citenamefont {Guo}, \citenamefont {Yang},\
  and\ \citenamefont {Wu}}]{liu-pra-2020}%
  \BibitemOpen
  \bibfield  {author} {\bibinfo {author} {\bibfnamefont {Y.}~\bibnamefont
  {Liu}}, \bibinfo {author} {\bibfnamefont {D.}~\bibnamefont {Wang}}, \bibinfo
  {author} {\bibfnamefont {S.}~\bibnamefont {Xue}}, \bibinfo {author}
  {\bibfnamefont {A.}~\bibnamefont {Huang}}, \bibinfo {author} {\bibfnamefont
  {X.}~\bibnamefont {Fu}}, \bibinfo {author} {\bibfnamefont {X.}~\bibnamefont
  {Qiang}}, \bibinfo {author} {\bibfnamefont {P.}~\bibnamefont {Xu}}, \bibinfo
  {author} {\bibfnamefont {H.-L.}\ \bibnamefont {Huang}}, \bibinfo {author}
  {\bibfnamefont {M.}~\bibnamefont {Deng}}, \bibinfo {author} {\bibfnamefont
  {C.}~\bibnamefont {Guo}}, \bibinfo {author} {\bibfnamefont {X.}~\bibnamefont
  {Yang}},\ and\ \bibinfo {author} {\bibfnamefont {J.}~\bibnamefont {Wu}},\
  }\bibfield  {title} {\bibinfo {title} {Variational quantum circuits for
  quantum state tomography},\ }\href
  {https://doi.org/10.1103/PhysRevA.101.052316} {\bibfield  {journal} {\bibinfo
   {journal} {Physical Review A}\ }\textbf {\bibinfo {volume} {101}},\ \bibinfo
  {pages} {052316} (\bibinfo {year} {2020})}\BibitemShut {NoStop}%
\bibitem [{\citenamefont {Granade}\ \emph {et~al.}(2016)\citenamefont
  {Granade}, \citenamefont {Combes},\ and\ \citenamefont
  {Cory}}]{granade-njp-2016}%
  \BibitemOpen
  \bibfield  {author} {\bibinfo {author} {\bibfnamefont {C.}~\bibnamefont
  {Granade}}, \bibinfo {author} {\bibfnamefont {J.}~\bibnamefont {Combes}},\
  and\ \bibinfo {author} {\bibfnamefont {D.~G.}\ \bibnamefont {Cory}},\
  }\bibfield  {title} {\bibinfo {title} {{Practical Bayesian tomography}},\
  }\href {https://doi.org/10.1088/1367-2630/18/3/033024} {\bibfield  {journal}
  {\bibinfo  {journal} {New Journal of Physics}\ }\textbf {\bibinfo {volume}
  {18}},\ \bibinfo {pages} {033024} (\bibinfo {year} {2016})}\BibitemShut
  {NoStop}%
\bibitem [{\citenamefont {Mondal}\ and\ \citenamefont
  {Dutta}(2023{\natexlab{b}})}]{Mondal-bayesian-2023}%
  \BibitemOpen
  \bibfield  {author} {\bibinfo {author} {\bibfnamefont {S.}~\bibnamefont
  {Mondal}}\ and\ \bibinfo {author} {\bibfnamefont {A.~K.}\ \bibnamefont
  {Dutta}},\ }\bibfield  {title} {\bibinfo {title} {{A Bayesian quantum state
  tomography along with adaptive frameworks based on linear minimum mean square
  error criterion}},\ }\href {https://doi.org/10.1088/1367-2630/ad0e49}
  {\bibfield  {journal} {\bibinfo  {journal} {New Journal of Physics}\ }\textbf
  {\bibinfo {volume} {25}},\ \bibinfo {pages} {123001} (\bibinfo {year}
  {2023}{\natexlab{b}})}\BibitemShut {NoStop}%
\bibitem [{\citenamefont {Quek}\ \emph {et~al.}(2021)\citenamefont {Quek},
  \citenamefont {Fort},\ and\ \citenamefont {Ng}}]{quek-npj-2021}%
  \BibitemOpen
  \bibfield  {author} {\bibinfo {author} {\bibfnamefont {Y.}~\bibnamefont
  {Quek}}, \bibinfo {author} {\bibfnamefont {S.}~\bibnamefont {Fort}},\ and\
  \bibinfo {author} {\bibfnamefont {H.~K.}\ \bibnamefont {Ng}},\ }\bibfield
  {title} {\bibinfo {title} {Adaptive quantum state tomography with neural
  networks},\ }\href {https://doi.org/10.1038/s41534-021-00436-9} {\bibfield
  {journal} {\bibinfo  {journal} {npj Quantum Information}\ }\textbf {\bibinfo
  {volume} {7}},\ \bibinfo {pages} {105} (\bibinfo {year} {2021})}\BibitemShut
  {NoStop}%
\bibitem [{\citenamefont {Gaikwad}\ \emph {et~al.}(2024)\citenamefont
  {Gaikwad}, \citenamefont {Bihani}, \citenamefont {Arvind},\ and\
  \citenamefont {Dorai}}]{gaikwad-pra-2024}%
  \BibitemOpen
  \bibfield  {author} {\bibinfo {author} {\bibfnamefont {A.}~\bibnamefont
  {Gaikwad}}, \bibinfo {author} {\bibfnamefont {O.}~\bibnamefont {Bihani}},
  \bibinfo {author} {\bibnamefont {Arvind}},\ and\ \bibinfo {author}
  {\bibfnamefont {K.}~\bibnamefont {Dorai}},\ }\bibfield  {title} {\bibinfo
  {title} {Neural-network-assisted quantum state and process tomography using
  limited data sets},\ }\href {https://doi.org/10.1103/PhysRevA.109.012402}
  {\bibfield  {journal} {\bibinfo  {journal} {Physical Review A}\ }\textbf
  {\bibinfo {volume} {109}},\ \bibinfo {pages} {012402} (\bibinfo {year}
  {2024})}\BibitemShut {NoStop}%
\bibitem [{\citenamefont {Lohani}\ \emph {et~al.}(2020)\citenamefont {Lohani},
  \citenamefont {Kirby}, \citenamefont {Brodsky}, \citenamefont {Danaci},\ and\
  \citenamefont {Glasser}}]{lohani-mlst-2020}%
  \BibitemOpen
  \bibfield  {author} {\bibinfo {author} {\bibfnamefont {S.}~\bibnamefont
  {Lohani}}, \bibinfo {author} {\bibfnamefont {B.~T.}\ \bibnamefont {Kirby}},
  \bibinfo {author} {\bibfnamefont {M.}~\bibnamefont {Brodsky}}, \bibinfo
  {author} {\bibfnamefont {O.}~\bibnamefont {Danaci}},\ and\ \bibinfo {author}
  {\bibfnamefont {R.~T.}\ \bibnamefont {Glasser}},\ }\bibfield  {title}
  {\bibinfo {title} {Machine learning assisted quantum state estimation},\
  }\href {https://doi.org/10.1088/2632-2153/ab9a21} {\bibfield  {journal}
  {\bibinfo  {journal} {Machine Learning: Science and Technology}\ }\textbf
  {\bibinfo {volume} {1}},\ \bibinfo {pages} {035007} (\bibinfo {year}
  {2020})}\BibitemShut {NoStop}%
\bibitem [{\citenamefont {Schmale}\ \emph {et~al.}(2022)\citenamefont
  {Schmale}, \citenamefont {Reh},\ and\ \citenamefont
  {G{\"a}rttner}}]{schmale-2022}%
  \BibitemOpen
  \bibfield  {author} {\bibinfo {author} {\bibfnamefont {T.}~\bibnamefont
  {Schmale}}, \bibinfo {author} {\bibfnamefont {M.}~\bibnamefont {Reh}},\ and\
  \bibinfo {author} {\bibfnamefont {M.}~\bibnamefont {G{\"a}rttner}},\
  }\bibfield  {title} {\bibinfo {title} {Efficient quantum state tomography
  with convolutional neural networks},\ }\href
  {https://doi.org/10.1038/s41534-022-00621-4} {\bibfield  {journal} {\bibinfo
  {journal} {npj Quantum Information}\ }\textbf {\bibinfo {volume} {8}},\
  \bibinfo {pages} {115} (\bibinfo {year} {2022})}\BibitemShut {NoStop}%
\bibitem [{\citenamefont {Ahmed}\ \emph
  {et~al.}(2021{\natexlab{a}})\citenamefont {Ahmed}, \citenamefont
  {{S{\'{a}}nchez Mu{\~{n}}oz}}, \citenamefont {Nori},\ and\ \citenamefont
  {Kockum}}]{Ahmed2021}%
  \BibitemOpen
  \bibfield  {author} {\bibinfo {author} {\bibfnamefont {S.}~\bibnamefont
  {Ahmed}}, \bibinfo {author} {\bibfnamefont {C.}~\bibnamefont {{S{\'{a}}nchez
  Mu{\~{n}}oz}}}, \bibinfo {author} {\bibfnamefont {F.}~\bibnamefont {Nori}},\
  and\ \bibinfo {author} {\bibfnamefont {A.~F.}\ \bibnamefont {Kockum}},\
  }\bibfield  {title} {\bibinfo {title} {{Quantum State Tomography with
  Conditional Generative Adversarial Networks}},\ }\href
  {https://doi.org/10.1103/PhysRevLett.127.140502} {\bibfield  {journal}
  {\bibinfo  {journal} {Physical Review Letters}\ }\textbf {\bibinfo {volume}
  {127}},\ \bibinfo {pages} {140502} (\bibinfo {year}
  {2021}{\natexlab{a}})}\BibitemShut {NoStop}%
\bibitem [{\citenamefont {Ahmed}\ \emph
  {et~al.}(2021{\natexlab{b}})\citenamefont {Ahmed}, \citenamefont
  {{S{\'{a}}nchez Mu{\~{n}}oz}}, \citenamefont {Nori},\ and\ \citenamefont
  {Kockum}}]{Ahmed2021a}%
  \BibitemOpen
  \bibfield  {author} {\bibinfo {author} {\bibfnamefont {S.}~\bibnamefont
  {Ahmed}}, \bibinfo {author} {\bibfnamefont {C.}~\bibnamefont {{S{\'{a}}nchez
  Mu{\~{n}}oz}}}, \bibinfo {author} {\bibfnamefont {F.}~\bibnamefont {Nori}},\
  and\ \bibinfo {author} {\bibfnamefont {A.~F.}\ \bibnamefont {Kockum}},\
  }\bibfield  {title} {\bibinfo {title} {{Classification and reconstruction of
  optical quantum states with deep neural networks}},\ }\href
  {https://doi.org/10.1103/PhysRevResearch.3.033278} {\bibfield  {journal}
  {\bibinfo  {journal} {Physical Review Research}\ }\textbf {\bibinfo {volume}
  {3}},\ \bibinfo {pages} {033278} (\bibinfo {year}
  {2021}{\natexlab{b}})}\BibitemShut {NoStop}%
\bibitem [{\citenamefont {Cha}\ \emph {et~al.}(2021)\citenamefont {Cha},
  \citenamefont {Ginsparg}, \citenamefont {Wu}, \citenamefont {Carrasquilla},
  \citenamefont {McMahon},\ and\ \citenamefont {Kim}}]{cha-mlst-2021}%
  \BibitemOpen
  \bibfield  {author} {\bibinfo {author} {\bibfnamefont {P.}~\bibnamefont
  {Cha}}, \bibinfo {author} {\bibfnamefont {P.}~\bibnamefont {Ginsparg}},
  \bibinfo {author} {\bibfnamefont {F.}~\bibnamefont {Wu}}, \bibinfo {author}
  {\bibfnamefont {J.}~\bibnamefont {Carrasquilla}}, \bibinfo {author}
  {\bibfnamefont {P.~L.}\ \bibnamefont {McMahon}},\ and\ \bibinfo {author}
  {\bibfnamefont {E.-A.}\ \bibnamefont {Kim}},\ }\bibfield  {title} {\bibinfo
  {title} {Attention-based quantum tomography},\ }\href
  {https://doi.org/10.1088/2632-2153/ac362b} {\bibfield  {journal} {\bibinfo
  {journal} {Machine learning: Science and Technology}\ }\textbf {\bibinfo
  {volume} {3}},\ \bibinfo {pages} {01LT01} (\bibinfo {year}
  {2021})}\BibitemShut {NoStop}%
\bibitem [{\citenamefont {Torlai}\ \emph {et~al.}(2018)\citenamefont {Torlai},
  \citenamefont {Mazzola}, \citenamefont {Carrasquilla}, \citenamefont
  {Troyer}, \citenamefont {Melko},\ and\ \citenamefont
  {Carleo}}]{torlai-np-2018}%
  \BibitemOpen
  \bibfield  {author} {\bibinfo {author} {\bibfnamefont {G.}~\bibnamefont
  {Torlai}}, \bibinfo {author} {\bibfnamefont {G.}~\bibnamefont {Mazzola}},
  \bibinfo {author} {\bibfnamefont {J.}~\bibnamefont {Carrasquilla}}, \bibinfo
  {author} {\bibfnamefont {M.}~\bibnamefont {Troyer}}, \bibinfo {author}
  {\bibfnamefont {R.}~\bibnamefont {Melko}},\ and\ \bibinfo {author}
  {\bibfnamefont {G.}~\bibnamefont {Carleo}},\ }\bibfield  {title} {\bibinfo
  {title} {Neural-network quantum state tomography},\ }\href
  {https://doi.org/10.1038/s41567-018-0048-5} {\bibfield  {journal} {\bibinfo
  {journal} {Nature Physics}\ }\textbf {\bibinfo {volume} {14}},\ \bibinfo
  {pages} {447} (\bibinfo {year} {2018})}\BibitemShut {NoStop}%
\bibitem [{\citenamefont {Glasser}\ \emph {et~al.}(2018)\citenamefont
  {Glasser}, \citenamefont {Pancotti}, \citenamefont {August}, \citenamefont
  {Rodriguez},\ and\ \citenamefont {Cirac}}]{glasser-prx-2018}%
  \BibitemOpen
  \bibfield  {author} {\bibinfo {author} {\bibfnamefont {I.}~\bibnamefont
  {Glasser}}, \bibinfo {author} {\bibfnamefont {N.}~\bibnamefont {Pancotti}},
  \bibinfo {author} {\bibfnamefont {M.}~\bibnamefont {August}}, \bibinfo
  {author} {\bibfnamefont {I.~D.}\ \bibnamefont {Rodriguez}},\ and\ \bibinfo
  {author} {\bibfnamefont {J.~I.}\ \bibnamefont {Cirac}},\ }\bibfield  {title}
  {\bibinfo {title} {{Neural-Network Quantum States, String-Bond States, and
  Chiral Topological States}},\ }\href
  {https://doi.org/10.1103/PhysRevX.8.011006} {\bibfield  {journal} {\bibinfo
  {journal} {Physical Review X}\ }\textbf {\bibinfo {volume} {8}},\ \bibinfo
  {pages} {011006} (\bibinfo {year} {2018})}\BibitemShut {NoStop}%
\bibitem [{\citenamefont {Lange}\ \emph {et~al.}(2023)\citenamefont {Lange},
  \citenamefont {Kebri{\v{c}}}, \citenamefont {Buser}, \citenamefont
  {Schollw{\"{o}}ck}, \citenamefont {Grusdt},\ and\ \citenamefont
  {Bohrdt}}]{lange-2023}%
  \BibitemOpen
  \bibfield  {author} {\bibinfo {author} {\bibfnamefont {H.}~\bibnamefont
  {Lange}}, \bibinfo {author} {\bibfnamefont {M.}~\bibnamefont {Kebri{\v{c}}}},
  \bibinfo {author} {\bibfnamefont {M.}~\bibnamefont {Buser}}, \bibinfo
  {author} {\bibfnamefont {U.}~\bibnamefont {Schollw{\"{o}}ck}}, \bibinfo
  {author} {\bibfnamefont {F.}~\bibnamefont {Grusdt}},\ and\ \bibinfo {author}
  {\bibfnamefont {A.}~\bibnamefont {Bohrdt}},\ }\bibfield  {title} {\bibinfo
  {title} {Adaptive {Q}uantum {S}tate {T}omography with {A}ctive {L}earning},\
  }\href {https://doi.org/10.22331/q-2023-10-09-1129} {\bibfield  {journal}
  {\bibinfo  {journal} {{Quantum}}\ }\textbf {\bibinfo {volume} {7}},\ \bibinfo
  {pages} {1129} (\bibinfo {year} {2023})}\BibitemShut {NoStop}%
\bibitem [{\citenamefont {Xin}\ \emph {et~al.}(2019)\citenamefont {Xin},
  \citenamefont {Lu}, \citenamefont {Cao}, \citenamefont {Anikeeva},
  \citenamefont {Lu}, \citenamefont {Li}, \citenamefont {Long},\ and\
  \citenamefont {Zeng}}]{xin-npj-2019}%
  \BibitemOpen
  \bibfield  {author} {\bibinfo {author} {\bibfnamefont {T.}~\bibnamefont
  {Xin}}, \bibinfo {author} {\bibfnamefont {S.}~\bibnamefont {Lu}}, \bibinfo
  {author} {\bibfnamefont {N.}~\bibnamefont {Cao}}, \bibinfo {author}
  {\bibfnamefont {G.}~\bibnamefont {Anikeeva}}, \bibinfo {author}
  {\bibfnamefont {D.}~\bibnamefont {Lu}}, \bibinfo {author} {\bibfnamefont
  {J.}~\bibnamefont {Li}}, \bibinfo {author} {\bibfnamefont {G.}~\bibnamefont
  {Long}},\ and\ \bibinfo {author} {\bibfnamefont {B.}~\bibnamefont {Zeng}},\
  }\bibfield  {title} {\bibinfo {title} {Local-measurement-based quantum state
  tomography via neural networks},\ }\href
  {https://doi.org/10.1038/s41534-019-0222-3} {\bibfield  {journal} {\bibinfo
  {journal} {npj Quantum Information}\ }\textbf {\bibinfo {volume} {5}},\
  \bibinfo {pages} {109} (\bibinfo {year} {2019})}\BibitemShut {NoStop}%
\bibitem [{\citenamefont {Yu}\ \emph {et~al.}(2019)\citenamefont {Yu},
  \citenamefont {Albarr{\'a}n-Arriagada}, \citenamefont {Retamal},
  \citenamefont {Wang}, \citenamefont {Liu}, \citenamefont {Ke}, \citenamefont
  {Meng}, \citenamefont {Li}, \citenamefont {Tang}, \citenamefont {Solano},
  \citenamefont {Lamata}, \citenamefont {Li},\ and\ \citenamefont
  {Guo}}]{yu-2019}%
  \BibitemOpen
  \bibfield  {author} {\bibinfo {author} {\bibfnamefont {S.}~\bibnamefont
  {Yu}}, \bibinfo {author} {\bibfnamefont {F.}~\bibnamefont
  {Albarr{\'a}n-Arriagada}}, \bibinfo {author} {\bibfnamefont {J.~C.}\
  \bibnamefont {Retamal}}, \bibinfo {author} {\bibfnamefont {Y.-T.}\
  \bibnamefont {Wang}}, \bibinfo {author} {\bibfnamefont {W.}~\bibnamefont
  {Liu}}, \bibinfo {author} {\bibfnamefont {Z.-J.}\ \bibnamefont {Ke}},
  \bibinfo {author} {\bibfnamefont {Y.}~\bibnamefont {Meng}}, \bibinfo {author}
  {\bibfnamefont {Z.-P.}\ \bibnamefont {Li}}, \bibinfo {author} {\bibfnamefont
  {J.-S.}\ \bibnamefont {Tang}}, \bibinfo {author} {\bibfnamefont
  {E.}~\bibnamefont {Solano}}, \bibinfo {author} {\bibfnamefont
  {L.}~\bibnamefont {Lamata}}, \bibinfo {author} {\bibfnamefont {C.-F.}\
  \bibnamefont {Li}},\ and\ \bibinfo {author} {\bibfnamefont {G.-C.}\
  \bibnamefont {Guo}},\ }\bibfield  {title} {\bibinfo {title} {{Reconstruction
  of a Photonic Qubit State with Reinforcement Learning}},\ }\href
  {https://doi.org/https://doi.org/10.1002/qute.201800074} {\bibfield
  {journal} {\bibinfo  {journal} {Advanced Quantum Technologies}\ }\textbf
  {\bibinfo {volume} {2}},\ \bibinfo {pages} {1800074} (\bibinfo {year}
  {2019})}\BibitemShut {NoStop}%
\bibitem [{\citenamefont {Neugebauer}\ \emph {et~al.}(2020)\citenamefont
  {Neugebauer}, \citenamefont {Fischer}, \citenamefont {J\"ager}, \citenamefont
  {Czischek}, \citenamefont {Jochim}, \citenamefont {Weidem\"uller},\ and\
  \citenamefont {G\"arttner}}]{neugebauer-pra-2020}%
  \BibitemOpen
  \bibfield  {author} {\bibinfo {author} {\bibfnamefont {M.}~\bibnamefont
  {Neugebauer}}, \bibinfo {author} {\bibfnamefont {L.}~\bibnamefont {Fischer}},
  \bibinfo {author} {\bibfnamefont {A.}~\bibnamefont {J\"ager}}, \bibinfo
  {author} {\bibfnamefont {S.}~\bibnamefont {Czischek}}, \bibinfo {author}
  {\bibfnamefont {S.}~\bibnamefont {Jochim}}, \bibinfo {author} {\bibfnamefont
  {M.}~\bibnamefont {Weidem\"uller}},\ and\ \bibinfo {author} {\bibfnamefont
  {M.}~\bibnamefont {G\"arttner}},\ }\bibfield  {title} {\bibinfo {title}
  {Neural-network quantum state tomography in a two-qubit experiment},\ }\href
  {https://doi.org/10.1103/PhysRevA.102.042604} {\bibfield  {journal} {\bibinfo
   {journal} {Physical Review A}\ }\textbf {\bibinfo {volume} {102}},\ \bibinfo
  {pages} {042604} (\bibinfo {year} {2020})}\BibitemShut {NoStop}%
\bibitem [{\citenamefont {Ghosh}\ \emph {et~al.}(2021)\citenamefont {Ghosh},
  \citenamefont {Opala}, \citenamefont {Matuszewski}, \citenamefont {Paterek},\
  and\ \citenamefont {Liew}}]{ghosh-2021}%
  \BibitemOpen
  \bibfield  {author} {\bibinfo {author} {\bibfnamefont {S.}~\bibnamefont
  {Ghosh}}, \bibinfo {author} {\bibfnamefont {A.}~\bibnamefont {Opala}},
  \bibinfo {author} {\bibfnamefont {M.}~\bibnamefont {Matuszewski}}, \bibinfo
  {author} {\bibfnamefont {T.}~\bibnamefont {Paterek}},\ and\ \bibinfo {author}
  {\bibfnamefont {T.~C.~H.}\ \bibnamefont {Liew}},\ }\bibfield  {title}
  {\bibinfo {title} {{Reconstructing Quantum States With Quantum Reservoir
  Networks}},\ }\href {https://doi.org/10.1109/TNNLS.2020.3009716} {\bibfield
  {journal} {\bibinfo  {journal} {IEEE Transactions on Neural Networks and
  Learning Systems}\ }\textbf {\bibinfo {volume} {32}},\ \bibinfo {pages}
  {3148} (\bibinfo {year} {2021})}\BibitemShut {NoStop}%
\bibitem [{\citenamefont {Koutn\'y}\ \emph {et~al.}(2022)\citenamefont
  {Koutn\'y}, \citenamefont {Motka}, \citenamefont {Hradil}, \citenamefont
  {\ifmmode \check{R}\else \v{R}\fi{}eh\'a\ifmmode~\check{c}\else
  \v{c}\fi{}ek},\ and\ \citenamefont {S\'anchez-Soto}}]{kauth-pra-2024}%
  \BibitemOpen
  \bibfield  {author} {\bibinfo {author} {\bibfnamefont {D.}~\bibnamefont
  {Koutn\'y}}, \bibinfo {author} {\bibfnamefont {L.}~\bibnamefont {Motka}},
  \bibinfo {author} {\bibfnamefont {Z.}~\bibnamefont {Hradil}}, \bibinfo
  {author} {\bibfnamefont {J.}~\bibnamefont {\ifmmode \check{R}\else
  \v{R}\fi{}eh\'a\ifmmode~\check{c}\else \v{c}\fi{}ek}},\ and\ \bibinfo
  {author} {\bibfnamefont {L.~L.}\ \bibnamefont {S\'anchez-Soto}},\ }\bibfield
  {title} {\bibinfo {title} {Neural-network quantum state tomography},\ }\href
  {https://doi.org/10.1103/PhysRevA.106.012409} {\bibfield  {journal} {\bibinfo
   {journal} {Physical Review A}\ }\textbf {\bibinfo {volume} {106}},\ \bibinfo
  {pages} {012409} (\bibinfo {year} {2022})}\BibitemShut {NoStop}%
\bibitem [{\citenamefont {Innan}\ \emph {et~al.}(2024)\citenamefont {Innan},
  \citenamefont {Siddiqui}, \citenamefont {Arora}, \citenamefont {Ghosh},
  \citenamefont {Ko{\c{c}}ak}, \citenamefont {Paragas}, \citenamefont {Galib},
  \citenamefont {Khan},\ and\ \citenamefont {Bennai}}]{innan-2024}%
  \BibitemOpen
  \bibfield  {author} {\bibinfo {author} {\bibfnamefont {N.}~\bibnamefont
  {Innan}}, \bibinfo {author} {\bibfnamefont {O.~I.}\ \bibnamefont {Siddiqui}},
  \bibinfo {author} {\bibfnamefont {S.}~\bibnamefont {Arora}}, \bibinfo
  {author} {\bibfnamefont {T.}~\bibnamefont {Ghosh}}, \bibinfo {author}
  {\bibfnamefont {Y.~P.}\ \bibnamefont {Ko{\c{c}}ak}}, \bibinfo {author}
  {\bibfnamefont {D.}~\bibnamefont {Paragas}}, \bibinfo {author} {\bibfnamefont
  {A.~A.~O.}\ \bibnamefont {Galib}}, \bibinfo {author} {\bibfnamefont
  {M.~A.-Z.}\ \bibnamefont {Khan}},\ and\ \bibinfo {author} {\bibfnamefont
  {M.}~\bibnamefont {Bennai}},\ }\bibfield  {title} {\bibinfo {title} {Quantum
  state tomography using quantum machine learning},\ }\href
  {https://doi.org/10.1007/s42484-024-00162-3} {\bibfield  {journal} {\bibinfo
  {journal} {Quantum Machine Intelligence}\ }\textbf {\bibinfo {volume} {6}},\
  \bibinfo {pages} {28} (\bibinfo {year} {2024})}\BibitemShut {NoStop}%
\bibitem [{\citenamefont {Palmieri}\ \emph {et~al.}(2024)\citenamefont
  {Palmieri}, \citenamefont {M{\"{u}}ller-Rigat}, \citenamefont {Srivastava},
  \citenamefont {Lewenstein}, \citenamefont {Rajchel-Mieldzio{\'{c}}},\ and\
  \citenamefont {P{\l}odzie{\'{n}}}}]{Palmieri2024}%
  \BibitemOpen
  \bibfield  {author} {\bibinfo {author} {\bibfnamefont {A.~M.}\ \bibnamefont
  {Palmieri}}, \bibinfo {author} {\bibfnamefont {G.}~\bibnamefont
  {M{\"{u}}ller-Rigat}}, \bibinfo {author} {\bibfnamefont {A.~K.}\ \bibnamefont
  {Srivastava}}, \bibinfo {author} {\bibfnamefont {M.}~\bibnamefont
  {Lewenstein}}, \bibinfo {author} {\bibfnamefont {G.}~\bibnamefont
  {Rajchel-Mieldzio{\'{c}}}},\ and\ \bibinfo {author} {\bibfnamefont
  {M.}~\bibnamefont {P{\l}odzie{\'{n}}}},\ }\bibfield  {title} {\bibinfo
  {title} {{Enhancing quantum state tomography via resource-efficient
  attention-based neural networks}},\ }\href
  {https://doi.org/10.1103/PhysRevResearch.6.033248} {\bibfield  {journal}
  {\bibinfo  {journal} {Physical Review Research}\ }\textbf {\bibinfo {volume}
  {6}},\ \bibinfo {pages} {033248} (\bibinfo {year} {2024})}\BibitemShut
  {NoStop}%
\bibitem [{\citenamefont {Ahmed}\ \emph {et~al.}(2023)\citenamefont {Ahmed},
  \citenamefont {Quijandr{\'{i}}a},\ and\ \citenamefont {Kockum}}]{Ahmed2023}%
  \BibitemOpen
  \bibfield  {author} {\bibinfo {author} {\bibfnamefont {S.}~\bibnamefont
  {Ahmed}}, \bibinfo {author} {\bibfnamefont {F.}~\bibnamefont
  {Quijandr{\'{i}}a}},\ and\ \bibinfo {author} {\bibfnamefont {A.~F.}\
  \bibnamefont {Kockum}},\ }\bibfield  {title} {\bibinfo {title}
  {{Gradient-Descent Quantum Process Tomography by Learning Kraus Operators}},\
  }\href {https://doi.org/10.1103/PhysRevLett.130.150402} {\bibfield  {journal}
  {\bibinfo  {journal} {Physical Review Letters}\ }\textbf {\bibinfo {volume}
  {130}},\ \bibinfo {pages} {150402} (\bibinfo {year} {2023})}\BibitemShut
  {NoStop}%
\bibitem [{\citenamefont {Kervinen}\ \emph {et~al.}(2024)\citenamefont
  {Kervinen}, \citenamefont {Ahmed}, \citenamefont {Kudra}, \citenamefont
  {Eriksson}, \citenamefont {Quijandr{\'{i}}a}, \citenamefont {Kockum},
  \citenamefont {Delsing},\ and\ \citenamefont {Gasparinetti}}]{Kervinen2024}%
  \BibitemOpen
  \bibfield  {author} {\bibinfo {author} {\bibfnamefont {M.}~\bibnamefont
  {Kervinen}}, \bibinfo {author} {\bibfnamefont {S.}~\bibnamefont {Ahmed}},
  \bibinfo {author} {\bibfnamefont {M.}~\bibnamefont {Kudra}}, \bibinfo
  {author} {\bibfnamefont {A.}~\bibnamefont {Eriksson}}, \bibinfo {author}
  {\bibfnamefont {F.}~\bibnamefont {Quijandr{\'{i}}a}}, \bibinfo {author}
  {\bibfnamefont {A.~F.}\ \bibnamefont {Kockum}}, \bibinfo {author}
  {\bibfnamefont {P.}~\bibnamefont {Delsing}},\ and\ \bibinfo {author}
  {\bibfnamefont {S.}~\bibnamefont {Gasparinetti}},\ }\bibfield  {title}
  {\bibinfo {title} {{Extended quantum process tomography of logical operations
  on an encoded bosonic qubit}},\ }\href
  {https://doi.org/10.1103/PhysRevA.110.L020401} {\bibfield  {journal}
  {\bibinfo  {journal} {Physical Review A}\ }\textbf {\bibinfo {volume}
  {110}},\ \bibinfo {pages} {L020401} (\bibinfo {year} {2024})}\BibitemShut
  {NoStop}%
\bibitem [{\citenamefont {Bolduc}\ \emph {et~al.}(2017)\citenamefont {Bolduc},
  \citenamefont {Knee}, \citenamefont {Gauger},\ and\ \citenamefont
  {Leach}}]{bolduc-npj-2017}%
  \BibitemOpen
  \bibfield  {author} {\bibinfo {author} {\bibfnamefont {E.}~\bibnamefont
  {Bolduc}}, \bibinfo {author} {\bibfnamefont {G.~C.}\ \bibnamefont {Knee}},
  \bibinfo {author} {\bibfnamefont {E.~M.}\ \bibnamefont {Gauger}},\ and\
  \bibinfo {author} {\bibfnamefont {J.}~\bibnamefont {Leach}},\ }\bibfield
  {title} {\bibinfo {title} {Projected gradient descent algorithms for quantum
  state tomography},\ }\href {https://doi.org/10.1038/s41534-017-0043-1}
  {\bibfield  {journal} {\bibinfo  {journal} {npj Quantum Information}\
  }\textbf {\bibinfo {volume} {3}},\ \bibinfo {pages} {44} (\bibinfo {year}
  {2017})}\BibitemShut {NoStop}%
\bibitem [{\citenamefont {Ferrie}(2014)}]{ferrie-prl-2014}%
  \BibitemOpen
  \bibfield  {author} {\bibinfo {author} {\bibfnamefont {C.}~\bibnamefont
  {Ferrie}},\ }\bibfield  {title} {\bibinfo {title} {{Self-Guided Quantum
  Tomography}},\ }\href {https://doi.org/10.1103/PhysRevLett.113.190404}
  {\bibfield  {journal} {\bibinfo  {journal} {Physical Review Letters}\
  }\textbf {\bibinfo {volume} {113}},\ \bibinfo {pages} {190404} (\bibinfo
  {year} {2014})}\BibitemShut {NoStop}%
\bibitem [{\citenamefont {Hsu}\ \emph {et~al.}(2024)\citenamefont {Hsu},
  \citenamefont {Kuo}, \citenamefont {Yu}, \citenamefont {Cai},\ and\
  \citenamefont {Hsieh}}]{hsu-prl-2024}%
  \BibitemOpen
  \bibfield  {author} {\bibinfo {author} {\bibfnamefont {M.-C.}\ \bibnamefont
  {Hsu}}, \bibinfo {author} {\bibfnamefont {E.-J.}\ \bibnamefont {Kuo}},
  \bibinfo {author} {\bibfnamefont {W.-H.}\ \bibnamefont {Yu}}, \bibinfo
  {author} {\bibfnamefont {J.-F.}\ \bibnamefont {Cai}},\ and\ \bibinfo {author}
  {\bibfnamefont {M.-H.}\ \bibnamefont {Hsieh}},\ }\bibfield  {title} {\bibinfo
  {title} {{Quantum State Tomography via Nonconvex Riemannian Gradient
  Descent}},\ }\href {https://doi.org/10.1103/PhysRevLett.132.240804}
  {\bibfield  {journal} {\bibinfo  {journal} {Physical Review Letters}\
  }\textbf {\bibinfo {volume} {132}},\ \bibinfo {pages} {240804} (\bibinfo
  {year} {2024})}\BibitemShut {NoStop}%
\bibitem [{\citenamefont {Wang}\ \emph {et~al.}(2024)\citenamefont {Wang},
  \citenamefont {Liu}, \citenamefont {Cheng}, \citenamefont {Li},\ and\
  \citenamefont {Chen}}]{wang-prr-2024}%
  \BibitemOpen
  \bibfield  {author} {\bibinfo {author} {\bibfnamefont {Y.}~\bibnamefont
  {Wang}}, \bibinfo {author} {\bibfnamefont {L.}~\bibnamefont {Liu}}, \bibinfo
  {author} {\bibfnamefont {S.}~\bibnamefont {Cheng}}, \bibinfo {author}
  {\bibfnamefont {L.}~\bibnamefont {Li}},\ and\ \bibinfo {author}
  {\bibfnamefont {J.}~\bibnamefont {Chen}},\ }\bibfield  {title} {\bibinfo
  {title} {Efficient factored gradient descent algorithm for quantum state
  tomography},\ }\href {https://doi.org/10.1103/PhysRevResearch.6.033034}
  {\bibfield  {journal} {\bibinfo  {journal} {Physical Review Research}\
  }\textbf {\bibinfo {volume} {6}},\ \bibinfo {pages} {033034} (\bibinfo {year}
  {2024})}\BibitemShut {NoStop}%
\bibitem [{\citenamefont {Flammia}\ and\ \citenamefont
  {Liu}(2011)}]{flammia-prl-2011}%
  \BibitemOpen
  \bibfield  {author} {\bibinfo {author} {\bibfnamefont {S.~T.}\ \bibnamefont
  {Flammia}}\ and\ \bibinfo {author} {\bibfnamefont {Y.-K.}\ \bibnamefont
  {Liu}},\ }\bibfield  {title} {\bibinfo {title} {{Direct Fidelity Estimation
  from Few Pauli Measurements}},\ }\href
  {https://doi.org/10.1103/PhysRevLett.106.230501} {\bibfield  {journal}
  {\bibinfo  {journal} {Physical Review Letters}\ }\textbf {\bibinfo {volume}
  {106}},\ \bibinfo {pages} {230501} (\bibinfo {year} {2011})}\BibitemShut
  {NoStop}%
\bibitem [{\citenamefont {da~Silva}\ \emph {et~al.}(2011)\citenamefont
  {da~Silva}, \citenamefont {Landon-Cardinal},\ and\ \citenamefont
  {Poulin}}]{silva-prl-2011}%
  \BibitemOpen
  \bibfield  {author} {\bibinfo {author} {\bibfnamefont {M.~P.}\ \bibnamefont
  {da~Silva}}, \bibinfo {author} {\bibfnamefont {O.}~\bibnamefont
  {Landon-Cardinal}},\ and\ \bibinfo {author} {\bibfnamefont {D.}~\bibnamefont
  {Poulin}},\ }\bibfield  {title} {\bibinfo {title} {Practical characterization
  of quantum devices without tomography},\ }\href
  {https://doi.org/10.1103/PhysRevLett.107.210404} {\bibfield  {journal}
  {\bibinfo  {journal} {Physical Review Letters}\ }\textbf {\bibinfo {volume}
  {107}},\ \bibinfo {pages} {210404} (\bibinfo {year} {2011})}\BibitemShut
  {NoStop}%
\bibitem [{\citenamefont {Spall}(1992)}]{spall-ieee-1992}%
  \BibitemOpen
  \bibfield  {author} {\bibinfo {author} {\bibfnamefont {J.}~\bibnamefont
  {Spall}},\ }\bibfield  {title} {\bibinfo {title} {Multivariate stochastic
  approximation using a simultaneous perturbation gradient approximation},\
  }\href {https://doi.org/10.1109/9.119632} {\bibfield  {journal} {\bibinfo
  {journal} {IEEE Transactions on Automatic Control}\ }\textbf {\bibinfo
  {volume} {37}},\ \bibinfo {pages} {332} (\bibinfo {year} {1992})}\BibitemShut
  {NoStop}%
\bibitem [{gd-()}]{gd-qst-python}%
  \BibitemOpen
  \href@noop {} {\bibinfo {title} {{Python code for GD-QST}}},\ \bibinfo
  {howpublished} {\url{https://github.com/mstorresh/GD-QST}}\BibitemShut
  {NoStop}%
\bibitem [{\citenamefont {Rodionov}\ \emph {et~al.}(2014)\citenamefont
  {Rodionov}, \citenamefont {Veitia}, \citenamefont {Barends}, \citenamefont
  {Kelly}, \citenamefont {Sank}, \citenamefont {Wenner}, \citenamefont
  {Martinis}, \citenamefont {Kosut},\ and\ \citenamefont
  {Korotkov}}]{rod-prb-2014}%
  \BibitemOpen
  \bibfield  {author} {\bibinfo {author} {\bibfnamefont {A.~V.}\ \bibnamefont
  {Rodionov}}, \bibinfo {author} {\bibfnamefont {A.}~\bibnamefont {Veitia}},
  \bibinfo {author} {\bibfnamefont {R.}~\bibnamefont {Barends}}, \bibinfo
  {author} {\bibfnamefont {J.}~\bibnamefont {Kelly}}, \bibinfo {author}
  {\bibfnamefont {D.}~\bibnamefont {Sank}}, \bibinfo {author} {\bibfnamefont
  {J.}~\bibnamefont {Wenner}}, \bibinfo {author} {\bibfnamefont {J.~M.}\
  \bibnamefont {Martinis}}, \bibinfo {author} {\bibfnamefont {R.~L.}\
  \bibnamefont {Kosut}},\ and\ \bibinfo {author} {\bibfnamefont {A.~N.}\
  \bibnamefont {Korotkov}},\ }\bibfield  {title} {\bibinfo {title} {Compressed
  sensing quantum process tomography for superconducting quantum gates},\
  }\href {https://doi.org/10.1103/PhysRevB.90.144504} {\bibfield  {journal}
  {\bibinfo  {journal} {Physical Review B}\ }\textbf {\bibinfo {volume} {90}},\
  \bibinfo {pages} {144504} (\bibinfo {year} {2014})}\BibitemShut {NoStop}%
\bibitem [{\citenamefont {Toh}\ \emph {et~al.}(2004)\citenamefont {Toh},
  \citenamefont {Tutuncu},\ and\ \citenamefont {Todd}}]{sdpt-2004}%
  \BibitemOpen
  \bibfield  {author} {\bibinfo {author} {\bibfnamefont {K.}~\bibnamefont
  {Toh}}, \bibinfo {author} {\bibfnamefont {R.}~\bibnamefont {Tutuncu}},\ and\
  \bibinfo {author} {\bibfnamefont {M.}~\bibnamefont {Todd}},\ }\bibfield
  {title} {\bibinfo {title} {{On the implementation of SDPT3 (version 3.1) - a
  MATLAB software package for semidefinite-quadratic-linear programming}},\
  }in\ \href {https://doi.org/10.1109/CACSD.2004.1393891} {\emph {\bibinfo
  {booktitle} {2004 IEEE International Conference on Robotics and Automation
  (IEEE Cat. No.04CH37508)}}}\ (\bibinfo {year} {2004})\ pp.\ \bibinfo {pages}
  {290--296}\BibitemShut {NoStop}%
\bibitem [{cvx()}]{cvx-solvers}%
  \BibitemOpen
  \href@noop {} {}\bibinfo {howpublished}
  {\url{https://www.cvxpy.org/tutorial/solvers/index.html}}\BibitemShut
  {NoStop}%
\bibitem [{\citenamefont {Toh}\ \emph {et~al.}(1999)\citenamefont {Toh},
  \citenamefont {Todd},\ and\ \citenamefont
  {T{\"u}t{\"u}nc{\"u}}}]{toh-sdpt-1999}%
  \BibitemOpen
  \bibfield  {author} {\bibinfo {author} {\bibfnamefont {K.~C.}\ \bibnamefont
  {Toh}}, \bibinfo {author} {\bibfnamefont {M.~J.}\ \bibnamefont {Todd}},\ and\
  \bibinfo {author} {\bibfnamefont {R.~H.}\ \bibnamefont
  {T{\"u}t{\"u}nc{\"u}}},\ }\bibfield  {title} {\bibinfo {title} {{SDPT3 - A
  Matlab software package for semidefinite programming, Version 1.3}},\ }\href
  {https://doi.org/10.1080/10556789908805762} {\bibfield  {journal} {\bibinfo
  {journal} {Optimization Methods and Software}\ }\textbf {\bibinfo {volume}
  {11}},\ \bibinfo {pages} {545} (\bibinfo {year} {1999})}\BibitemShut
  {NoStop}%
\bibitem [{\citenamefont {Hou}\ \emph {et~al.}(2016)\citenamefont {Hou},
  \citenamefont {Zhong}, \citenamefont {Tian}, \citenamefont {Dong},
  \citenamefont {Qi}, \citenamefont {Li}, \citenamefont {Wang}, \citenamefont
  {Nori}, \citenamefont {Xiang}, \citenamefont {Li},\ and\ \citenamefont
  {Guo}}]{hou-njp-2016}%
  \BibitemOpen
  \bibfield  {author} {\bibinfo {author} {\bibfnamefont {Z.}~\bibnamefont
  {Hou}}, \bibinfo {author} {\bibfnamefont {H.-S.}\ \bibnamefont {Zhong}},
  \bibinfo {author} {\bibfnamefont {Y.}~\bibnamefont {Tian}}, \bibinfo {author}
  {\bibfnamefont {D.}~\bibnamefont {Dong}}, \bibinfo {author} {\bibfnamefont
  {B.}~\bibnamefont {Qi}}, \bibinfo {author} {\bibfnamefont {L.}~\bibnamefont
  {Li}}, \bibinfo {author} {\bibfnamefont {Y.}~\bibnamefont {Wang}}, \bibinfo
  {author} {\bibfnamefont {F.}~\bibnamefont {Nori}}, \bibinfo {author}
  {\bibfnamefont {G.-Y.}\ \bibnamefont {Xiang}}, \bibinfo {author}
  {\bibfnamefont {C.-F.}\ \bibnamefont {Li}},\ and\ \bibinfo {author}
  {\bibfnamefont {G.-C.}\ \bibnamefont {Guo}},\ }\bibfield  {title} {\bibinfo
  {title} {Full reconstruction of a 14-qubit state within four hours},\ }\href
  {https://doi.org/10.1088/1367-2630/18/8/083036} {\bibfield  {journal}
  {\bibinfo  {journal} {New Journal of Physics}\ }\textbf {\bibinfo {volume}
  {18}},\ \bibinfo {pages} {083036} (\bibinfo {year} {2016})}\BibitemShut
  {NoStop}%
\bibitem [{\citenamefont {Ruder}(2016)}]{ruder2016overview}%
  \BibitemOpen
  \bibfield  {author} {\bibinfo {author} {\bibfnamefont {S.}~\bibnamefont
  {Ruder}},\ }\href@noop {} {\bibinfo {title} {An overview of gradient descent
  optimization algorithms}} (\bibinfo {year} {2016}),\ \Eprint
  {https://arxiv.org/abs/1609.04747} {arXiv:1609.04747} \BibitemShut {NoStop}%
\bibitem [{\citenamefont {Bertsekas}\ and\ \citenamefont
  {Tsitsiklis}(2000)}]{dimitri-siam-2000}%
  \BibitemOpen
  \bibfield  {author} {\bibinfo {author} {\bibfnamefont {D.~P.}\ \bibnamefont
  {Bertsekas}}\ and\ \bibinfo {author} {\bibfnamefont {J.~N.}\ \bibnamefont
  {Tsitsiklis}},\ }\bibfield  {title} {\bibinfo {title} {{Gradient Convergence
  in Gradient methods with Errors}},\ }\href
  {https://doi.org/10.1137/S1052623497331063} {\bibfield  {journal} {\bibinfo
  {journal} {SIAM Journal on Optimization}\ }\textbf {\bibinfo {volume} {10}},\
  \bibinfo {pages} {627} (\bibinfo {year} {2000})}\BibitemShut {NoStop}%
\bibitem [{\citenamefont {Panageas}\ \emph {et~al.}(2019)\citenamefont
  {Panageas}, \citenamefont {Piliouras},\ and\ \citenamefont
  {Wang}}]{pana-book-2019}%
  \BibitemOpen
  \bibfield  {author} {\bibinfo {author} {\bibfnamefont {I.}~\bibnamefont
  {Panageas}}, \bibinfo {author} {\bibfnamefont {G.}~\bibnamefont
  {Piliouras}},\ and\ \bibinfo {author} {\bibfnamefont {X.}~\bibnamefont
  {Wang}},\ }\bibfield  {title} {\bibinfo {title} {First-order methods almost
  always avoid saddle points: The case of vanishing step-sizes},\ }in\ \href
  {https://proceedings.neurips.cc/paper_files/paper/2019/file/3fb04953d95a94367bb133f862402bce-Paper.pdf}
  {\emph {\bibinfo {booktitle} {Advances in Neural Information Processing
  Systems}}},\ Vol.~\bibinfo {volume} {32},\ \bibinfo {editor} {edited by\
  \bibinfo {editor} {\bibfnamefont {H.}~\bibnamefont {Wallach}}, \bibinfo
  {editor} {\bibfnamefont {H.}~\bibnamefont {Larochelle}}, \bibinfo {editor}
  {\bibfnamefont {A.}~\bibnamefont {Beygelzimer}}, \bibinfo {editor}
  {\bibfnamefont {F.}~\bibnamefont {d\textquotesingle Alch\'{e}-Buc}}, \bibinfo
  {editor} {\bibfnamefont {E.}~\bibnamefont {Fox}},\ and\ \bibinfo {editor}
  {\bibfnamefont {R.}~\bibnamefont {Garnett}}}\ (\bibinfo  {publisher} {Curran
  Associates, Inc.},\ \bibinfo {year} {2019})\BibitemShut {NoStop}%
\bibitem [{\citenamefont {Du}\ \emph {et~al.}(2019)\citenamefont {Du},
  \citenamefont {Zhai}, \citenamefont {Poczos},\ and\ \citenamefont
  {Singh}}]{simon-arxiv-2019}%
  \BibitemOpen
  \bibfield  {author} {\bibinfo {author} {\bibfnamefont {S.~S.}\ \bibnamefont
  {Du}}, \bibinfo {author} {\bibfnamefont {X.}~\bibnamefont {Zhai}}, \bibinfo
  {author} {\bibfnamefont {B.}~\bibnamefont {Poczos}},\ and\ \bibinfo {author}
  {\bibfnamefont {A.}~\bibnamefont {Singh}},\ }\href@noop {} {\bibinfo {title}
  {{Gradient Descent Provably Optimizes Over-parameterized Neural Networks}}}
  (\bibinfo {year} {2019}),\ \Eprint {https://arxiv.org/abs/1810.02054}
  {arXiv:1810.02054} \BibitemShut {NoStop}%
\bibitem [{\citenamefont {Hoshi}\ \emph {et~al.}(2025)\citenamefont {Hoshi},
  \citenamefont {Nagase}, \citenamefont {Kwon}, \citenamefont {Iyama},
  \citenamefont {Kamiya}, \citenamefont {Fujii}, \citenamefont {Mukai},
  \citenamefont {Ahmed}, \citenamefont {Kockum}, \citenamefont {Watabe},
  \citenamefont {Yoshihara},\ and\ \citenamefont {Tsai}}]{Hoshi2025}%
  \BibitemOpen
  \bibfield  {author} {\bibinfo {author} {\bibfnamefont {D.}~\bibnamefont
  {Hoshi}}, \bibinfo {author} {\bibfnamefont {T.}~\bibnamefont {Nagase}},
  \bibinfo {author} {\bibfnamefont {S.}~\bibnamefont {Kwon}}, \bibinfo {author}
  {\bibfnamefont {D.}~\bibnamefont {Iyama}}, \bibinfo {author} {\bibfnamefont
  {T.}~\bibnamefont {Kamiya}}, \bibinfo {author} {\bibfnamefont
  {S.}~\bibnamefont {Fujii}}, \bibinfo {author} {\bibfnamefont
  {H.}~\bibnamefont {Mukai}}, \bibinfo {author} {\bibfnamefont
  {S.}~\bibnamefont {Ahmed}}, \bibinfo {author} {\bibfnamefont {A.~F.}\
  \bibnamefont {Kockum}}, \bibinfo {author} {\bibfnamefont {S.}~\bibnamefont
  {Watabe}}, \bibinfo {author} {\bibfnamefont {F.}~\bibnamefont {Yoshihara}},\
  and\ \bibinfo {author} {\bibfnamefont {J.-S.}\ \bibnamefont {Tsai}},\
  }\bibfield  {title} {\bibinfo {title} {{Entangling Schr{\"{o}}dinger's cat
  states by bridging discrete- and continuous-variable encoding}},\ }\href
  {https://doi.org/10.1038/s41467-025-56503-8} {\bibfield  {journal} {\bibinfo
  {journal} {Nature Communications}\ }\textbf {\bibinfo {volume} {16}},\
  \bibinfo {pages} {1309} (\bibinfo {year} {2025})}\BibitemShut {NoStop}%
\bibitem [{\citenamefont {Tagare}(2011)}]{Tagare2011}%
  \BibitemOpen
  \bibfield  {author} {\bibinfo {author} {\bibfnamefont {H.~D.}\ \bibnamefont
  {Tagare}},\ }\bibfield  {title} {\bibinfo {title} {{Notes on optimization on
  Stiefel manifolds}},\ }\href@noop {} {\bibfield  {journal} {\bibinfo
  {journal} {Yale University, New Haven}\ } (\bibinfo {year}
  {2011})}\BibitemShut {NoStop}%
\bibitem [{\citenamefont {Boumal}(2023)}]{Boumal2023}%
  \BibitemOpen
  \bibfield  {author} {\bibinfo {author} {\bibfnamefont {N.}~\bibnamefont
  {Boumal}},\ }\href@noop {} {\emph {\bibinfo {title} {{An Introduction to
  Optimization on Smooth Manifolds}}}}\ (\bibinfo  {publisher} {Cambridge
  University Press},\ \bibinfo {year} {2023})\BibitemShut {NoStop}%
\bibitem [{\citenamefont {Wen}\ and\ \citenamefont {Yin}(2013)}]{wen-mp-1013}%
  \BibitemOpen
  \bibfield  {author} {\bibinfo {author} {\bibfnamefont {Z.}~\bibnamefont
  {Wen}}\ and\ \bibinfo {author} {\bibfnamefont {W.}~\bibnamefont {Yin}},\
  }\bibfield  {title} {\bibinfo {title} {A feasible method for optimization
  with orthogonality constraints},\ }\href
  {https://doi.org/10.1007/s10107-012-0584-1} {\bibfield  {journal} {\bibinfo
  {journal} {Mathematical Programming}\ }\textbf {\bibinfo {volume} {142}},\
  \bibinfo {pages} {397} (\bibinfo {year} {2013})}\BibitemShut {NoStop}%
\bibitem [{\citenamefont {Jiang}\ and\ \citenamefont
  {Dai}(2015)}]{jiang-mp-2015}%
  \BibitemOpen
  \bibfield  {author} {\bibinfo {author} {\bibfnamefont {B.}~\bibnamefont
  {Jiang}}\ and\ \bibinfo {author} {\bibfnamefont {Y.-H.}\ \bibnamefont
  {Dai}},\ }\bibfield  {title} {\bibinfo {title} {A framework of constraint
  preserving update schemes for optimization on stiefel manifold},\ }\href
  {https://doi.org/10.1007/s10107-014-0816-7} {\bibfield  {journal} {\bibinfo
  {journal} {Mathematical Programming}\ }\textbf {\bibinfo {volume} {153}},\
  \bibinfo {pages} {535} (\bibinfo {year} {2015})}\BibitemShut {NoStop}%
\bibitem [{\citenamefont {Brieger}\ \emph {et~al.}(2023)\citenamefont
  {Brieger}, \citenamefont {Roth},\ and\ \citenamefont
  {Kliesch}}]{martin-prx-quant-2023}%
  \BibitemOpen
  \bibfield  {author} {\bibinfo {author} {\bibfnamefont {R.}~\bibnamefont
  {Brieger}}, \bibinfo {author} {\bibfnamefont {I.}~\bibnamefont {Roth}},\ and\
  \bibinfo {author} {\bibfnamefont {M.}~\bibnamefont {Kliesch}},\ }\bibfield
  {title} {\bibinfo {title} {Compressive gate set tomography},\ }\href
  {https://doi.org/10.1103/PRXQuantum.4.010325} {\bibfield  {journal} {\bibinfo
   {journal} {PRX Quantum}\ }\textbf {\bibinfo {volume} {4}},\ \bibinfo {pages}
  {010325} (\bibinfo {year} {2023})}\BibitemShut {NoStop}%
\bibitem [{\citenamefont {Luchnikov}\ \emph {et~al.}(2021)\citenamefont
  {Luchnikov}, \citenamefont {Krechetov},\ and\ \citenamefont
  {Filippov}}]{Luchnikov2021}%
  \BibitemOpen
  \bibfield  {author} {\bibinfo {author} {\bibfnamefont {I.~A.}\ \bibnamefont
  {Luchnikov}}, \bibinfo {author} {\bibfnamefont {M.~E.}\ \bibnamefont
  {Krechetov}},\ and\ \bibinfo {author} {\bibfnamefont {S.~N.}\ \bibnamefont
  {Filippov}},\ }\bibfield  {title} {\bibinfo {title} {{Riemannian geometry and
  automatic differentiation for optimization problems of quantum physics and
  quantum technologies}},\ }\href {https://doi.org/10.1088/1367-2630/ac0b02}
  {\bibfield  {journal} {\bibinfo  {journal} {New Journal of Physics}\ }\textbf
  {\bibinfo {volume} {23}},\ \bibinfo {pages} {073006} (\bibinfo {year}
  {2021})}\BibitemShut {NoStop}%
\bibitem [{\citenamefont {Absil}\ \emph {et~al.}(2009)\citenamefont {Absil},
  \citenamefont {Mahony},\ and\ \citenamefont
  {Sepulchre}}]{absil2009optimization}%
  \BibitemOpen
  \bibfield  {author} {\bibinfo {author} {\bibfnamefont {P.-A.}\ \bibnamefont
  {Absil}}, \bibinfo {author} {\bibfnamefont {R.}~\bibnamefont {Mahony}},\ and\
  \bibinfo {author} {\bibfnamefont {R.}~\bibnamefont {Sepulchre}},\ }\href@noop
  {} {\emph {\bibinfo {title} {Optimization Algorithms on Matrix Manifolds}}}\
  (\bibinfo  {publisher} {Princeton University Press},\ \bibinfo {address}
  {Princeton, NJ},\ \bibinfo {year} {2009})\BibitemShut {NoStop}%
\bibitem [{\citenamefont {Li}\ \emph {et~al.}(2020)\citenamefont {Li},
  \citenamefont {Fuxin},\ and\ \citenamefont {Todorovic}}]{jun-arxiv-2020}%
  \BibitemOpen
  \bibfield  {author} {\bibinfo {author} {\bibfnamefont {J.}~\bibnamefont
  {Li}}, \bibinfo {author} {\bibfnamefont {L.}~\bibnamefont {Fuxin}},\ and\
  \bibinfo {author} {\bibfnamefont {S.}~\bibnamefont {Todorovic}},\ }\href@noop
  {} {\bibinfo {title} {{Efficient Riemannian Optimization on the Stiefel
  Manifold via the Cayley Transform}}} (\bibinfo {year} {2020}),\ \Eprint
  {https://arxiv.org/abs/2002.01113} {arXiv:2002.01113} \BibitemShut {NoStop}%
\bibitem [{\citenamefont {Levitin}\ and\ \citenamefont
  {Polyak}(1966)}]{levitin-1966}%
  \BibitemOpen
  \bibfield  {author} {\bibinfo {author} {\bibfnamefont {E.}~\bibnamefont
  {Levitin}}\ and\ \bibinfo {author} {\bibfnamefont {B.}~\bibnamefont
  {Polyak}},\ }\bibfield  {title} {\bibinfo {title} {Constrained minimization
  methods},\ }\href
  {https://doi.org/https://doi.org/10.1016/0041-5553(66)90114-5} {\bibfield
  {journal} {\bibinfo  {journal} {USSR Computational Mathematics and
  Mathematical Physics}\ }\textbf {\bibinfo {volume} {6}},\ \bibinfo {pages}
  {1} (\bibinfo {year} {1966})}\BibitemShut {NoStop}%
\bibitem [{\citenamefont {Bruck}(1977)}]{bruck-1977}%
  \BibitemOpen
  \bibfield  {author} {\bibinfo {author} {\bibfnamefont {R.~E.}\ \bibnamefont
  {Bruck}},\ }\bibfield  {title} {\bibinfo {title} {{On the weak convergence of
  an ergodic iteration for the solution of variational inequalities for
  monotone operators in Hilbert space}},\ }\href
  {https://doi.org/https://doi.org/10.1016/0022-247X(77)90152-4} {\bibfield
  {journal} {\bibinfo  {journal} {Journal of Mathematical Analysis and
  Applications}\ }\textbf {\bibinfo {volume} {61}},\ \bibinfo {pages} {159}
  (\bibinfo {year} {1977})}\BibitemShut {NoStop}%
\bibitem [{\citenamefont {Lvovsky}(2004)}]{imle}%
  \BibitemOpen
  \bibfield  {author} {\bibinfo {author} {\bibfnamefont {A.~I.}\ \bibnamefont
  {Lvovsky}},\ }\bibfield  {title} {\bibinfo {title} {Iterative
  maximum-likelihood reconstruction in quantum homodyne tomography},\ }\href
  {https://doi.org/10.1088/1464-4266/6/6/014} {\bibfield  {journal} {\bibinfo
  {journal} {Journal of Optics B: Quantum and Semiclassical Optics}\ }\textbf
  {\bibinfo {volume} {6}},\ \bibinfo {pages} {S556} (\bibinfo {year}
  {2004})}\BibitemShut {NoStop}%
\bibitem [{\citenamefont {Goodfellow}\ \emph {et~al.}(2014)\citenamefont
  {Goodfellow}, \citenamefont {Pouget-Abadie}, \citenamefont {Mirza},
  \citenamefont {Xu}, \citenamefont {Warde-Farley}, \citenamefont {Ozair},
  \citenamefont {Courville},\ and\ \citenamefont {Bengio}}]{ian-gan-2014}%
  \BibitemOpen
  \bibfield  {author} {\bibinfo {author} {\bibfnamefont {I.}~\bibnamefont
  {Goodfellow}}, \bibinfo {author} {\bibfnamefont {J.}~\bibnamefont
  {Pouget-Abadie}}, \bibinfo {author} {\bibfnamefont {M.}~\bibnamefont
  {Mirza}}, \bibinfo {author} {\bibfnamefont {B.}~\bibnamefont {Xu}}, \bibinfo
  {author} {\bibfnamefont {D.}~\bibnamefont {Warde-Farley}}, \bibinfo {author}
  {\bibfnamefont {S.}~\bibnamefont {Ozair}}, \bibinfo {author} {\bibfnamefont
  {A.}~\bibnamefont {Courville}},\ and\ \bibinfo {author} {\bibfnamefont
  {Y.}~\bibnamefont {Bengio}},\ }\bibfield  {title} {\bibinfo {title}
  {Generative adversarial nets},\ }in\ \href
  {https://proceedings.neurips.cc/paper_files/paper/2014/file/5ca3e9b122f61f8f06494c97b1afccf3-Paper.pdf}
  {\emph {\bibinfo {booktitle} {Advances in Neural Information Processing
  Systems}}},\ Vol.~\bibinfo {volume} {27},\ \bibinfo {editor} {edited by\
  \bibinfo {editor} {\bibfnamefont {Z.}~\bibnamefont {Ghahramani}}, \bibinfo
  {editor} {\bibfnamefont {M.}~\bibnamefont {Welling}}, \bibinfo {editor}
  {\bibfnamefont {C.}~\bibnamefont {Cortes}}, \bibinfo {editor} {\bibfnamefont
  {N.}~\bibnamefont {Lawrence}},\ and\ \bibinfo {editor} {\bibfnamefont
  {K.}~\bibnamefont {Weinberger}}}\ (\bibinfo  {publisher} {Curran Associates,
  Inc.},\ \bibinfo {year} {2014})\BibitemShut {NoStop}%
\bibitem [{\citenamefont {Mirza}\ and\ \citenamefont
  {Osindero}(2014)}]{mirza-arxiv-2014}%
  \BibitemOpen
  \bibfield  {author} {\bibinfo {author} {\bibfnamefont {M.}~\bibnamefont
  {Mirza}}\ and\ \bibinfo {author} {\bibfnamefont {S.}~\bibnamefont
  {Osindero}},\ }\href@noop {} {\bibinfo {title} {{Conditional Generative
  Adversarial Nets}}} (\bibinfo {year} {2014}),\ \Eprint
  {https://arxiv.org/abs/1411.1784} {arXiv:1411.1784} \BibitemShut {NoStop}%
\bibitem [{cga()}]{cgan-qst-python}%
  \BibitemOpen
  \href@noop {} {\bibinfo {title} {{Python code for QST-CGAN}}},\ \bibinfo
  {howpublished} {\url{https://zenodo.org/records/5105470}}\BibitemShut
  {NoStop}%
\bibitem [{\citenamefont {Johansson}\ \emph {et~al.}(2012)\citenamefont
  {Johansson}, \citenamefont {Nation},\ and\ \citenamefont
  {Nori}}]{Johansson2012}%
  \BibitemOpen
  \bibfield  {author} {\bibinfo {author} {\bibfnamefont {J.~R.}\ \bibnamefont
  {Johansson}}, \bibinfo {author} {\bibfnamefont {P.~D.}\ \bibnamefont
  {Nation}},\ and\ \bibinfo {author} {\bibfnamefont {F.}~\bibnamefont {Nori}},\
  }\bibfield  {title} {\bibinfo {title} {{QuTiP: An open-source Python
  framework for the dynamics of open quantum systems}},\ }\href
  {https://doi.org/10.1016/j.cpc.2012.02.021} {\bibfield  {journal} {\bibinfo
  {journal} {Computer Physics Communications}\ }\textbf {\bibinfo {volume}
  {183}},\ \bibinfo {pages} {1760} (\bibinfo {year} {2012})}\BibitemShut
  {NoStop}%
\bibitem [{\citenamefont {Johansson}\ \emph {et~al.}(2013)\citenamefont
  {Johansson}, \citenamefont {Nation},\ and\ \citenamefont
  {Nori}}]{Johansson2013}%
  \BibitemOpen
  \bibfield  {author} {\bibinfo {author} {\bibfnamefont {J.~R.}\ \bibnamefont
  {Johansson}}, \bibinfo {author} {\bibfnamefont {P.~D.}\ \bibnamefont
  {Nation}},\ and\ \bibinfo {author} {\bibfnamefont {F.}~\bibnamefont {Nori}},\
  }\bibfield  {title} {\bibinfo {title} {{QuTiP 2: A Python framework for the
  dynamics of open quantum systems}},\ }\href
  {https://doi.org/10.1016/j.cpc.2012.11.019} {\bibfield  {journal} {\bibinfo
  {journal} {Computer Physics Communications}\ }\textbf {\bibinfo {volume}
  {184}},\ \bibinfo {pages} {1234} (\bibinfo {year} {2013})}\BibitemShut
  {NoStop}%
\bibitem [{\citenamefont {Lambert}\ \emph {et~al.}(2024)\citenamefont
  {Lambert}, \citenamefont {Gigu\`ere}, \citenamefont {Menczel}, \citenamefont
  {Li}, \citenamefont {Hopf}, \citenamefont {Su\'arez}, \citenamefont {Gali},
  \citenamefont {Lishman}, \citenamefont {Gadhvi}, \citenamefont {Agarwal},
  \citenamefont {Galicia}, \citenamefont {Shammah}, \citenamefont {Nation},
  \citenamefont {Johansson}, \citenamefont {Ahmed}, \citenamefont {Cross},
  \citenamefont {Pitchford},\ and\ \citenamefont {Nori}}]{Lambert2024}%
  \BibitemOpen
  \bibfield  {author} {\bibinfo {author} {\bibfnamefont {N.}~\bibnamefont
  {Lambert}}, \bibinfo {author} {\bibfnamefont {E.}~\bibnamefont {Gigu\`ere}},
  \bibinfo {author} {\bibfnamefont {P.}~\bibnamefont {Menczel}}, \bibinfo
  {author} {\bibfnamefont {B.}~\bibnamefont {Li}}, \bibinfo {author}
  {\bibfnamefont {P.}~\bibnamefont {Hopf}}, \bibinfo {author} {\bibfnamefont
  {G.}~\bibnamefont {Su\'arez}}, \bibinfo {author} {\bibfnamefont
  {M.}~\bibnamefont {Gali}}, \bibinfo {author} {\bibfnamefont {J.}~\bibnamefont
  {Lishman}}, \bibinfo {author} {\bibfnamefont {R.}~\bibnamefont {Gadhvi}},
  \bibinfo {author} {\bibfnamefont {R.}~\bibnamefont {Agarwal}}, \bibinfo
  {author} {\bibfnamefont {A.}~\bibnamefont {Galicia}}, \bibinfo {author}
  {\bibfnamefont {N.}~\bibnamefont {Shammah}}, \bibinfo {author} {\bibfnamefont
  {P.}~\bibnamefont {Nation}}, \bibinfo {author} {\bibfnamefont {J.~R.}\
  \bibnamefont {Johansson}}, \bibinfo {author} {\bibfnamefont {S.}~\bibnamefont
  {Ahmed}}, \bibinfo {author} {\bibfnamefont {S.}~\bibnamefont {Cross}},
  \bibinfo {author} {\bibfnamefont {A.}~\bibnamefont {Pitchford}},\ and\
  \bibinfo {author} {\bibfnamefont {F.}~\bibnamefont {Nori}},\ }\href@noop {}
  {\bibinfo {title} {{QuTiP 5: The Quantum Toolbox in Python}}} (\bibinfo
  {year} {2024}),\ \Eprint {https://arxiv.org/abs/2412.04705}
  {arXiv:2412.04705} \BibitemShut {NoStop}%
\bibitem [{\citenamefont {Jozsa}(1994)}]{jozsa-jmp-1994}%
  \BibitemOpen
  \bibfield  {author} {\bibinfo {author} {\bibfnamefont {R.}~\bibnamefont
  {Jozsa}},\ }\bibfield  {title} {\bibinfo {title} {Fidelity for mixed quantum
  states},\ }\href {https://doi.org/10.1080/09500349414552171} {\bibfield
  {journal} {\bibinfo  {journal} {Journal of Modern Optics}\ }\textbf {\bibinfo
  {volume} {41}},\ \bibinfo {pages} {2315} (\bibinfo {year}
  {1994})}\BibitemShut {NoStop}%
\bibitem [{\citenamefont {Yang}\ \emph {et~al.}(2024)\citenamefont {Yang},
  \citenamefont {Khanahmadi}, \citenamefont {Strandberg}, \citenamefont
  {Gaikwad}, \citenamefont {Castillo-Moreno}, \citenamefont {Kockum},
  \citenamefont {Ullah}, \citenamefont {Johansson}, \citenamefont {Eriksson},\
  and\ \citenamefont {Gasparinetti}}]{yang-arxiv-2024}%
  \BibitemOpen
  \bibfield  {author} {\bibinfo {author} {\bibfnamefont {J.}~\bibnamefont
  {Yang}}, \bibinfo {author} {\bibfnamefont {M.}~\bibnamefont {Khanahmadi}},
  \bibinfo {author} {\bibfnamefont {I.}~\bibnamefont {Strandberg}}, \bibinfo
  {author} {\bibfnamefont {A.}~\bibnamefont {Gaikwad}}, \bibinfo {author}
  {\bibfnamefont {C.}~\bibnamefont {Castillo-Moreno}}, \bibinfo {author}
  {\bibfnamefont {A.~F.}\ \bibnamefont {Kockum}}, \bibinfo {author}
  {\bibfnamefont {M.~A.}\ \bibnamefont {Ullah}}, \bibinfo {author}
  {\bibfnamefont {G.}~\bibnamefont {Johansson}}, \bibinfo {author}
  {\bibfnamefont {A.~M.}\ \bibnamefont {Eriksson}},\ and\ \bibinfo {author}
  {\bibfnamefont {S.}~\bibnamefont {Gasparinetti}},\ }\href@noop {} {\bibinfo
  {title} {Deterministic generation of frequency-bin-encoded microwave
  photons}} (\bibinfo {year} {2024}),\ \Eprint
  {https://arxiv.org/abs/2410.23202} {arXiv:2410.23202} \BibitemShut {NoStop}%
\bibitem [{apg()}]{apg-mle-code}%
  \BibitemOpen
  \href@noop {} {\bibinfo {title} {{MATLAB routines for APG-MLE QST}}},\
  \bibinfo {howpublished} {\url{https://github.com/qMLE/qMLE}}\BibitemShut
  {NoStop}%
\bibitem [{\citenamefont {Kingma}(2014)}]{diederik2014adam}%
  \BibitemOpen
  \bibfield  {author} {\bibinfo {author} {\bibfnamefont {D.~P.}\ \bibnamefont
  {Kingma}},\ }\href@noop {} {\bibinfo {title} {{Adam: A method for stochastic
  optimization}}} (\bibinfo {year} {2014}),\ \Eprint
  {https://arxiv.org/abs/1412.6980} {arXiv:1412.6980} \BibitemShut {NoStop}%
\bibitem [{\citenamefont {Hessel}\ \emph {et~al.}(2020)\citenamefont {Hessel},
  \citenamefont {Budden}, \citenamefont {Viola}, \citenamefont {Rosca},
  \citenamefont {Sezener},\ and\ \citenamefont {Hennigan}}]{optax-2020}%
  \BibitemOpen
  \bibfield  {author} {\bibinfo {author} {\bibfnamefont {M.}~\bibnamefont
  {Hessel}}, \bibinfo {author} {\bibfnamefont {D.}~\bibnamefont {Budden}},
  \bibinfo {author} {\bibfnamefont {F.}~\bibnamefont {Viola}}, \bibinfo
  {author} {\bibfnamefont {M.}~\bibnamefont {Rosca}}, \bibinfo {author}
  {\bibfnamefont {E.}~\bibnamefont {Sezener}},\ and\ \bibinfo {author}
  {\bibfnamefont {T.}~\bibnamefont {Hennigan}},\ }\href
  {http://github.com/deepmind/optax} {\bibinfo {title} {Optax: composable
  gradient transformation and optimisation, in jax!}} (\bibinfo {year}
  {2020})\BibitemShut {NoStop}%
\bibitem [{\citenamefont {Bradbury}\ \emph {et~al.}(2018)\citenamefont
  {Bradbury}, \citenamefont {Frostig}, \citenamefont {Hawkins}, \citenamefont
  {Johnson}, \citenamefont {Leary}, \citenamefont {Maclaurin}, \citenamefont
  {Necula}, \citenamefont {Paszke}, \citenamefont {Vander{P}las}, \citenamefont
  {Wanderman-{M}ilne},\ and\ \citenamefont {Zhang}}]{jax-github}%
  \BibitemOpen
  \bibfield  {author} {\bibinfo {author} {\bibfnamefont {J.}~\bibnamefont
  {Bradbury}}, \bibinfo {author} {\bibfnamefont {R.}~\bibnamefont {Frostig}},
  \bibinfo {author} {\bibfnamefont {P.}~\bibnamefont {Hawkins}}, \bibinfo
  {author} {\bibfnamefont {M.~J.}\ \bibnamefont {Johnson}}, \bibinfo {author}
  {\bibfnamefont {C.}~\bibnamefont {Leary}}, \bibinfo {author} {\bibfnamefont
  {D.}~\bibnamefont {Maclaurin}}, \bibinfo {author} {\bibfnamefont
  {G.}~\bibnamefont {Necula}}, \bibinfo {author} {\bibfnamefont
  {A.}~\bibnamefont {Paszke}}, \bibinfo {author} {\bibfnamefont
  {J.}~\bibnamefont {Vander{P}las}}, \bibinfo {author} {\bibfnamefont
  {S.}~\bibnamefont {Wanderman-{M}ilne}},\ and\ \bibinfo {author}
  {\bibfnamefont {Q.}~\bibnamefont {Zhang}},\ }\href
  {http://github.com/jax-ml/jax} {\bibinfo {title} {{JAX}: composable
  transformations of {P}ython+{N}um{P}y programs}} (\bibinfo {year}
  {2018})\BibitemShut {NoStop}%
\end{thebibliography}%

\end{document}